\documentclass[journal]{IEEEtran}

\usepackage{graphicx}
\usepackage{amsmath}
\usepackage{amssymb}
\usepackage[caption=false,font=footnotesize]{subfig}
\usepackage[noadjust]{cite}
\usepackage{float}
\usepackage{algorithm}
\usepackage{bibentry}
\usepackage{balance}
\usepackage{algorithmicx}
\usepackage{algpseudocode}
\usepackage{array}
\usepackage{hyperref}
\hypersetup{
    colorlinks=true,
    linkcolor=blue,
    filecolor=magenta,      
    urlcolor=cyan,
}

\usepackage{booktabs}
\usepackage{threeparttable} 
\usepackage{tabularx}
\usepackage{adjustbox}
\usepackage{dblfloatfix}
\usepackage{makecell} 
\usepackage{verbatim}
\usepackage{url}
\usepackage{pifont}
\usepackage[table,xcdraw]{xcolor}

\newcommand{\cellmark}[3]{%
\begin{tabular}{@{}c@{}}
\textcolor{#1}{\textbf{#2}}\\[-1pt]
\footnotesize (#3)
\end{tabular}
}

\begin{document}

\title{Curated Wireless Datasets for Aerial Network Research}

\author{
Amir Hossein Fahim Raouf,
\thanks{Corresponding author. Corresponding author email: \href{mailto:amirh.fraouf@ieee.org}{amirh.fraouf@ieee.org}}
Donggu Lee,
Mushfiqur Rahman,
Saad Masrur,
Gautham Reddy,
Cole Dickerson,
Md Sharif Hossen,
Sergio Vargas Villar,
An\i l G\"{u}rses,
Simran Singh,
Sung Joon Maeng,
Martins Ezuma,
Christopher Roberts,
Mohamed Rabeek Sarbudeen,
Thomas J. Zajkowski,
Magreth Mushi,
Ozgur Ozdemir,
Ram Asokan,
Ismail Guvenc,
Mihail L. Sichitiu,
and Rudra Dutta%
\thanks{This work was supported in part by the National Science Foundation under Grants CNS-2332835 and CNS-1939334; in part by the Idaho National Laboratory (INL) Laboratory Directed Research and Development (LDRD) Program under BMC No. 264247, Release No. 26 on BEA's Prime Contract No. DE-AC07-05ID14517; and in part by the National Science Foundation Graduate Research Fellowship Program under Grant DGE-2137100.}
\thanks{
Amir Hossein Fahim Raouf, Donggu Lee, Mushfiqur Rahman, Saad Masrur, Gautham Reddy, Cole Dickerson, Md Sharif Hossen, Sergio Vargas Villar, An\i l G\"{u}rses, Simran Singh, Christopher Roberts, Mohamed Rabeek Sarbudeen, Thomas J. Zajkowski, Magreth Mushi, Ozgur Ozdemir, Ismail Guvenc, Mihail L. Sichitiu, and Rudra Dutta are with the Department of Electrical and Computer Engineering, North Carolina State University, Raleigh, NC, USA.}%
\thanks{Sung Joon Maeng is with Hanyang University, Ansan 15588, South Korea.}%
\thanks{Martins Ezuma is with Amazon Lab126, Sunnyvale, CA, USA.}%
\thanks{Ram Asokan is with Wireless Research Center, Wake Forest, NC, USA.}%
}

\maketitle

\begin{abstract}
This Review consolidates publicly available aerial wireless measurement datasets collected using AERPAW. We organize signal-level, power-level, and KPI-level datasets under a unified taxonomy, harmonize metadata, and provide verified access with reproducible post-processing scripts. The curated catalog supports propagation modeling, machine learning, localization, and system-level evaluation for 5G-Advanced and emerging 6G aerial networks.
\end{abstract}

\section{Introduction}\label{sec:intro}
As the global demand for seamless and high-capacity wireless connectivity continues to grow, aerial platforms, particularly unmanned aerial vehicles~(UAVs) have emerged as a promising complement to terrestrial infrastructure in 5G-Advanced and future 6G networks~\cite{geraci2022will, guidotti2022path}. UAVs offer unique opportunities for agile deployment, rapid coverage extension, and spectrum monitoring in disaster-stricken or hard-to-reach environments. Their altitude advantage over terrestrial transmitters and receivers enables line-of-sight~(LoS) links, supporting applications such as aerial base stations~(BSs), edge caching, environmental monitoring, and emergency response~\cite{mozaffari2019tutorial}. Advances in sensing, perception, and decision-making have enabled UAVs with autonomous navigation capabilities to operate with minimal human intervention in complex and dynamic environments.

Recognizing this potential, regulatory and standardization bodies have begun laying the groundwork for the integration of aerial users into cellular networks. For instance, the Federal Communications Commission~(FCC) has examined the use of licensed mid-band spectrum for UAV command-and-control~(C2) links and emphasized the role of 5G in enabling unmanned aircraft systems~(UAS) integration within cellular networks~\cite{fccuas2021, fcc22uas}. In parallel, the 3rd Generation Partnership Project~(3GPP) has introduced UAV-specific enhancements in Release 15 and subsequent releases to address challenges such as interference mitigation, mobility management, and flight-relevant key performance indicators~(KPIs) for aerial user equipments~(UEs)~\cite{3gpp_tr_36_777}. The International Telecommunication Union~(ITU) has similarly recognized the potential of UAVs in advancing global broadband connectivity and enhancing disaster response capabilities~\cite{itu2291}. In addition to these regulatory and standardization efforts, industry stakeholders, e.g.,  Ericsson, Samsung, and Qualcomm, have submitted joint contributions to 3GPP Radio Access Network~(RAN) Working Group 4 (RAN4) aimed at evaluating interference scenarios between aerial and terrestrial users in the shared spectrum bands, highlighting the critical role of empirical measurements in guiding standards development and system design (e.g., see~\cite{3gpp_ran4_uav_contribs}).

This shift toward aerial connectivity has underscored the growing need for open, well-documented, and reproducible wireless datasets captured from airborne platforms. Such datasets serve as a critical enabler for validating theoretical models, training machine learning algorithms, and guiding system-level simulations, protocol design, and deployment studies. Recent airspace-integration efforts further reinforce this need. For example, the Federal Aviation Administration~(FAA) electronic conspicuity white paper outlines near-term pathways that leverage existing terrestrial cellular infrastructure to support UAS functions, such as command-and-control and electronic conspicuity, and highlights multiple open research questions that require operationally representative testing rather than purely simulation-based evaluation~\cite{faa_ec_whitepaper}. Consequently, curated datasets collected in real environments are increasingly necessary to validate propagation assumptions, quantify KPI behavior at altitude, and inform deployment-relevant designs.

Moreover, the availability of publicly releasable aerial wireless datasets remains limited due to the practical challenges of conducting controlled airborne measurements, including wide-area interference considerations, spectrum coordination requirements, and the need for experimental authorizations. These constraints make it difficult to perform repeatable, large-scale measurement campaigns and highlight the importance of dedicated experimental platforms designed to safely and systematically support real-world aerial data collection.

To tackle these challenges, platforms such as the Aerial Experimentation and Research Platform for Advanced Wireless~(AERPAW) have emerged as critical enablers of aerial wireless research. AERPAW integrates a diverse suite of aerial data collection systems, including drones, helikites, and software-programmable radios, equipped with synchronized GPS modules, wideband radio frequency~(RF) sensing capabilities, and Signal Metadata Format~(SigMF)-compliant data logging infrastructure. This multi-modal framework facilitates reproducible and scalable experimentation across a broad range of wireless scenarios. These efforts have produced high-resolution datasets that span I/Q signals, received power, and commercial network KPIs across a wide altitude range, helping researchers model three-dimensional~(3D) propagation environments and evaluate LoS and non-line-of-sight~(NLoS) transitions at scale~\cite{aerpaw_datasets}.

It is emphasized that this manuscript is intended as a dataset descriptor paper. The contribution does not lie in reporting new experimental findings or previously unobserved channel behaviors. Instead, the novelty resides in the systematic integration, harmonization, and documentation of aerial wireless datasets that have already been collected and publicly released. The paper consolidates these datasets under a unified structure, standardizes their metadata and access mechanisms, and provides reproducible post-processing scripts to facilitate reuse by the research community. As such, the primary objective is to lower the barrier to entry for aerial wireless research by making diverse datasets more findable, accessible, interoperable, and reusable.

The remainder of this paper is organized as follows.
The literature review and contributions discuss prior aerial wireless datasets and measurement efforts and identify key gaps motivating this work.
The challenges for generating datasets with aerial wireless systems summarize the technical, logistical, and regulatory challenges of collecting reproducible aerial wireless data.
The integrated dataset catalog and access presents a unified dataset taxonomy and a consolidated catalog with verified access information.
The dataset description sections present the curated AERPAW datasets in a standardized manner, highlighting measurement scope, representative trends, and supported use cases, with detailed file-level information provided in Appendix~\ref{app:file_structure}.
The synthesis, impact, and research outlook consolidates cross-dataset insights, demonstrated research impact, and future research directions.
Finally, the paper concludes with concluding remarks.

\section{Literature Review and Contributions}\label{sec:lit_rev}

There have been only some sporadic efforts on systematically capturing and disseminating aerial cellular datasets, with many studies constrained by narrow deployment scenarios, proprietary data formats, or limited reproducibility. For example, Mozny \textit{et al.} conducted an extensive performance evaluation of Long-Term Evolution~(LTE) and 5G networks for UAV services, highlighting significant degradation in downlink performance at higher altitudes, yet the corresponding data remains unavailable for public use~\cite{mozny2023experimental}. Similarly, Braunfelds \textit{et al.} presented controlled drone flight measurements over a commercial LTE network, analyzing KPIs such as Reference Signal Received Power~(RSRP), Reference Signal Received Quality~(RSRQ), and Signal-to-Interference-plus-Noise Ratio~(SINR) as functions of altitude, but without providing standardized or reusable datasets~\cite{braunfelds2024experimental}.

Other experimental studies have explored software-defined radio~(SDR)-enabled aerial testbeds for deploying LTE BSs on UAVs, demonstrating the feasibility of airborne infrastructure for enhanced connectivity~\cite{zhang2024empowering}. Separately, Ruseno \textit{et al.} analyzed 4G signal quality in the context of UAV Remote ID systems, leveraging machine learning techniques to model signal performance under varying conditions~\cite{ruseno2024analysis}. While these studies contribute valuable architectural and performance insights, they do not offer altitude-resolved, metadata-rich datasets necessary for reproducible research. Furthermore, Zulkifley \textit{et al.} evaluated the feasibility of LTE-based connectivity for small UAVs, showing that increasing altitude leads to substantial degradation in signal quality and increased latency due to non-optimized terrestrial deployments~\cite{zulkifley2021mobile}. In a complementary study, Kov\'acs \textit{et al.} performed aerial measurements over live LTE networks and analyzed interference patterns, highlighting the challenges posed by sidelobe reception and elevated interference levels in UAV operations~\cite{kovacs2017interference}. 

In parallel, studies on terrestrial and regional spectrum occupancy, such as that by Chennamsetty \textit{et al.}, have demonstrated the utility of passive spectrum monitoring across 4G and 5G bands, though such efforts remain disconnected from the aerial domain~\cite{chennamsetty2023real}. While most prior efforts have focused on targeted aerial tests or single-purpose deployments, a few have investigated broader spectrum usage trends under real-world operational conditions. For instance, Kuester \textit{et al.} presented a comprehensive study of radio spectrum occupancy during the COVID-19 pandemic, capturing temporal and spatial variations in 4G and 5G usage across multiple urban and suburban locations using passive monitoring equipment~\cite{kuester2022radio}. Although not UAV-specific, such work underscores the value of large-scale, reproducible spectrum datasets in understanding wireless dynamics and informs the design of future aerial deployments.

Ensuring future progress in aerial connectivity research requires sustained support for measurement platforms that emphasize openness, documentation, and multi-modal sensing capabilities. The community would benefit from initiatives that align with FAIR data principles (\underline{F}indable, \underline{A}ccessible, \underline{I}nteroperable, and \underline{R}eusable)~\cite{wilkinson2016fair}, enabling standardized benchmarking and accelerating advancements toward 5G-Advanced and 6G network deployments. In this work, the presented datasets follow these principles by being findable through publicly indexed repositories with persistent citations, accessible through open download links and accompanying documentation, interoperable via widely used data formats and analysis tools such as MATLAB and Python, and reusable through detailed metadata, experiment descriptions, and provided processing and simulation scripts.

In contrast to prior surveys that focus primarily on analytical models or isolated experimental campaigns, this work complements the state of the art by emphasizing empirical data resources themselves. By situating existing datasets within a unified taxonomy and access framework, the manuscript provides a data-centric perspective on the evolution of aerial wireless research, highlighting trends toward cross-layer measurement, heterogeneous platforms, and reproducible experimentation.

\subsection{Summary of Contributions}\label{subsec:contribution}

The main contributions of this dataset descriptor are summarized as follows:

\begin{itemize}
    \item Integrated dataset taxonomy: A unified categorization of otherwise fragmented aerial wireless datasets into signal-level, power-level, and key performance indicator-level data records, enabling consistent interpretation and cross-dataset comparison across heterogeneous measurement campaigns.

    \item Metadata harmonization: Standardized and consistent documentation of measurement parameters, including platform altitude, geographic context, carrier frequency, bandwidth, sampling configuration, and calibration status, addressing common interoperability and reproducibility limitations in existing public datasets.

    \item Accessible and reproducible data access: Consolidation of publicly available datasets into a single reference with verified access links or persistent repositories, accompanied by open-source post-processing and visualization scripts to support reproducible and transparent reuse.

    \item Validation and usage guidance: Representative data visualizations and analysis examples provided as integrity checks and practical guidance, clarifying intended use cases and reuse potential for propagation modeling, system evaluation, localization, and data-driven research.

    \item FAIR-oriented dataset design: Alignment with the principles of findability, accessibility, interoperability, and reusability to facilitate long-term benchmarking and comparative research within the aerial wireless community.
\end{itemize}


\section{Challenges for Generating Datasets With Aerial Wireless Systems}\label{sec:challenges}

There are substantial difficulties in collecting high-quality datasets for aerial wireless systems. In what follows, we will detail a few challenges that were encountered and overcome by the AERPAW team. Some of the challenges are common across all the platforms for advanced wireless research~(PAWR), while some are unique to AERPAW. 

\subsection{Programmable Radios}

Many of the existing datasets available for aerial wireless systems used commercial off-the-shelf~(COTS) equipment to collect wireless data; for example, wireless phones, or even a small 4G/5G modem connected to a single-board computer~(SBC) like a Raspberry Pi can be used to collect some KPIs like RSRP, RSRQ, and cell ID. However, while light and portable, such a setup can {\em only} collect those KPIs for those particular networking technologies. The situation is similar for Wi-Fi, LoRa, or other COTS equipment.

In contrast, in AERPAW we decided to build our radio system around some of the best SDRs available, namely the universal software radio peripherals~(USRPs) from National Instruments~(NI). The main advantage of a USRP setup is that they can impersonate {\em any} of the technologies that a COTS radio can; furthermore, the USRPs can transmit and receive custom waveforms for which there is no equivalent COTS radio, thus allowing an unprecedented degree of flexibility and programmability, and hence a broad range of wireless experiments.

The main challenge of using URSPs is the relatively large size of the resulting portable node: although the USRPs themselves can be relatively small and light (especially true for the B200 series of USRPs), the {\em supporting} hardware is large and heavy. In order to drive the USRPs a relatively powerful computer needs to be employed (a seventh-generation Intel NUC in our case), filters and power amplifiers on the front ends, and even a custom-made GPS-DO for tight frequency and time synchronization. This increases the size of a portable node by almost an order of magnitude (from a few hundred grams for a COTS portable node to 3.5~kg for a B210 node which requires separate front ends for each of its channels).

\subsection{Outdoor Radio Infrastructure and Spectrum}

Supporting outdoor experiments with drones and wireless technologies requires access to towers equipped with diverse radio systems and to experimental spectrum bands for testing new waveforms and protocols. While some studies and datasets rely on commercial cellular networks, such approaches limit the scope and flexibility of experiments. The AERPAW platform addresses these challenges through the deployment of five towers, two rooftop sites, and one purpose-built light pole to host USRPs and other COTS wireless devices. All fixed nodes are fiber-connected and dedicated exclusively to AERPAW experimentation. Furthermore, AERPAW is designated as one of the four FCC Innovation Zones in the United States, providing access to specific frequency bands for wireless experimentation with drones~\cite{FCC_FAA_Regulations_AERPAW}. Because airborne transmissions from drones can create significant interference on the ground and with incumbent spectrum users, obtaining experimental frequencies is particularly challenging. AERPAW has secured access to the 900~MHz Industrial, Scientific, and Medical~(ISM) band, 1.7/2.1~GHz, and 3.3-3.45~GHz bands to support experiments involving USRPs, commercial 4G/5G equipment, and UAVs, and continues to expand its available experimental spectrum.

\subsection{Programmable Drones}

The immediate consequence of having large portable node based on USRPs is that the drones required to fly the portable nodes have to be much larger than a drone designed to carry portable nodes based on COTSs UEs. In turn, the larger drones are more expensive, and more difficult to design and implement than smaller drones. In AERPAW, we designed and implemented our large drones from first principles. AERPAW could have used COTS drones (or at least COTS frames) for their drones, but instead chose to design and implement them from readily available materials like carbon fiber plates and carbon fiber tubes. The main advantage of this approach is the reproducibility of the AERPAW frames: all the COTS frames we initially considered (including the DJI Matrix 600) are currently discontinued.

Another important choice for the AERPAW drones is the open-source software stack employed: in the interest of a fully programmable drone, the AERPAW vehicle control software is fully open: the autopilot firmware is ArduPilot~\cite{ArduPilot}, and the software is based on the MAVLink open protocol~\cite{MAVLink}. The ground control station~(GCS) used both in development and operations is QGroundControl~\cite{QGroundControl}. This allows for relying on a large base of existing software while developing software that can be reused by other researchers. The software employed by the AERPAW drones allows experimenters both high-level (e.g., preplanned trajectories) as well as low-level (e.g., off-board control) of the AERPAW drones, allowing for highly customized experiments.

\subsection{Development in Digital Twins}
A unique requirement for AERPAW, among all other PAWR platforms, is its digital twin~(DT). In particular, the use of autonomous vehicles in AERPAW and the safety requirements for these vehicles make the development of vehicle software in the testbed itself a major challenge. Instead, for all canonical experiments, all experimenters have to develop their experiments in a custom-made DT of the physical testbed. The AERPAW DT is deployed in the AERPAW data-compute store, which can host several hundred instances of the DT (the exact number depends on the complexity of the experiment instantiated). In the DT, all the software of the real testbed is preserved while simulating three main hardware components of the real testbed: the frames of the drones, the USRPs, and the propagation between the USRPs. The virtual USRPs operate at I/Q sample level, thus allowing for the development of realistic channel models, including antenna patterns, Multiple-Input Multiple-Output~(MIMO) radios, reflections, and Doppler shifts. The drone emulation includes a virtual machine running the same firmware as the drones on the autopilots of the drones in the testbed, resulting in identical responses to commands of both the drones in the testbed and in the DT.

The use of AERPAW DT allows experimenters to develop all their radio and drone software fully remotely, at their own pace, without needing to access any radio or drone hardware. Once the experimenters develop and test their software in the AERPAW DT, they can be deployed quickly in the real-world testbed environment. AERPAW supports a large variety of sample experiments that are tested in the DT~\cite{AERPAW_Sample_Experiments} and can be accessed by experimenters to quickly initiate baseline experiments.  

\subsection{Precise Localization and Roll/Yaw/Pitch Information}

Finally, AERPAW developed an infrastructure that achieves high levels of precision for capturing high-quality datasets. For example, to achieve centimeter-level accuracy, we have deployed a Real‑Time Kinematic~(RTK) BS at one of the fixed nodes, and we performed the precise point positioning~(PPP) procedure, resulting in an accuracy of a few millimeters for this BS. RTK updates are then fed online for each of out vehicles (drones, rovers, and the helikite), ensuring that all collected geographical information is captured with sub-centimeter accuracy. 

Additionally, each fixed node and several portable nodes are equipped with GPS receivers providing both time and frequency corrections to both the USRPs as well as the fixed and portable nodes, allowing for tight time synchronization, which in turn results in testbed-wide synchronized logs. The logs are being generated from multiple sources at each node: the vehicles generate vehicle information (e.g., latitude, longitude, altitude, roll, pitch, yaw, velocities, etc.), the low-level radio software (e.g., srsRAN) generates radio KPIs (e.g., RSRP, RSRQ, I/Q samples), and traffic software generates traffic KPIs (e.g., throughput, delay, error rates). All these logs are time-stamped with a testbed-wide synchronized time-stamp, allowing for coherent post-processing.

\begin{table*}[t]
\centering
\caption{Unified and verified summary of publicly available aerial wireless datasets~\cite{aerpaw_datasets}, including persistent DOIs or access links, data formats, platforms, frequency bands, and representative prior usage to support reproducible research.}
\label{tab:datasets_summary}

\renewcommand{\arraystretch}{0.95}
\setlength{\tabcolsep}{3pt}

\begin{tabularx}{\textwidth}{
  >{\raggedright\arraybackslash}p{3cm}
  >{\raggedright\arraybackslash}p{1.6cm}
  X
  X
  X
  >{\raggedright\arraybackslash}p{1.4cm}
  >{\raggedright\arraybackslash}p{1.5cm}
}
\toprule
\textbf{Name} &
\textbf{DOI / Access Link} &
\textbf{Focus of Dataset} &
\textbf{Data Types and Formats} &
\textbf{Platforms} &
\textbf{Frequency} &
\textbf{Published Papers} \\
\midrule

Wireless I/Q Datasets (Signal-Level) &
\cite{sungJoon_dataport,maeng2023lte_codeocean} &
LTE I/Q sample collection for A2G propagation, channel estimation, spectrum occupancy, and UAV localization &
SigMF raw I/Q samples; GPS logs; antenna pattern metadata &
UAV with NI USRP B205mini; fixed AERPAW LTE eNB &
LTE &
\cite{maeng2023aeriq,maeng2024kriging,maeng2023impact} \\
\addlinespace[1pt]\hline\addlinespace[1pt]

Wireless Spectrum Datasets (Power-Level) &
\cite{hossein2024packapalooza2024,hossein2024aerpaw,maeng2023aerpaw,maeng2023packapalooza,maeng2023packapalooza2023,raouf2025aerpaw} &
Wideband aerial spectrum monitoring for propagation analysis and model calibration &
PSD (dBm); frequency-tagged logs; GPS metadata &
Helikite with dual USRP B205mini-i; Intel NUC &
87\,MHz--6\,GHz &
\cite{raouf2023cellular,raouf2023spectrum} \\
\addlinespace[1pt]\hline\addlinespace[1pt]

5G NSA Wireless KPI Datasets (KPI-Level) &
\cite{asokan2025ericsson5gnsa, singh2023packapaloozadryad,singh2024lakewheelerandroidhorizontalsweeps} &
Aerial LTE/NR KPI measurements on a 5G-NSA network using UAV-mounted nodes &
KPI logs (RSRP, RSRQ, SINR, throughput); GPS and UAV telemetry (CSV/JSON) &
AERPAW SAM UAV; Quectel 5G modem; Ericsson 5G RAN &
n77 &
\cite{asokan2024aerial} \\
\addlinespace[1pt]\hline\addlinespace[1pt]

LoRa Propagation Datasets (Power-Level) &
\cite{sergio2024lorawanperformance} &
Aerial and ground LoRaWAN propagation for IoT coverage and latency analysis &
RSSI/SNR; packet metadata; latency metrics; GPS/IMU logs &
UAV, tethered helikite, ground vehicles; LoRaWAN gateways &
915\,MHz &
--- \\
\addlinespace[1pt]\hline\addlinespace[1pt]

Multipath Propagation Datasets (Signal-Level) &
\cite{channel_sounder_github} &
A2G multipath channel characterization via synchronized channel sounding &
Raw I/Q with metadata; CIRs; path-loss traces; notebooks &
UAV and fixed node with USRP B210; GNSS-disciplined oscillators &
2.4\,GHz &
\cite{gurses_sichitiu_a2g_uav} \\
\addlinespace[1pt]\hline\addlinespace[1pt]

Wireless Localization Datasets (TDOA-Based) (Signal-Level) &
\cite{dickerson2025aerpaw} &
UAV localization and tracking using distributed RF sensors &
TDOA estimates; GPS ground truth; LoS/NLoS labels; error metrics &
Keysight RF sensors; UAV-mounted SDR &
2.4--3.3\,GHz &
\cite{dickerson2025tdoa} \\
\addlinespace[1pt]\hline\addlinespace[1pt]

AFAR RF Source Localization Dataset (Power-Level) &
\cite{gurses2024afar} &
UAV/UGV RF source localization across digital twin and testbed environments &
RSS/RSQ; UAV navigation; CIR-derived metrics; metadata &
UAV and UGV with USRP B205mini; AERPAW digital twin and testbed &
Cellular &
\cite{masrur2025bridging,kudyba2024bayesian,masrur2025collection} \\
\addlinespace[1pt]\hline\addlinespace[1pt]

UAV Signal Classification Datasets (Signal-Level) &
\cite{UAV_RF_Dataset_IEEEPortal} &
RF-based detection and classification of UAV controller signals &
RF captures (MAT); spectrograms; labels; MATLAB scripts &
Keysight oscilloscope; parabolic antenna; RF front-end &
2.4\,GHz &
\cite{UAV_RF_detection_classification,microUAV_RF_Detection} \\
\addlinespace[1pt]\hline\addlinespace[1pt]

UAV Trajectory, RSRP, and Throughput Dataset (KPI-Level) &
\cite{UAVSimFramework2025,Hossen2025Dryad} &
Trajectory-aware KPI emulation and validation for LTE SISO links &
Time-stamped CSV; RSRP; SNR; throughput; UAV telemetry &
UAV LTE SISO; AERPAW digital twin; MATLAB simulation &
LTE &
\cite{Hossen2025} \\
\addlinespace[1pt]\hline\addlinespace[1pt]

Ray-Tracing Measurement Comparison Dataset (Power-Level) &
\cite{Donggu_dryad_dataset} &
Comparison of measured and ray-tracing-simulated RSS along UAV trajectories &
Measured/simulated RSS; path coefficients; delays; MATLAB scripts &
UAV transmitter; fixed receivers; NVIDIA Sionna RT &
3.3\,GHz &
\cite{Donggu_OJVT} \\
\bottomrule
\end{tabularx}

\end{table*}

\section{Integrated Dataset Catalog and Access}\label{sec:dataset_catalog}
This work consolidates multiple aerial wireless datasets that were collected across different campaigns, platforms, and experimental objectives, as summarized in Table~\ref{tab:datasets_summary}. To facilitate integration and reuse, all datasets are organized under a unified taxonomy based on the level of abstraction at which the measurements are recorded.

At the lowest level, signal-level datasets provide raw in-phase and quadrature samples suitable for detailed physical-layer analysis and custom receiver processing. Power-level datasets capture received signal strength and path loss behavior as functions of altitude, distance, and environment. Key performance indicator-level datasets contain higher-layer metrics such as throughput, latency, and packet error rate, enabling system-level evaluation.

Each dataset is accompanied by a standardized metadata record that includes measurement location, altitude range, carrier frequency, bandwidth, antenna configuration, and temporal resolution. Where applicable, data are provided in widely adopted formats to promote interoperability, and conversion scripts are supplied when proprietary or campaign-specific formats are used.

All datasets and associated scripts are hosted in publicly accessible repositories with persistent links. This centralized catalog structure enables users to efficiently identify relevant datasets, understand their characteristics, and integrate them into their own analysis pipelines.

To facilitate informed dataset selection and comparative evaluation, Table~\ref{tab:dataset_comparison_tasks} provides a task-oriented comparison of the surveyed datasets. This comparison highlights practical considerations that emerge from the analysis, including the importance of matching dataset abstraction level to the intended research objective, recognizing trade-offs between raw signal fidelity and system-level scalability, and accounting for dataset-specific limitations such as spatial coverage, technology dependence, and environmental specificity when interpreting results or combining datasets. For instance, we explicitly recommend utilizing Signal-Level datasets for physical layer algorithm development where phase information is critical, while advising the use of KPI-Level datasets for scalable network orchestration studies where storage efficiency and processing speed are paramount.

Building on this unified catalog and comparative perspective, we next turn to the individual datasets in detail. The following sections present a set of aerial wireless datasets collected at the NSF AERPAW, enabled by the experimental capabilities and constraints described earlier.
 For each dataset, we summarize the measurement objectives, hardware and software configuration, dataset format, and representative results. Public repository links and post-processing scripts are provided to support reproducibility and reuse, and selected application examples are included to illustrate relevance to propagation analysis, system evaluation, and data-driven aerial wireless research.

\begin{table*}[t]
\centering
\caption{Task-oriented comparative analysis of aerial wireless datasets, with dataset categories annotated by manuscript section labels to support cross-dataset selection and reuse.}
\label{tab:dataset_comparison_tasks}

\renewcommand{\arraystretch}{0.9}
\setlength{\tabcolsep}{3pt}

\begin{adjustbox}{max width=\linewidth}
\begin{tabularx}{\linewidth}{|
  >{\raggedright\arraybackslash}p{2.2cm}|
  >{\centering\arraybackslash}p{2.7cm}|
  >{\centering\arraybackslash}p{2.7cm}|
  >{\centering\arraybackslash}p{2.1cm}|
  >{\centering\arraybackslash}p{2.6cm}|
  >{\raggedright\arraybackslash}X|}
\hline
\rowcolor{black!10}
\textbf{Dataset Category (Section)} &
\textbf{Propagation / Channel} &
\textbf{ML / Data-Driven} &
\textbf{Localization / Sensing} &
\textbf{KPI / System-Level} &
\textbf{Key Limitations} \\
\hline

Wireless I/Q Datasets  &
\cellmark{black}{\checkmark}{high-fidelity I/Q} &
\cellmark{black}{\checkmark}{feature extraction} &
\cellmark{black}{$\triangle$}{indirect} &
\textcolor{black}{\textbf{$\times$}} &
Large data volume; requires synchronization and advanced signal processing \\
\hline

Wireless Spectrum Datasets  &
\cellmark{black}{\checkmark}{power statistics} &
\cellmark{black}{$\triangle$}{coarse features} &
\textcolor{black}{\textbf{$\times$}} &
\textcolor{black}{\textbf{$\times$}} &
Power-level abstraction; no phase or timing information \\
\hline

Multipath Propagation Datasets  &
\cellmark{black}{\checkmark}{delay/Doppler} &
\cellmark{black}{$\triangle$}{model fitting} &
\textcolor{black}{\textbf{$\times$}} &
\textcolor{black}{\textbf{$\times$}} &
Scenario-specific; limited network-layer metrics \\
\hline

LoRa Propagation Datasets  &
\cellmark{black}{\checkmark}{long-range A2G} &
\cellmark{black}{$\triangle$}{technology-specific} &
\textcolor{black}{\textbf{$\times$}} &
\cellmark{black}{$\triangle$}{IoT metrics} &
Narrowband characteristics limit generalization to broadband systems \\
\hline

Ray-Tracing Comparison Datasets  &
\cellmark{black}{\checkmark}{measurement vs.\ RT} &
\cellmark{black}{$\triangle$}{hybrid modeling} &
\textcolor{black}{\textbf{$\times$}} &
\textcolor{black}{\textbf{$\times$}} &
Environment-specific assumptions; accuracy depends on scene modeling \\
\hline

Wireless Localization Datasets (TDOA-Based)  &
\cellmark{black}{$\triangle$}{signal-derived} &
\cellmark{black}{$\triangle$}{feature learning} &
\cellmark{black}{\checkmark}{primary focus} &
\textcolor{black}{\textbf{$\times$}} &
Localization-oriented; limited general channel characterization \\
\hline

AFAR RF Source Localization Dataset  &
\cellmark{black}{$\triangle$}{signal-derived} &
\cellmark{black}{\checkmark}{supervised learning} &
\cellmark{black}{\checkmark}{primary focus} &
\textcolor{black}{\textbf{$\times$}} &
Challenge-driven scenarios; conclusions depend on task design and environment \\
\hline

UAV Signal Classification Datasets  &
\textcolor{black}{\textbf{$\times$}} &
\cellmark{black}{\checkmark}{RF fingerprinting} &
\textcolor{black}{\textbf{$\times$}} &
\textcolor{black}{\textbf{$\times$}} &
Focused on controller classification rather than propagation analysis \\
\hline

5G NSA Wireless KPI Datasets  &
\cellmark{black}{$\triangle$}{aggregate trends} &
\cellmark{black}{$\triangle$}{KPI learning} &
\textcolor{black}{\textbf{$\times$}} &
\cellmark{black}{\checkmark}{network evaluation} &
No raw signal access; performance depends on operator configuration \\
\hline

Trajectory--RSRP--Throughput Datasets  &
\cellmark{black}{$\triangle$}{link-level trends} &
\cellmark{black}{$\triangle$}{KPI prediction} &
\textcolor{black}{\textbf{$\times$}} &
\cellmark{black}{\checkmark}{throughput analysis} &
Includes emulated components; lacks raw I/Q measurements \\
\hline

\end{tabularx}
\end{adjustbox}

\vspace{1mm}
\textit{Legend:}
\textcolor{black}{\textbf{\checkmark}}\;well suited,\;
\textcolor{black}{\textbf{$\triangle$}}\;partially suited,\;
\textcolor{black}{\textbf{$\times$}}\;not suited.
\end{table*}

\section{Wireless I/Q Dataset}\label{sec:wireless_IQ}

I/Q datasets represent the most fundamental layer of wireless measurements, providing a versatile foundation for supporting research across a wide range of topics. From these raw signals, numerous KPIs can be derived, making them especially valuable for both modeling and experimental studies. In this section, we present I/Q datasets collected from an LTE network at different UAV flight heights over the Lake Wheeler Field Labs.

\subsection{Description of Hardware and Software}

\begin{figure}[!t]
	\centering
        \includegraphics[width=0.8\linewidth]{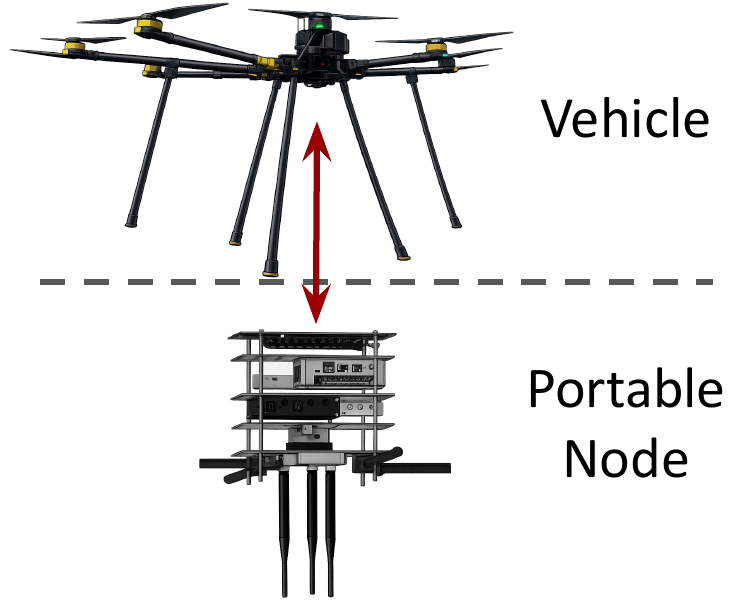}
	\caption{Illustration of the AERPAW large multirotor-type UAV setup for the experiment, where the UAV carries a portable node.}
    \label{fig:UAV_PN}
\end{figure}

The large multirotor-type UAV from AERPAW was used to collect I/Q samples during the experiment. As shown in Fig.~\ref{fig:UAV_PN}, the UAV carries a portable payload that includes an NI USRP B205mini SDR. Python scripts run on the SDR and GPS module to collect I/Q samples at the desired center frequency, sampling rate, and interval, and to record the UAV's location and position. The UAV is equipped with a dipole-type antenna~(SA-1400-5900). Before the experiment, the UAV's flight path, navigation speed, and position were pre-planned by placing waypoints on the map, enabling automatic control of the UAV and repeatable experiments. 
The released dataset includes the recorded I/Q samples and the corresponding GPS logs. The raw data are stored in MATLAB~(.mat) format, while the GPS information is provided as text files~(.txt) to facilitate post-processing.

An AERPAW fixed radio node at the Lake Wheeler Road Field Labs~(LWRFL) site (see Fig.~\ref{fig:geo} and Fig.~\ref{fig:bs}) is configured as an LTE Evolved Node B~(eNB) to transmit the LTE downlink signal. The srsRAN open-source SDR software is used to realize the LTE eNB, where the transmitter antenna gain, center frequency, and number of resource blocks are configurable. A USRP B205mini SDR with a wideband antenna~(RM-WB1) is installed at the fixed radio node.

\subsection{Dataset Format}
\begin{figure}[!t]
    \centering

    \subfloat[Campaign Location]{
        \includegraphics[width=0.48\linewidth,trim={0cm 0cm 0cm 0cm},clip]{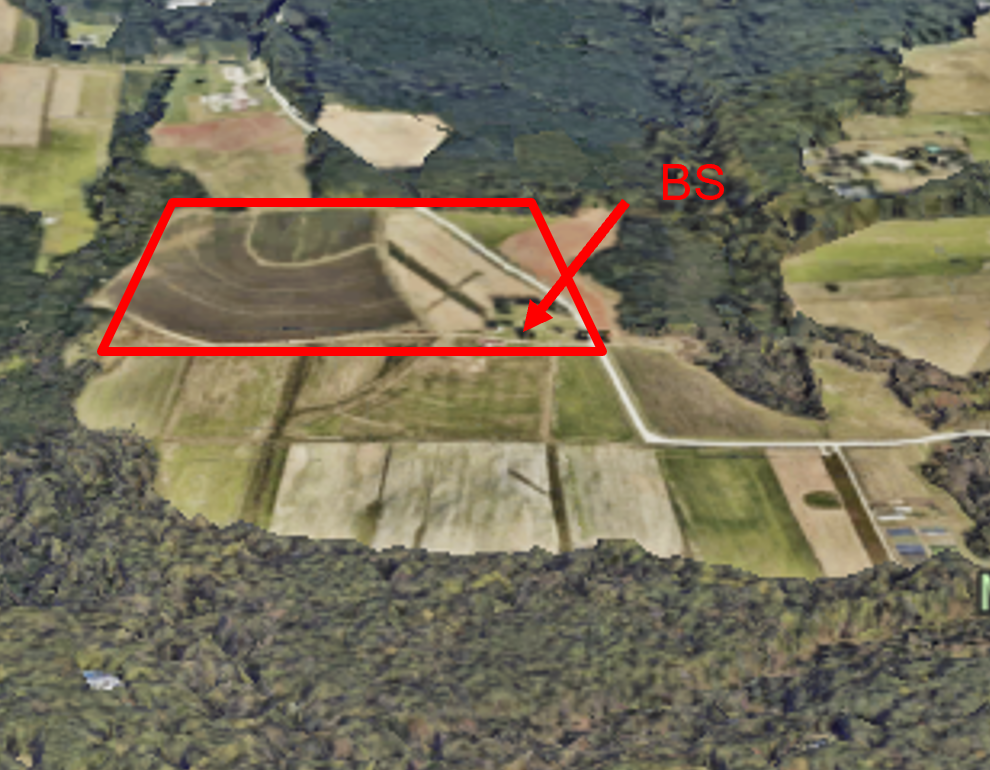}
        \label{fig:geo}
    }
    \subfloat[Base Station]{
        \includegraphics[width=0.48\linewidth,trim={0cm 0cm 0cm 0cm},clip]{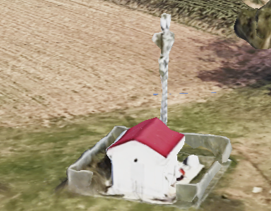}
        \label{fig:bs}
    }

    \vspace{-0.3cm}
    \subfloat[AERPAW UAV]{
        \includegraphics[width=0.48\linewidth,trim={0cm 0cm 0cm 0cm},clip]{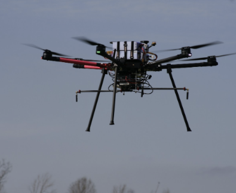}
        \label{fig:uav}
    }
    \subfloat[Trajectory]{
        \includegraphics[width=0.48\linewidth,trim={0cm 0cm 0cm 0cm},clip]{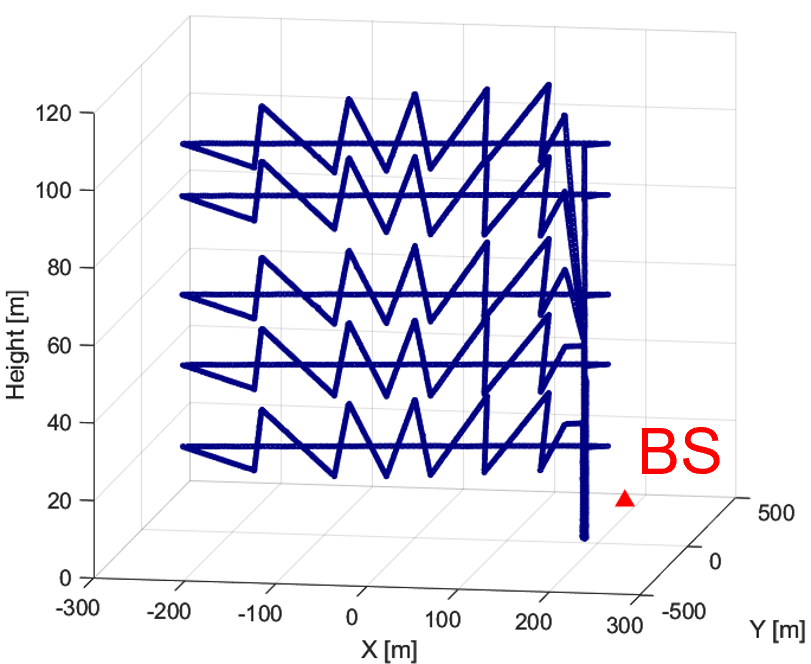}
        \label{fig:traj}
    }
    \hfill
    \caption{Campaign environment and UAV trajectory for the I/Q measurement dataset: \textbf{(a)} Google Earth view of the site, \textbf{(b)} BS or transmitter, \textbf{(c)} pre-planned UAV trajectory, and \textbf{(d)} AERPAW UAV for I/Q signal reception.}
    \label{fig:lte_iq_meas}
    \vspace{-4mm}
\end{figure}

The I/Q measurement campaign was conducted using an LTE base station and a UAV platform (see Fig.~\ref{fig:uav}), where the UAV followed a zigzag trajectory at five fixed altitudes ranging from $30$~m to $110$~m in $20$~m increments, as illustrated in Fig.~\ref{fig:traj}. The resulting wireless I/Q samples and synchronized UAV GPS logs are publicly available on IEEE Dataport~\cite{sungJoon_dataport}, providing a high-fidelity resource for air-to-ground (A2G) propagation analysis, channel estimation, and signal-level modeling.

All measurements are released in SigMF format~\cite{sigmf_spec}, including standardized metadata that captures recording parameters, timing information, and measurement context. The I/Q recordings consist of consecutive $20$~ms snapshots acquired at a $2$~MHz sampling rate, while GPS logs provide time-stamped latitude, longitude, and altitude information at one-second intervals. These data enable precise alignment between signal-level observations and UAV mobility.

For reproducibility and ease of reuse, the dataset is accompanied by documentation and post-processing utilities that support conversion to commonly used analysis formats such as MATLAB and CSV. Detailed directory hierarchy, file naming conventions, and conversion script descriptions are provided in Appendix~\ref{app:file_structure}.

\begin{figure*}[!t]
    \centering
    \subfloat[LTE Resource Grid]{
    \includegraphics[width=0.32\linewidth,trim={0cm 0cm 0cm 0cm},clip]{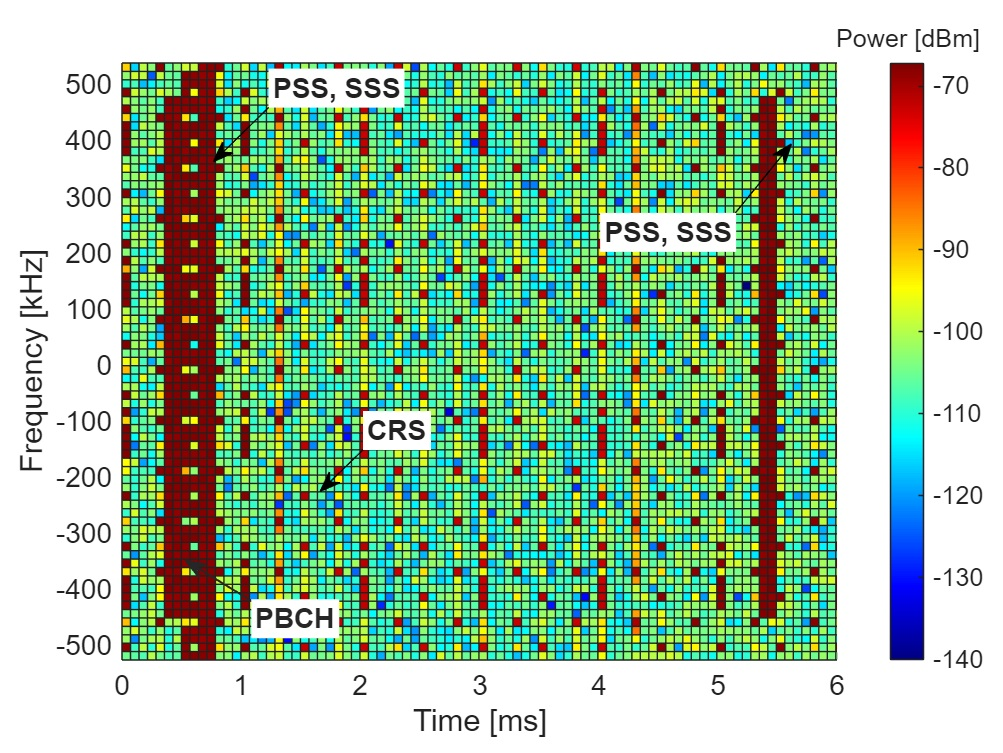}\label{fig:lte_res_grid}
    }
    \subfloat[Estimated time and frequency selective channel.]{
        \includegraphics[width=0.32\linewidth,trim={0cm 0cm 0cm 0cm},clip]{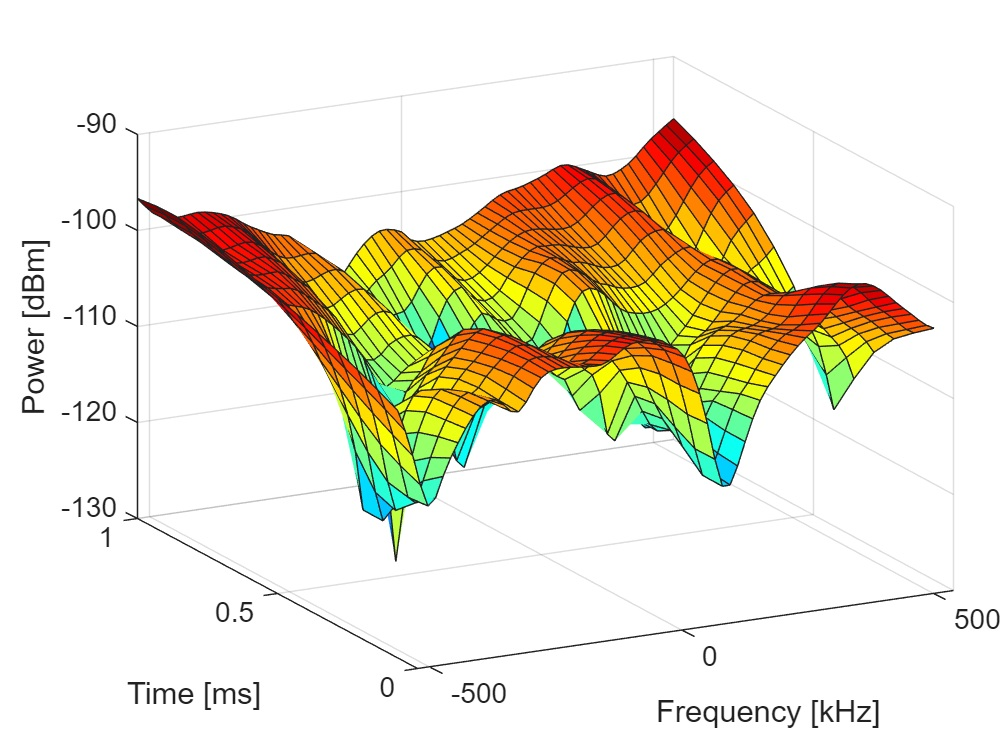}\label{fig:lte_channel}
    }
    \subfloat[RSRP at five altitudes.]{
        \includegraphics[width=0.32\linewidth,trim={0cm 0cm 0cm 0cm},clip]{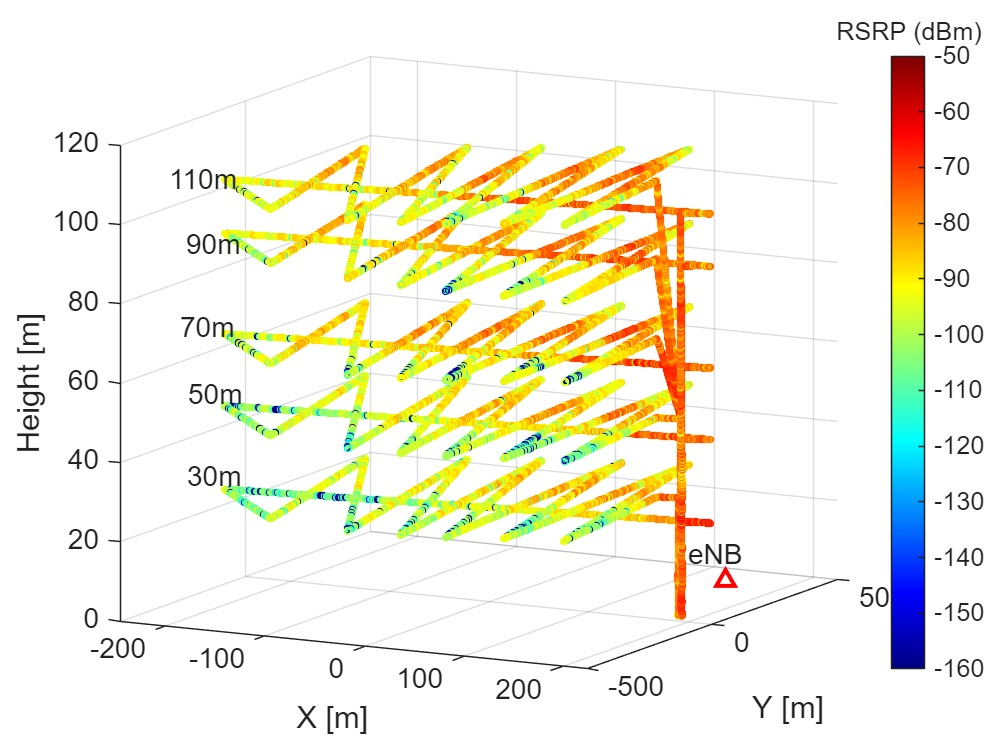}\label{fig:lte_rsrp}
    }
    \vspace{-3mm}\\
    \subfloat[Time vs 3D Distance]{
        \includegraphics[width=0.32\linewidth,trim={0cm 0cm 0cm 0cm},clip]{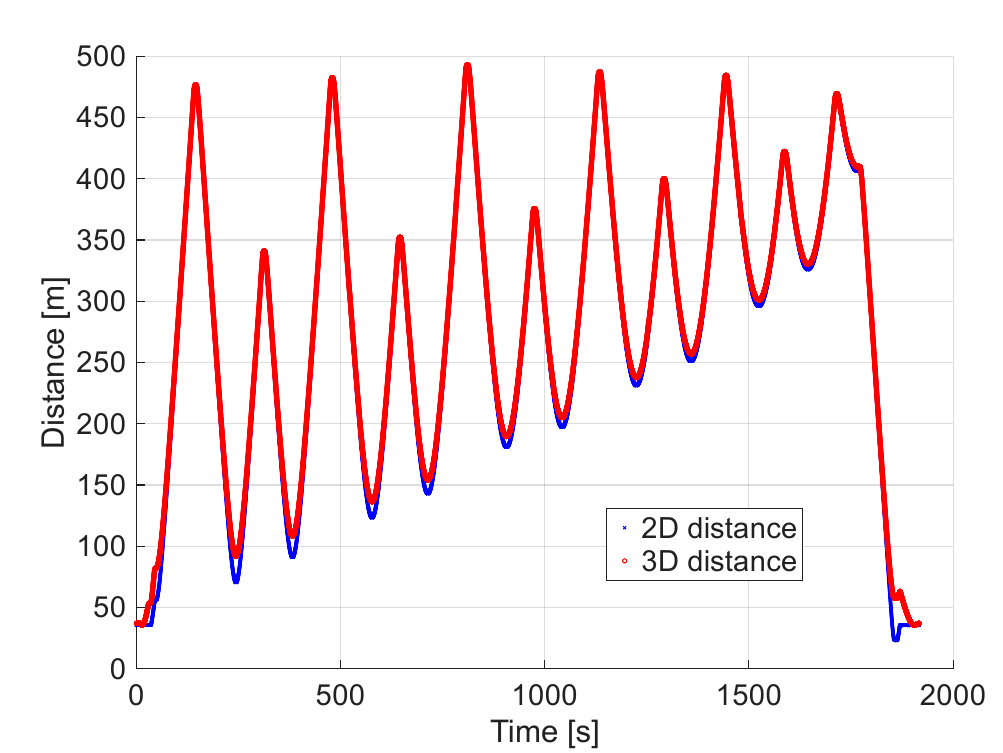}\label{fig:time_distance}
    }
    \subfloat[Time vs UAV Speed]{
        \includegraphics[width=0.32\linewidth,trim={0cm 0cm 0cm 0cm},clip]{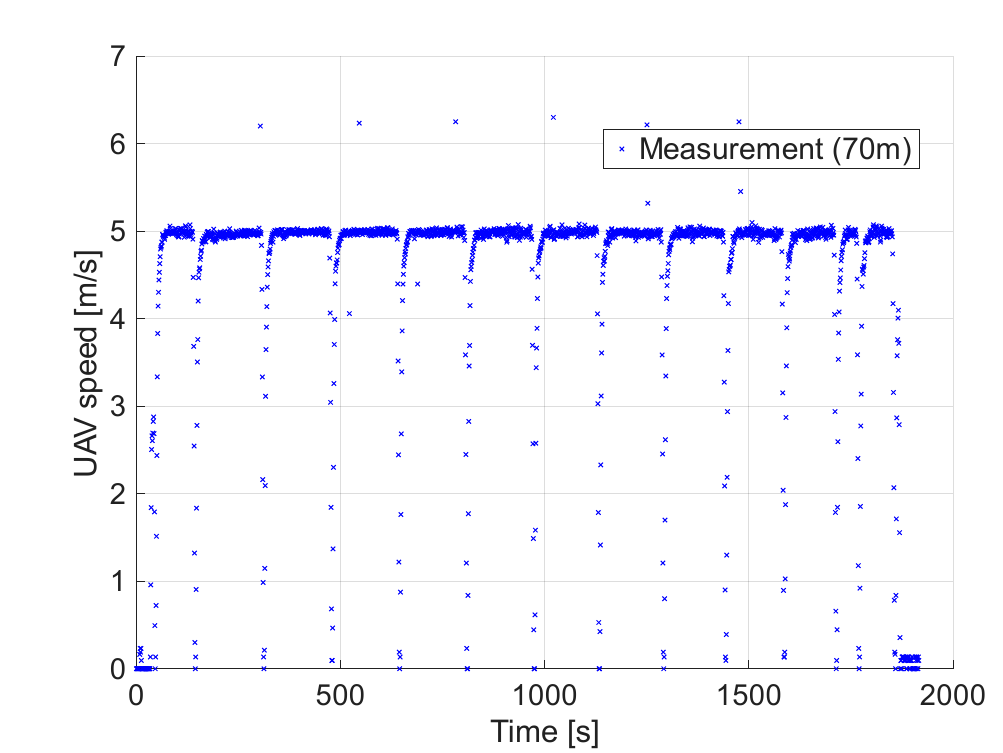}\label{fig:time_speed}
    }
    \subfloat[Measured and modeled RSRP vs. 3D distance.]{
        \includegraphics[width=0.32\linewidth,trim={0cm 0cm 0cm 0cm},clip]{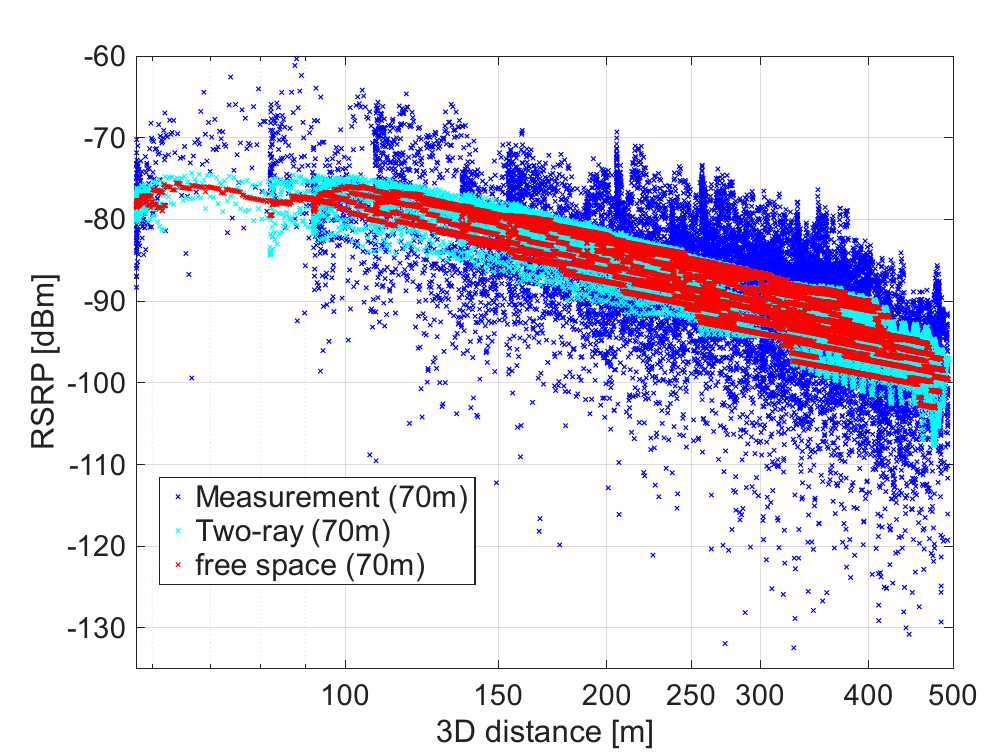}\label{fig:distance_rsrp}
    }

    \caption{Representative results from the Wireless I/Q dataset: \textbf{(a)} LTE resource grid, \textbf{(b)} estimated channel, \textbf{(c)} RSRP along UAV trajectory, \textbf{(d)} time vs. 3D distance (UAV altitude: \(70\,\mathrm{m}\)), \textbf{(e)} time vs. UAV speed (UAV altitude: \(70\,\mathrm{m}\)), and \textbf{(f)} 3D distance vs. RSRP with path loss fitting (UAV altitude: \(70\,\mathrm{m}\)). The figure shows that RSRP decreases as cumulative distance increases, while the speed and distance traces confirm measurements were taken under well-controlled mobility, supporting statistically consistent path-loss and fading analysis.}
    \label{fig:lte_iq_process}
    \vspace{-4mm}
\end{figure*}

\subsection{Representative Results}

In this section, we present several representative results using the published post-processing scripts~\cite{maeng2023lte_codeocean,maeng2023aeriq}.
Since the I/Q samples in this dataset are LTE waveforms and the measurement window is $20$~milliseconds, we can decode the LTE frame start time and extract a full $10$~millisecond LTE frame from each I/Q measurement MATLAB file. We can then plot the LTE resource grid and estimate the channel using the reference signals, as shown in Fig.~\ref{fig:lte_res_grid} and Fig.~\ref{fig:lte_channel}, respectively. From each I/Q measurement file, we can also obtain the RSRP value, and by matching the Unix timestamp with the GPS logs, we can determine the corresponding 3D location of the UAV and plot the RSRP over the UAV trajectory, as shown in Fig.~\ref{fig:lte_rsrp}. Additionally, from the GPS logs, we can plot the 3D distance from the BS to the UAV over time and calculate the UAV's speed by differentiating the position, as shown in Fig.~\ref{fig:time_distance} and Fig.~\ref{fig:time_speed}, respectively. Finally, Fig.~\ref{fig:distance_rsrp} accumulates the GPS logs and RSRP data along the Unix timestamp to plot the RSRP across the 3D distance, where the fitted free-space and two-ray path loss models incorporating 3D antenna patterns are aligned with the measured RSRP behavior.

\subsection{Possible Uses of Dataset}

\noindent\textbf{A2G Propagation Modeling:} 
Artificial intelligence (AI)-based A2G propagation models can be developed using the provided I/Q datasets. AI-based training and testing models can incorporate multiple features, such as the 3D antenna pattern, UAV altitude and position, BS tower height, communication frequency and bandwidth, UAV speed. Note that our dataset~\cite{IEEEDataPort_antenna} includes 3D radiation pattern measurements of both the transmit and receive antennas, obtained in an anechoic chamber.

\noindent\textbf{UAV Receiver Algorithm Design:}
The provided I/Q dataset can facilitate the development of practical time and frequency synchronization algorithms, cell search, channel estimation, and decoding techniques optimized for UAV communication systems.

\noindent\textbf{Spectrum Occupancy and Interpolation:} 
For spectrum sharing and coexistence between terrestrial and aerial networks, UAV-based spectrum monitoring and interpolation techniques have been widely investigated. Our I/Q dataset includes a single zigzag trajectory at multiple altitudes, enabling altitude-dependent spectrum analysis and the study of spectrum interpolation techniques. In our preliminary works~\cite{maeng2024kriging, rahman20243d}, we propose spectrum interpolation approaches based on the 3D Kriging~\cite{maeng2024kriging} and matrix completion~\cite{rahman20243d} using the I/Q dataset.

\noindent\textbf{UAV Localization and Tracking:} 
The detection, localization, and tracking of signal sources by UAVs are key techniques for ensuring privacy and enabling network coexistence. By classifying malicious UAVs or incumbent signal sources, UAV-based systems can enhance situational awareness and support secure and reliable spectrum operations. Our I/Q dataset provides received signal strength~(RSS) measurements along the UAV trajectory across the experiment site, enabling the development and evaluation of source localization and tracking algorithms. In our preliminary work, we propose a UAV localization technique based on the two-ray path loss model, incorporating 3D antenna radiation patterns~\cite{maeng2023impact}.

\section{Wireless Spectrum Dataset}\label{sec:wireless_spectrum}
Empirical spectrum measurements are essential for analyzing wireless channel behavior under practical conditions. Such datasets offer insight into signal variations across frequency, altitude, and environment, and serve as a basis for validating analytical and simulation models. This section describes the measurement setup, including the hardware platform and data collection procedures used to obtain the reported results. Currently, five distinct spectrum measurement datasets are available on the AERPAW dataset page~\cite{aerpaw_datasets} and Dryad~\cite{maeng2023aerpaw, hossein2024aerpaw, maeng2023packapalooza, maeng2023packapalooza2023, hossein2024packapalooza2024}. In this work, we focus on the Packapalooza 2024 dataset~\cite{hossein2024packapalooza2024}; however, the other datasets follow the same file format and directory structure.

\subsection{Description of Hardware and Software}
The measurement setup consists of several components selected to ensure reliable data acquisition and analysis. The main platform is a Helikite equipped with an SDR system. The Helikite provides stable flight at altitudes up to 300 meters, enabling spectrum measurements over a wide area with limited obstruction. The SDR system comprises a USRP device and an antenna, forming the core of the data collection unit.

The portable nodes utilize the NI USRP B205mini-i~\cite{ettusB205mini}, the smallest USRP featuring a Xilinx Spartan 6 XC6SLX150 FPGA~\cite{amdSpartan6} and an Analog Devices AD9361 RFIC direct-conversion transceiver~\cite{adAD9361}. This device supports frequencies from 70~MHz to 6~GHz and offers up to 56~MHz of instantaneous bandwidth\footnote{Instantaneous bandwidth is the maximum width of a frequency band that the device can receive or transmit without retuning.} (61.44~MS/s quadrature) for full-duplex operation. As the core of our SDR platform, the USRP B205mini-i provides high sensitivity and selectivity across most commercial wireless bands, including LTE, 5G NR, and ISM.

\begin{figure}[!t]
	\centering
        \subfloat[Setup of the portable node on the helikite.
        ]{\includegraphics[width=0.8\linewidth]{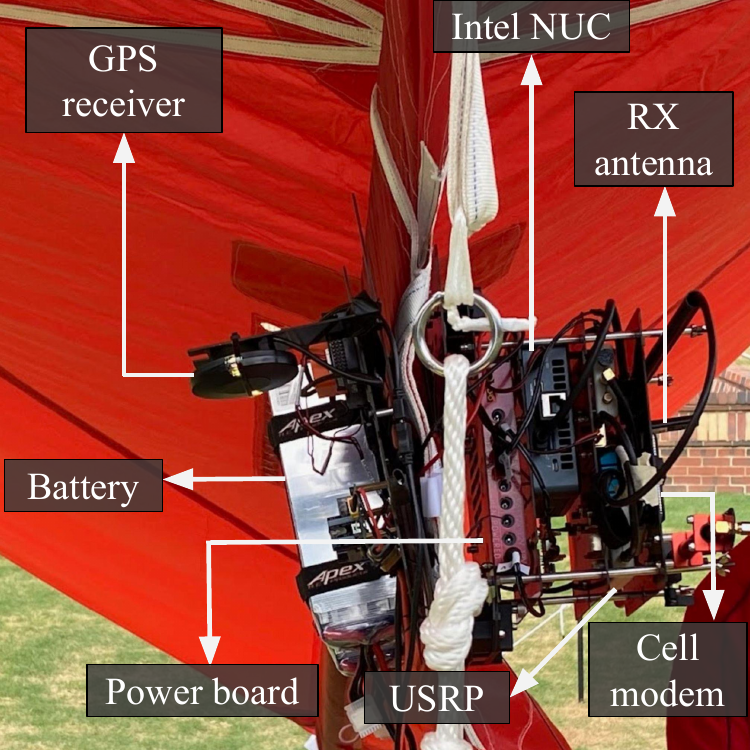}\label{fig:helikite}}
        \vspace{-2mm}
        \subfloat[Spectrum sweep procedure.
        ]{\includegraphics[width=0.98\linewidth]{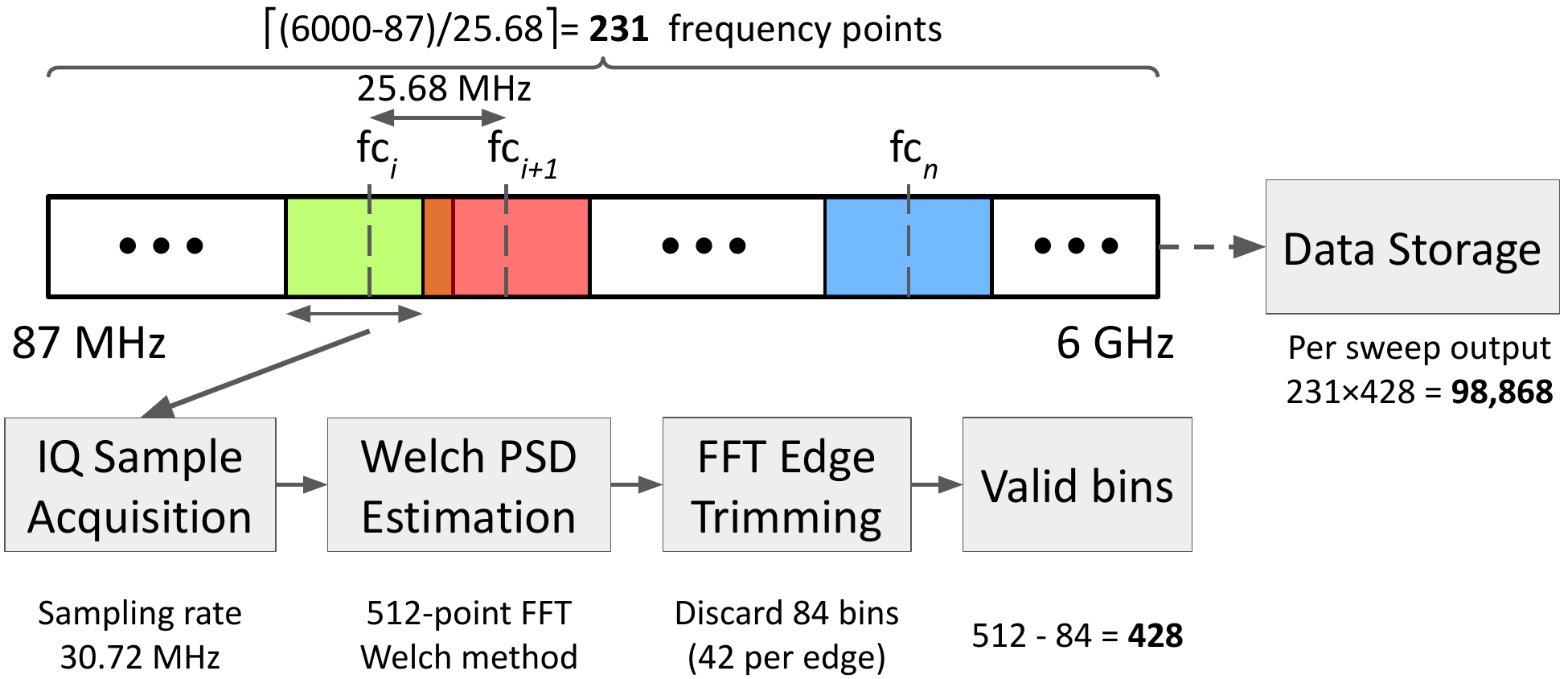}\label{fig:sweep}}
        \vspace{-2mm}
	\caption{Measurement setup and procedure for spectrum data collection using the helikite-mounted portable node.
\textbf{(a)} Experimental setup of the portable node on the tethered helikite.
\textbf{(b)} Spectrum sweep procedure. The procedure indicates that spectrum measurements are collected using repeatable sweeps under fixed hardware settings, making the resulting power statistics suitable for comparative altitude- and time-based analysis.}
\label{fig:helikite_sweep}\vspace{-0.1in}
\end{figure}

Fig.~\ref{fig:helikite} illustrates the helikite-mounted measurement configuration, including the adapter housing the battery, GPS logger, and portable node. The GPS logger records the platform location using information from the vehicle autopilot while the platform is externally controlled.
The GPS logger is connected via a USB cable to a companion computer~(Intel NUC) housed within the portable node. The Intel NUC manages two B205mini-i SDRs; one is equipped with a 3.5~GHz front end while the other operates without a specific front end.

The portable node is designed to operate on a 19~V supply from the battery, which powers not only the companion computer but also the active components of the front end, such as the Low Noise Amplifier~(LNA). Both USRPs are connected to identical but separate receive antennas and are configured only to receive signals; there is no transmission functionality in this setup. 
The measured data for each sweep requires approximately 15 seconds to be stored. The spectrum sweep procedure used in the experiments is depicted in Fig.~\ref{fig:sweep}, where the center frequency shift and sampling rate are 25.68~MHz and 30.72~MHz, respectively.

The primary software components consist of custom Python scripts that automate data collection and initial processing stages, ensuring consistency and efficiency in long-duration measurement campaigns. For additional information regarding the spectrum monitoring experiment, please refer to~\cite{AERPAW_spec_measure}.

\subsection{Dataset Format}
The wireless spectrum measurement dataset is publicly available through the Dryad Digital Repository and is designed to support wideband spectrum analysis, propagation studies, and data-driven wireless research. The dataset consists of time-stamped power spectral measurements collected during aerial experiments, together with synchronized UAV location information.

All spectrum and positioning data are released in SigMF format, accompanied by standardized metadata that document the measurement methodology, equipment configuration, frequency settings, and acquisition context. Power spectrum records provide frequency-domain power measurements expressed in dBm over the monitored bandwidth, while GPS logs include time-aligned latitude, longitude, altitude, and Unix timestamp information. This structure enables precise association between spectrum observations and UAV mobility.

To facilitate reproducible analysis and interoperability with commonly used signal processing tools, the dataset includes supporting documentation and utilities for conversion into MATLAB and CSV formats. Detailed directory hierarchy, file naming conventions, and conversion script descriptions are provided in Appendix~\ref{app:file_structure}.

\subsection{Representative Results}
This section presents representative results from spectrum monitoring experiments conducted in both urban and rural environments. Utilizing advanced aerial platforms such as helikites, these experiments offer valuable insights into how environmental factors and topographical features influence wireless signal propagation and distribution. By comparing the results from densely populated urban areas during the Packapalooza event with those from the more open and sparse rural areas near Lake Wheeler, we aim to highlight the distinct challenges and dynamics encountered in different settings.

\begin{figure}[!t]
	\centering
        \subfloat[Packapalooza 2024.
        ]{\includegraphics[width=0.98\linewidth]{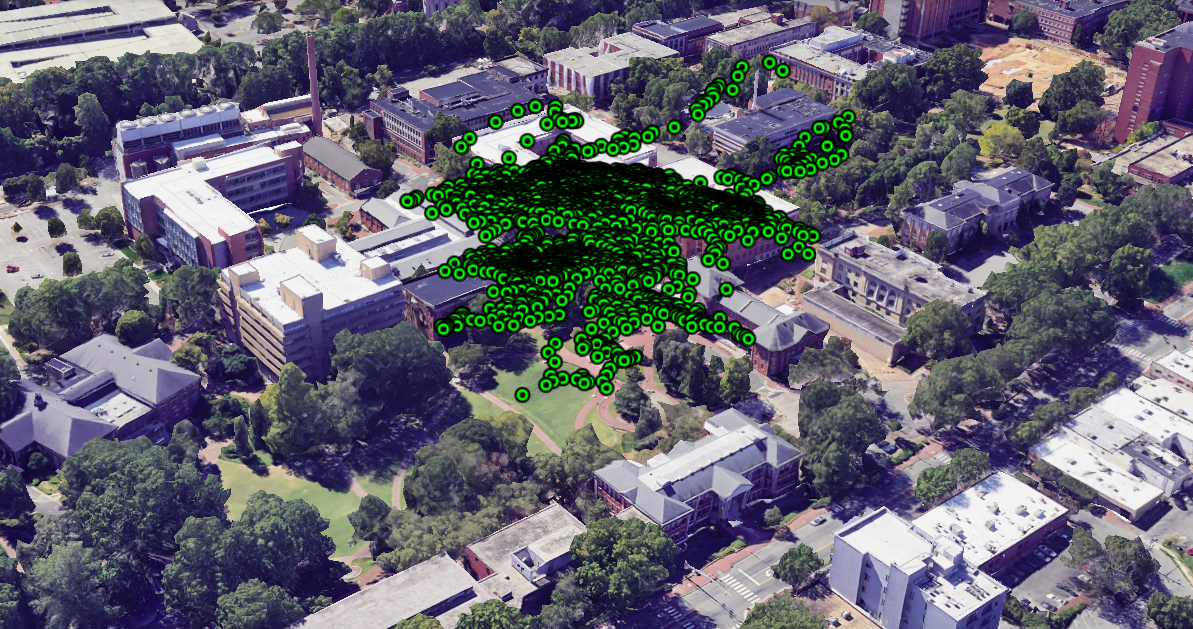}\label{fig:helikite_loc_3d}}
        \vspace{-0.02in}
        \subfloat[Lake Wheeler 2024.
        ]{\includegraphics[width=0.98\linewidth]{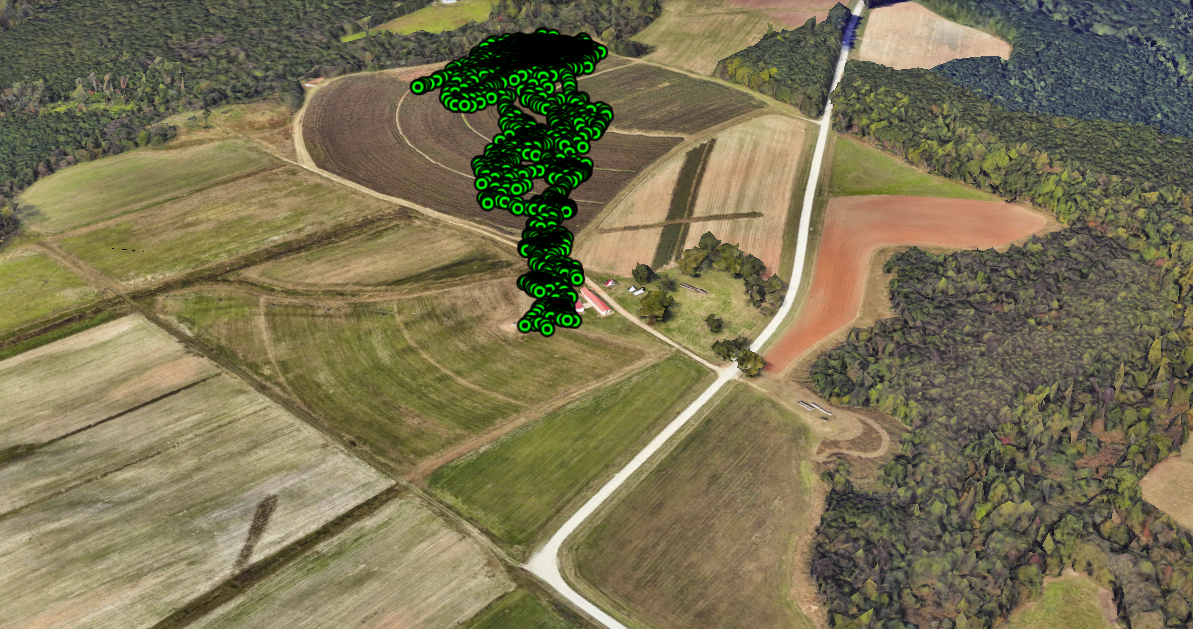}\label{fig:helikite_loc_3d_wheeler}}
        \vspace{-0.01in}
	\caption{Helikite location for spectrum measurements in \textbf{(a)} Packapalooza 2024 and \textbf{(b)} Lake Wheeler 2024. The two locations illustrate that the dataset captures environment-dependent spectrum statistics, enabling practical comparison between less congested and more interference-prone RF environments.}
\label{fig:3d_loc_map}\vspace{-0.1in}
\end{figure}

\begin{figure}[!t]
	\centering
        \subfloat[Packapalooza 2024.
        ]{\includegraphics[width=0.98\linewidth]{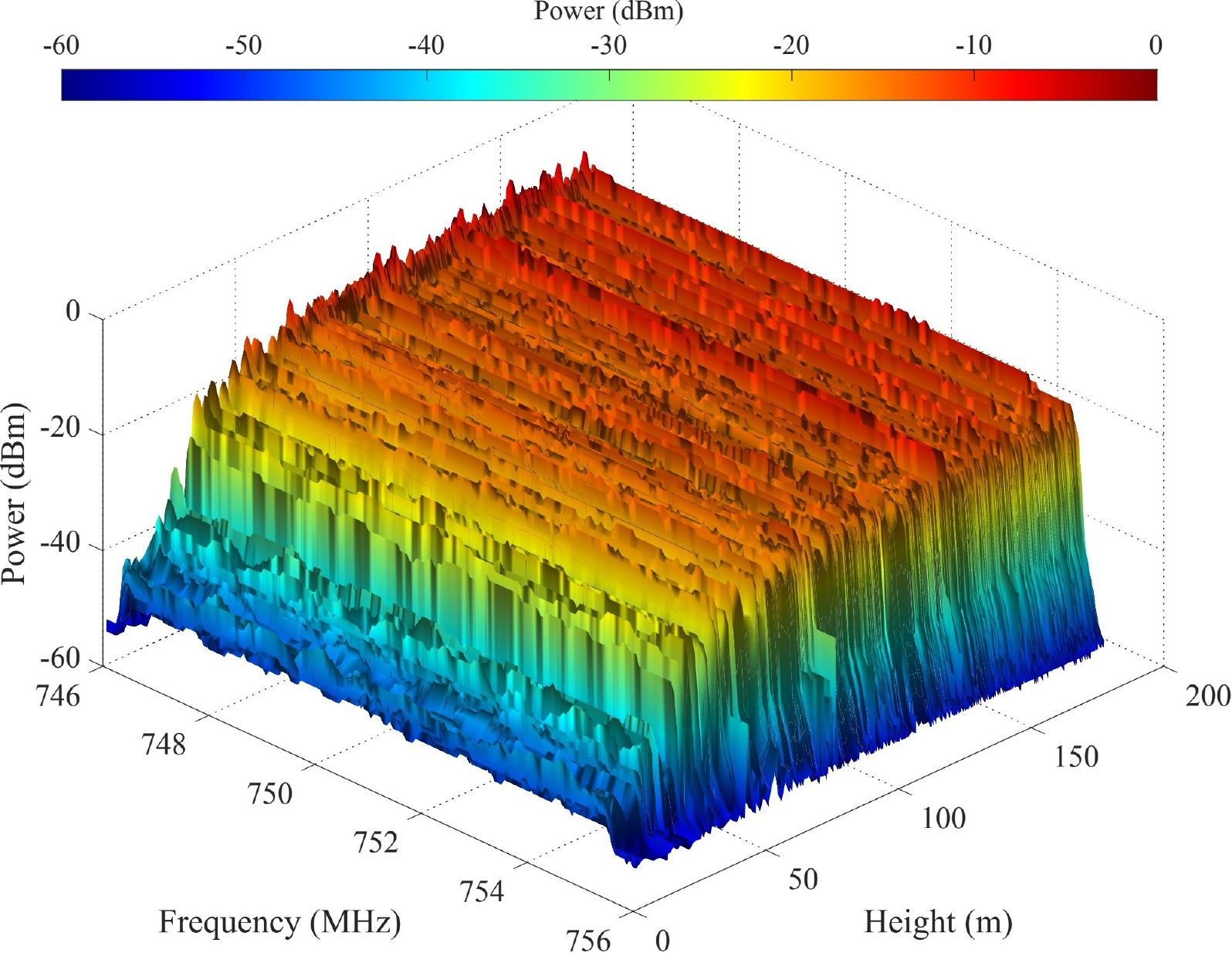}\label{fig:dl_13_pow_vs_alt_pack24}}
        \vspace{-0.02in}
        \subfloat[Lake Wheeler 2024.
        ]{\includegraphics[width=0.98\linewidth]{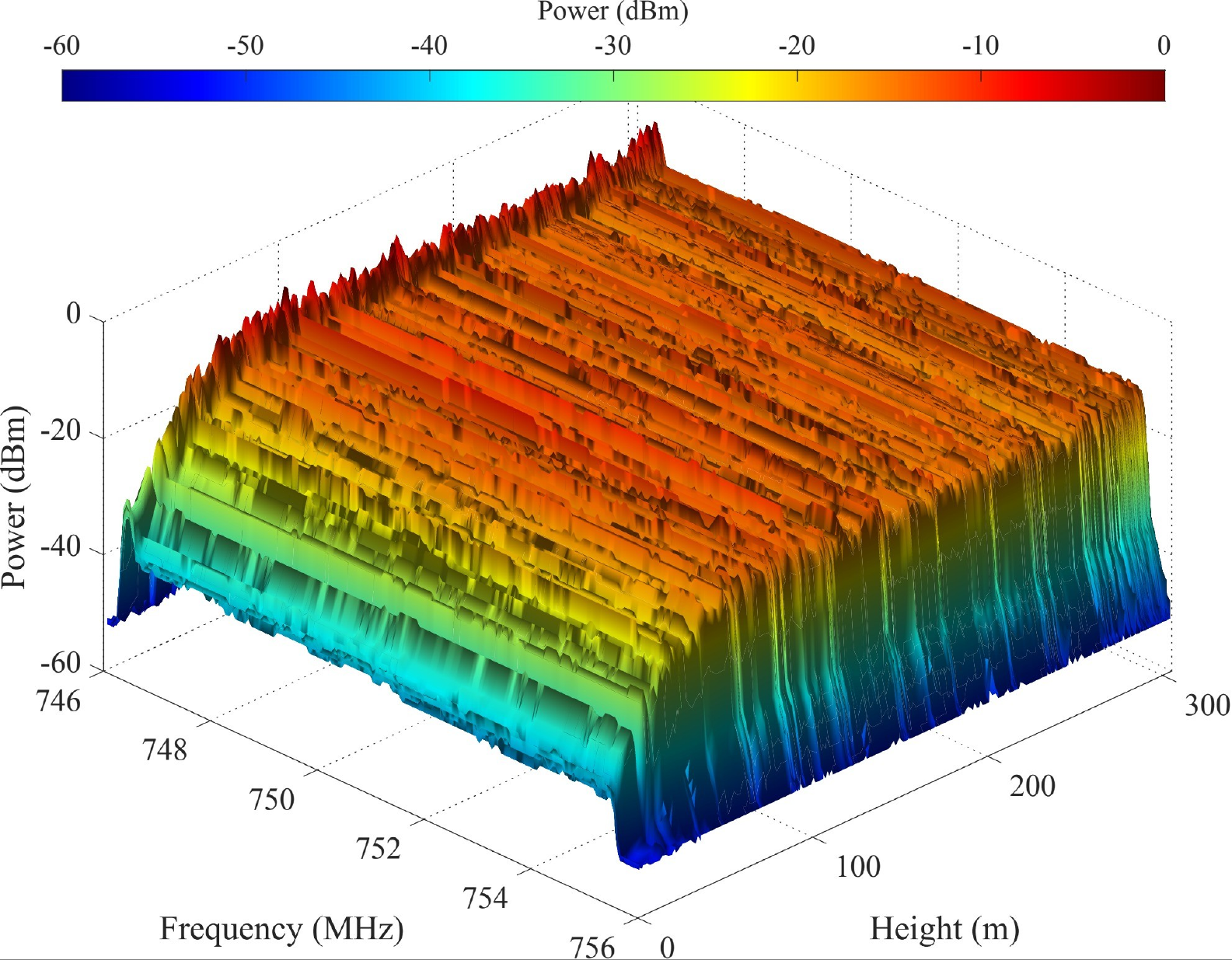}\label{fig:dl_13_pow_vs_alt_wheeler24}}
        \vspace{-0.01in}
	\caption{Altitude dependent power in the $746-756$ spectrum (LTE DL Band-13) in \textbf{(a)} Packapalooza 2024 and \textbf{(b)} Lake Wheeler 2024. The plot shows systematic power fluctuations with altitude, indicating that elevation is a statistically significant factor in received spectrum power for aerial platforms.}
\label{fig:dl_13_pow_vs_alt}\vspace{-0.1in}
\end{figure}

\begin{figure}[!t]
    \centering     
    {\includegraphics[width=0.4\textwidth]{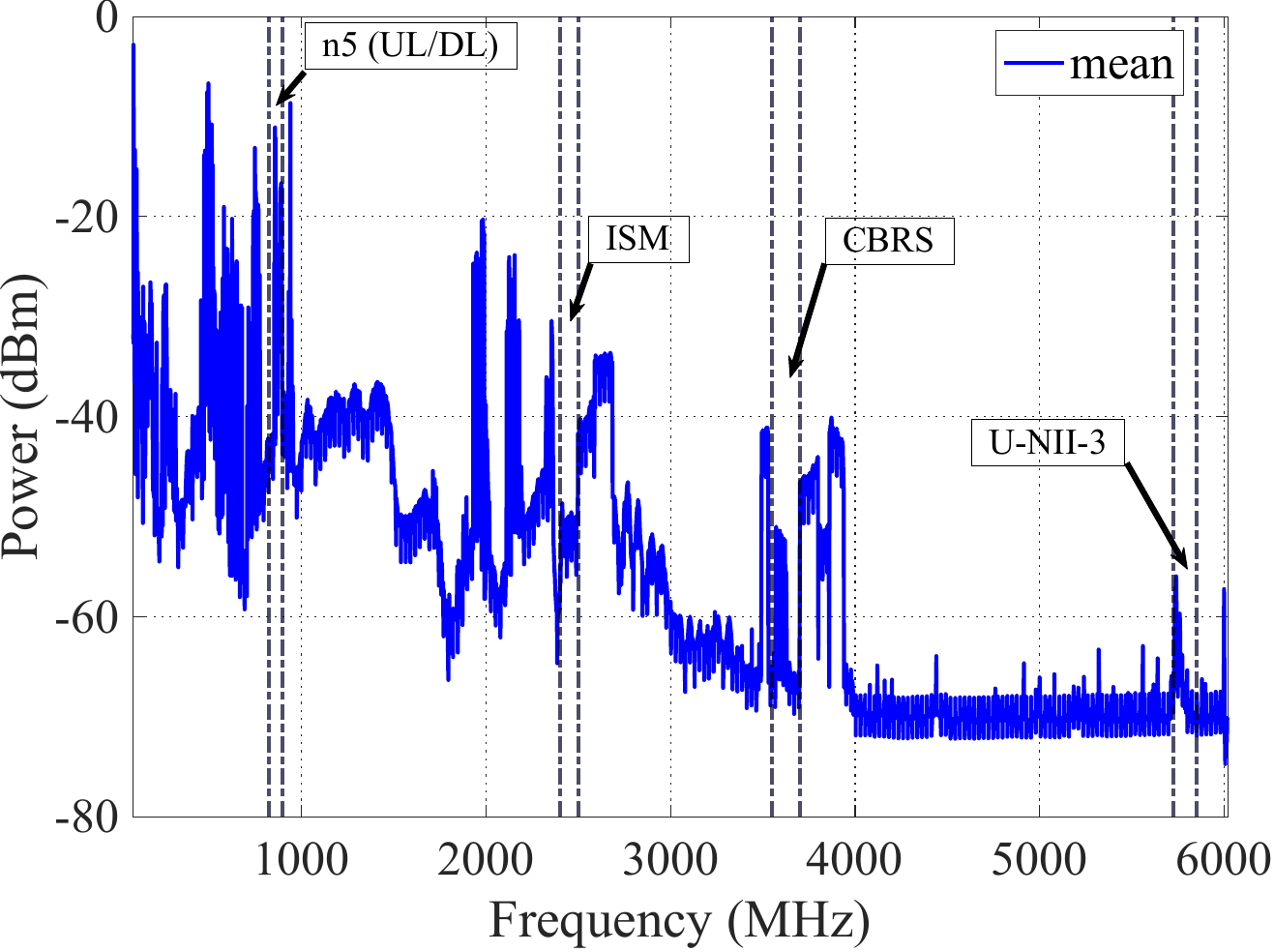}}
	\caption{Representative results on the mean received power versus frequency averaged across all altitudes for the Packapalooza 2024 dataset.}
\label{fig:mean_freq}
\end{figure}

\noindent\textbf{Helikite Trajectory Analysis:}
In urban settings during the Packapalooza event, the helikite's trajectory, as depicted in Fig.~\ref{fig:helikite_loc_3d}, showcases its path above a densely populated area with significant deviations caused by complex wind interactions with urban structures. This erratic movement, indicated by a red trace, potentially affects spectrum measurements due to variable altitudes and obstructions. Conversely, in the rural landscape near Lake Wheeler as shown in Fig.~\ref{fig:helikite_loc_3d_wheeler}, the helikite exhibits a more stable and elongated flight path across open fields, suggesting more consistent data collection due to fewer obstructions and a steadier altitude control.

\noindent\textbf{Spectrum Analysis:}
In Fig.~\ref{fig:dl_13_pow_vs_alt}, the 3D plots for Band 13 downlink (746 - 756~MHz) reveal distinct variations in signal power across different environments. In urban settings, Fig.~\ref{fig:dl_13_pow_vs_alt_pack24}, power levels fluctuate significantly with altitude due to multipath effects and obstructions, showing a trend toward stabilization as altitude increases. In contrast, rural settings in Fig.~\ref{fig:dl_13_pow_vs_alt_wheeler24} display a more uniform increase in power levels at higher altitudes, indicating clearer signal paths and fewer obstructions. Fig.~\ref{fig:mean_freq} illustrates the mean received power as a function of frequency across altitudes for the Packapalooza 2024 dataset, with several active United States radio bands also indicated. The results show that the mean received power is significantly higher below 1~GHz, where many LTE and NR network bands are located, compared to the remainder of the sub-6~GHz spectrum.

\subsection{Possible Uses of Dataset}
Our helikite-based spectrum monitoring dataset provides calibrated received power measurements across a wide frequency range (87~MHz--6~GHz) with corresponding GPS coordinates, altitude, and timestamp. While it does not include raw I/Q samples or power spectral density~(PSD) estimates, the dataset remains valuable for many practical wireless research and regulatory applications.

\noindent\textbf{Spectrum Allocation Analysis:}
The dataset enables spatial and altitudinal characterization of spectrum utilization across urban and rural environments. For example, comparisons of received power in the Citizens Broadband Radio Service~(CBRS) and Television White Spaces~(TVWS) bands can reveal underutilized areas or high-demand regions. In urban settings, it allows researchers to measure outdoor signal levels in the 6~GHz unlicensed band to assess potential interference from indoor Wi-Fi 6E deployments. The availability of time-stamped data also permits the exploration of temporal usage trends, such as peak usage periods or band-specific congestion.

\noindent\textbf{Calibration of Analytical Models:}
The dataset supports calibration of analytical and simulation models, including stochastic geometry and empirical path loss frameworks. The relationship between received power and altitude can be used to validate altitude-aware propagation assumptions. Differences observed between rural and urban measurements can help refine clutter loss models. In addition, LoS probability models can be empirically evaluated using elevation-dependent signal trends.

\noindent\textbf{Propagation Model Tuning:}
Researchers can use the dataset to develop or refine radio propagation models. The location-tagged power measurements support construction of empirical path loss curves for a range of frequency bands. Comparing measurements taken in urban versus rural environments helps to characterize the impact of buildings, vegetation, and other obstructions. Moreover, differences between low-frequency bands like FM and higher bands like 3.5~GHz can be used to study frequency-dependent attenuation.

\noindent\textbf{Machine Learning for Signal Estimation:}
The dataset is suitable for developing machine learning models that estimate received power from spatial and environmental features. Inputs such as latitude, longitude, altitude, and frequency can be used to train regressors for power prediction. The labeled nature of rural and urban environments supports classification tasks, such as identifying the type of environment based on observed signal levels. The data can also help delineate signal boundaries or approximate coverage maps through supervised learning.

\noindent\textbf{Anomaly and Interference Detection:}
The received power measurements allow for basic anomaly detection techniques. Sudden spikes or dips in power levels may indicate unauthorized transmissions or interference events. Statistical properties such as skewness or variance can be used to detect deviations from normal signal patterns. When the data is tracked over time, researchers can analyze signal disruptions or temporal anomalies in specific frequency bands.

\noindent\textbf{Interpolation and Coverage Mapping:}
Despite the absence of I/Q or PSD data, the dataset is well-suited for generating radio environment maps. The geolocation and altitude information associated with each measurement can be used for spatial interpolation techniques such as Kriging or inverse distance weighting. By interpolating the power values, researchers can generate two-dimensional~(2D) or 3D signal coverage maps for individual bands. The multi-band nature of the dataset further allows for frequency-aware coverage visualizations across the monitored spectrum.

\section{5G NSA Wireless KPI Dataset}\label{sec:5G_KPI}

\begin{figure*}
    \centering
    \includegraphics[width=0.9\linewidth]{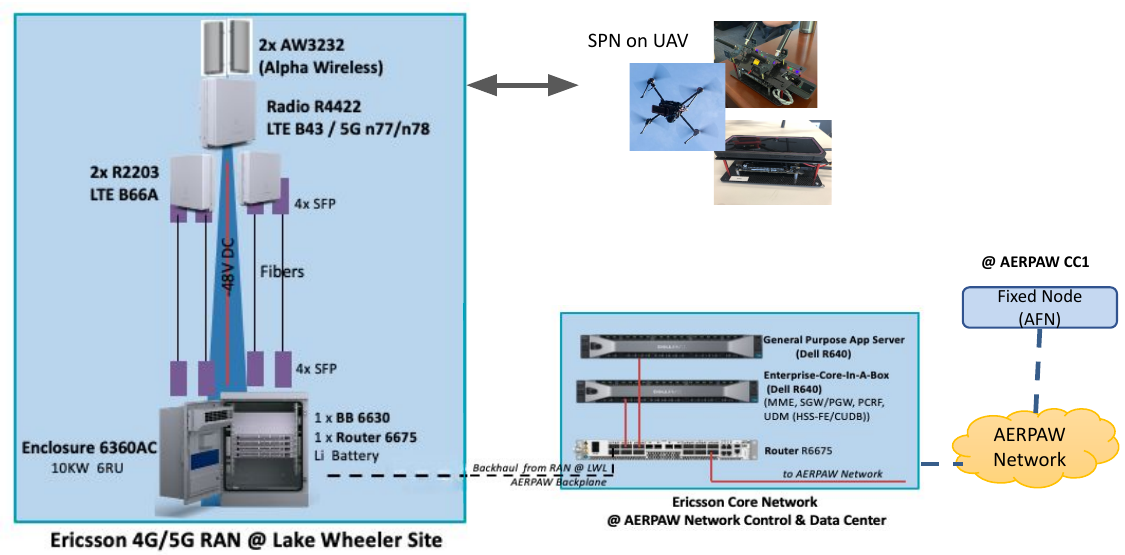}
   \vspace{2mm}
    \caption{5G-NSA AERPAW Infrastructure using Ericsson network.}
   \label{fig:Ericsson_HW}
\end{figure*}

As the demand for using advanced communications to support
various use cases of UAVs rapidly increases, understanding the
performance of 5G terrestrial networks in the 3D spatial domain
becomes critical. In this section, we present datasets for 5G wireless KPIs and the corresponding experimental results from aerial measurements using a 5G-enabled UAV on a 5G non-standalone (5G-NSA) network in C-band in the AERPAW testbed. In particular, the datasets provide the field-measured RF and physical~(PHY) layer parameters of LTE and NR carriers on the 5G-NSA network using three different types of small portable nodes~(SPN) based on Quectel 5G modem, a modified Android phone with Nemo software, or a COTS Android device with PawPrints, a custom App that relies on open-source Android application programming interfaces~(APIs).

\subsection{Description of Hardware and Software}

AERPAW provides multiple wireless radio access platforms for 4G and 5G experimentation, including a commercial-grade Ericsson 5G network that serves as the primary infrastructure for these datasets.
AERPAW infrastructure for these experiments involves a) Ericsson 5G network with RAN and Core, b) SPN, c) AERPAW fixed node~(AFN) as application server of user plane traffic, and d) the UAV to carry the SPN during aerial experimentation (Fig.~\ref{fig:Ericsson_HW}). The RAN of this 5G system is deployed at the AERPAW LWRFL, which is a rural agricultural area (see Fig.~\ref{fig:bs}), predominantly an open aerial field with some vegetation on the ground.

To characterize the aerial performance of a 5G system, we used a 5G NSA system with overlaid NR and LTE sectors. The LTE anchor carrier is in band $66$~($1.7/2.1$~GHz) with $5$~MHz channel bandwidth and an NR carrier at $3.4$~GHz in band n77 with $100$~MHz of channel bandwidth. The sectors use a pair of dual +/-45 deg cross polarized directional antennas with $120$~degrees of azimuth beam width facing the north-west direction from the BS tower. LTE employs 2×2 MIMO on the downlink, whereas NR uses 4×4 MIMO. For this experiment, both LTE and NR carriers are set at 5 watts of transmit power per antenna port. 

\begin{figure}[t!]
    \centering
    \subfloat[SAM mounted with portable node (SPN)]{
        \includegraphics[width=0.9\linewidth]{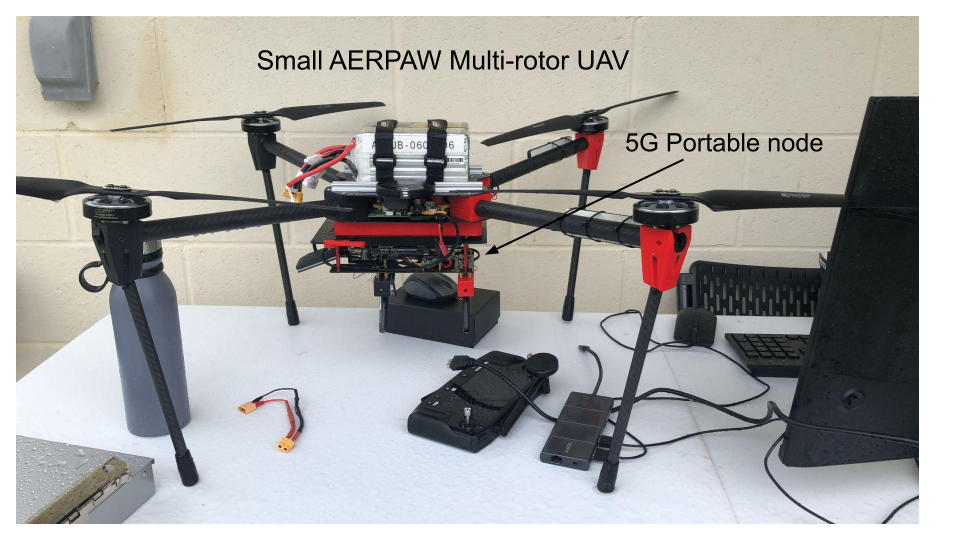}
        \label{fig:sam}}    
    \vspace{-0.02in}
    \subfloat[Small portable node (SPN) with Quectel 5G modem]{
        \includegraphics[width=0.9\linewidth]{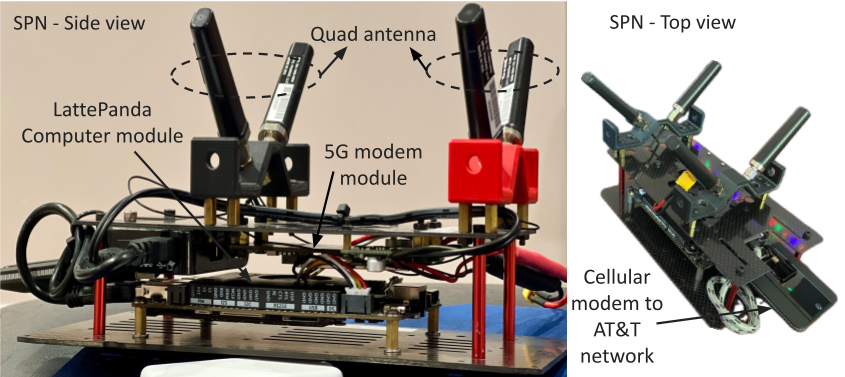}
        \label{fig:spn}}
    \vspace{-0.01in}
    \caption{SAM mounted with a portable radio node used in A2G measurements. \textbf{(a)} SAM mounted with SPN and \textbf{(b)} SPN with Quectel 5G modem.}
    \label{fig:SAMspn}\vspace{-3mm}
\end{figure}

\begin{figure}
    \centering
    \includegraphics[width=1\linewidth]{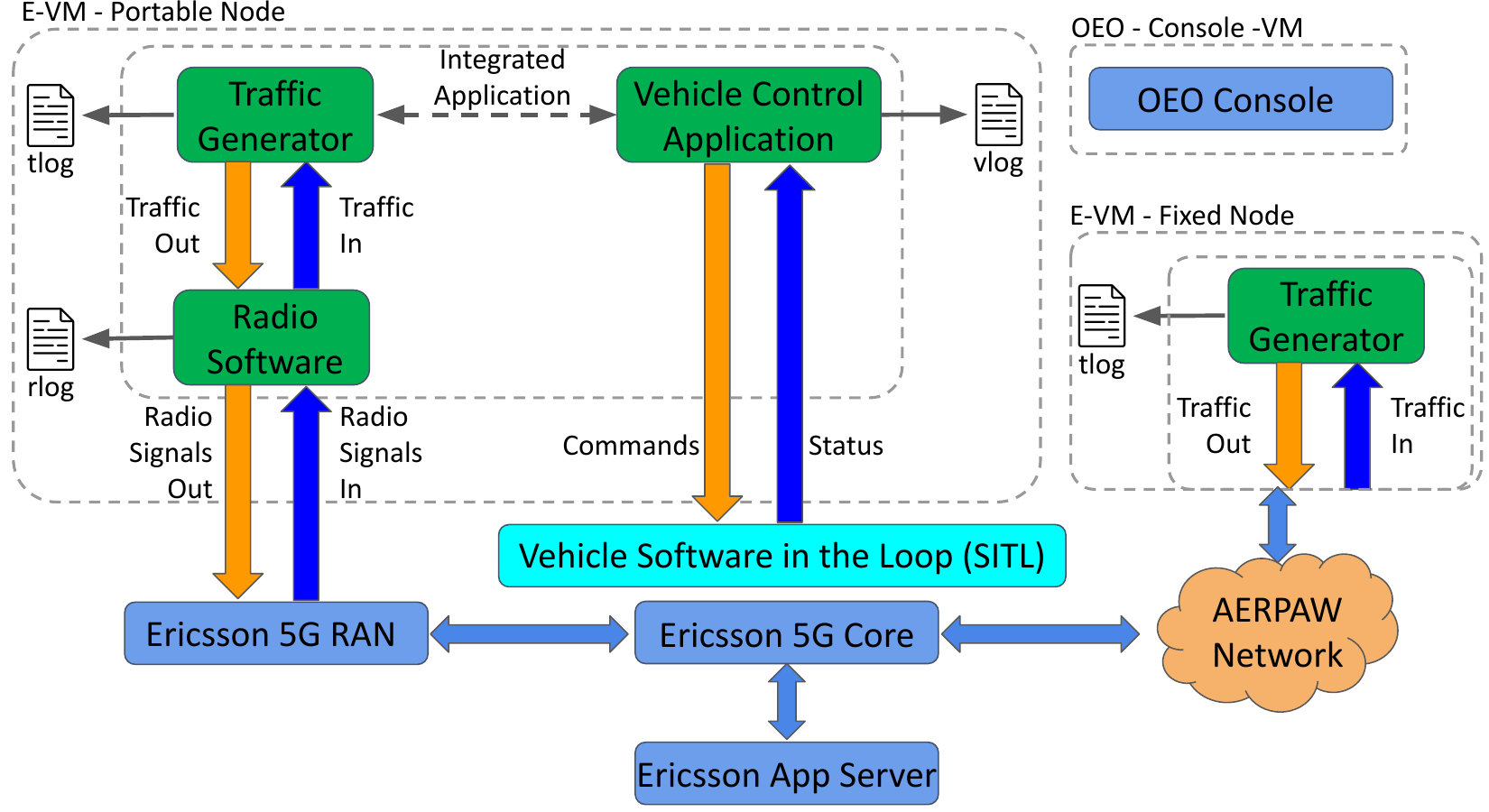}
   \vspace{2mm}
    \caption{Software functional workflow with 5G-NSA network. The workflow highlights that KPIs are derived from time-synchronized radio, traffic, and UAV telemetry, which is essential for statistically reliable spatial performance mapping.}
    \label{fig:Ericsson_SW}
\end{figure}

For aerial experiments, three types of portable nodes were used as 5G user equipment carried by UAVs: a modem-based node, a Keysight Nemo device, and a COTS Android device running PawPrints software.
The portable node based on the 5G modem as shown in Fig.~\ref{fig:SAMspn} uses a Quectel 5G module to connect to the Ericsson 5G network, a LattePanda module as a companion computer to interface with the UAV, and an ATT 5G modem for C2. The portable node that is used as UE is mounted on a small AERPAW Multi-rotor~(SAM) UAV. 

A high-level end-to-end software architecture is given in Fig.~\ref{fig:Ericsson_SW}. There are three main software functions at the portable node, namely radio software, traffic software, and vehicle control software. These software modules run on the E-VM of the portable node and generate real-time radio, traffic, and vehicle logs during the experiment. The E-VM at the AFN provides the other end point of the server-client model for the user plane data through a traffic software and logging.

\begin{figure*}[!t]
    \centering
    \subfloat[RSRP of LTE and NR signals versus distance and time.]{
        \includegraphics[width=0.48\linewidth,trim={0cm 0cm 0cm 0cm},clip]{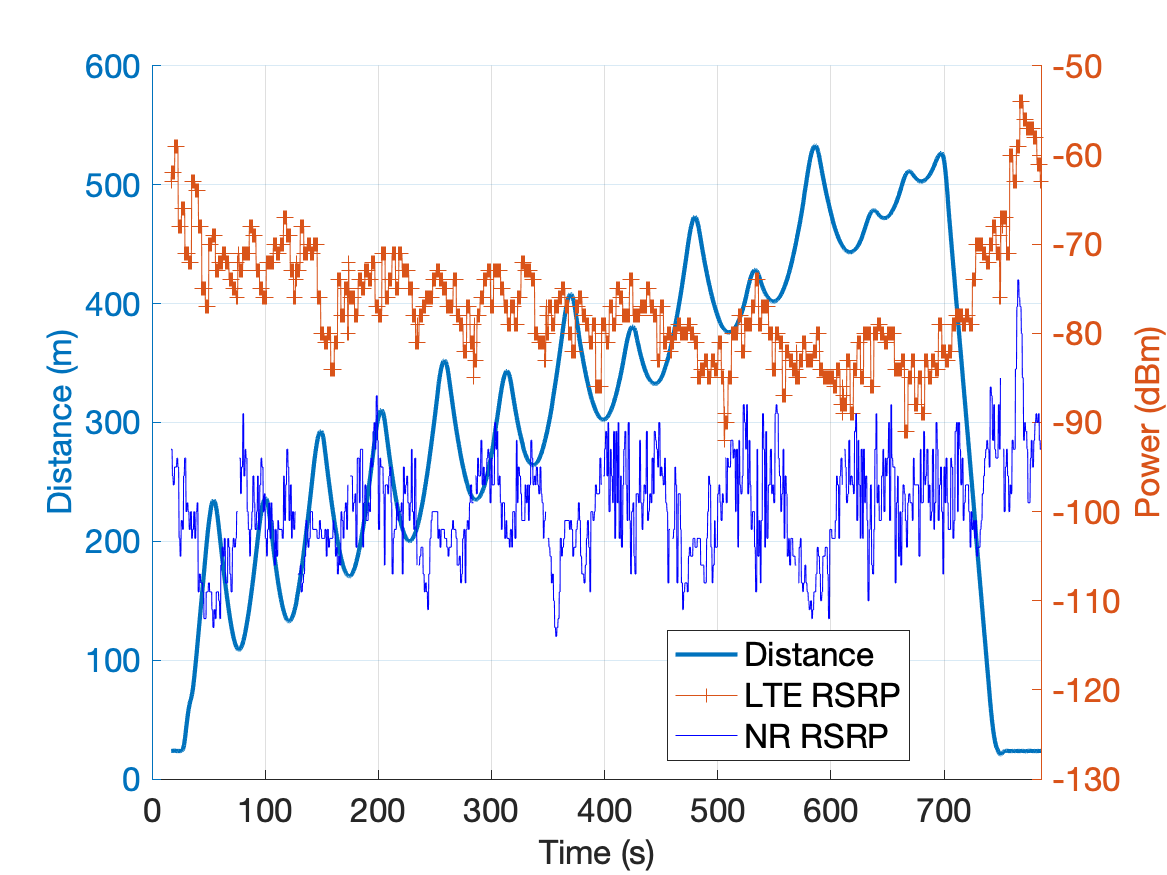}
        \label{fig:rsrp}
    }
    \subfloat[SINR of LTE and NR signals versus distance and time.]{
        \includegraphics[width=0.48\linewidth,trim={0cm 0cm 0cm 0cm},clip]{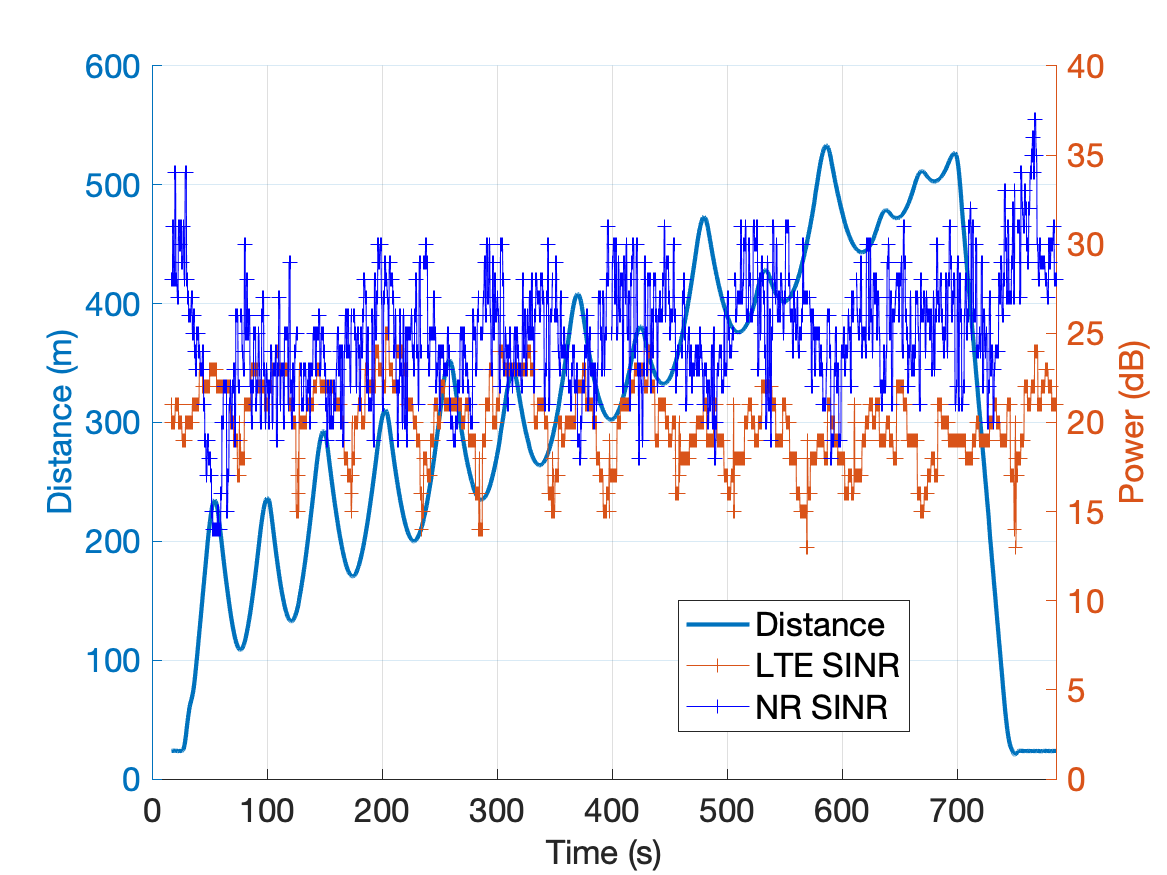}
        \label{fig:sinr}
    }
    \vspace{-0.3cm}

    \subfloat[RSRP of NR signals over 3D trajectory.]{
        \includegraphics[width=0.48\linewidth,trim={0cm 0cm 0cm 0cm},clip]{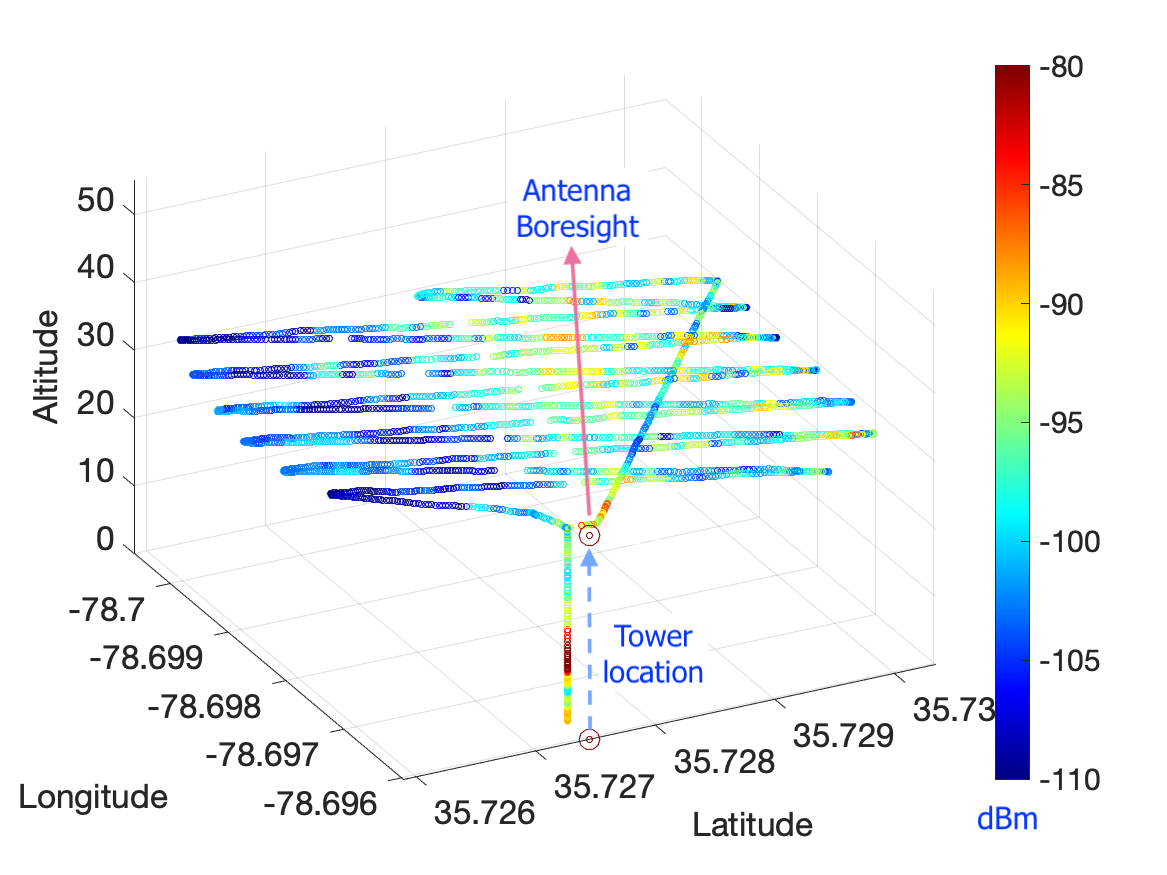}
        \label{fig:rsrp3d}
    }
    \subfloat[SINR of NR carrier]{
        \includegraphics[width=0.48\linewidth,trim={0cm 0cm 0cm 0cm},clip]{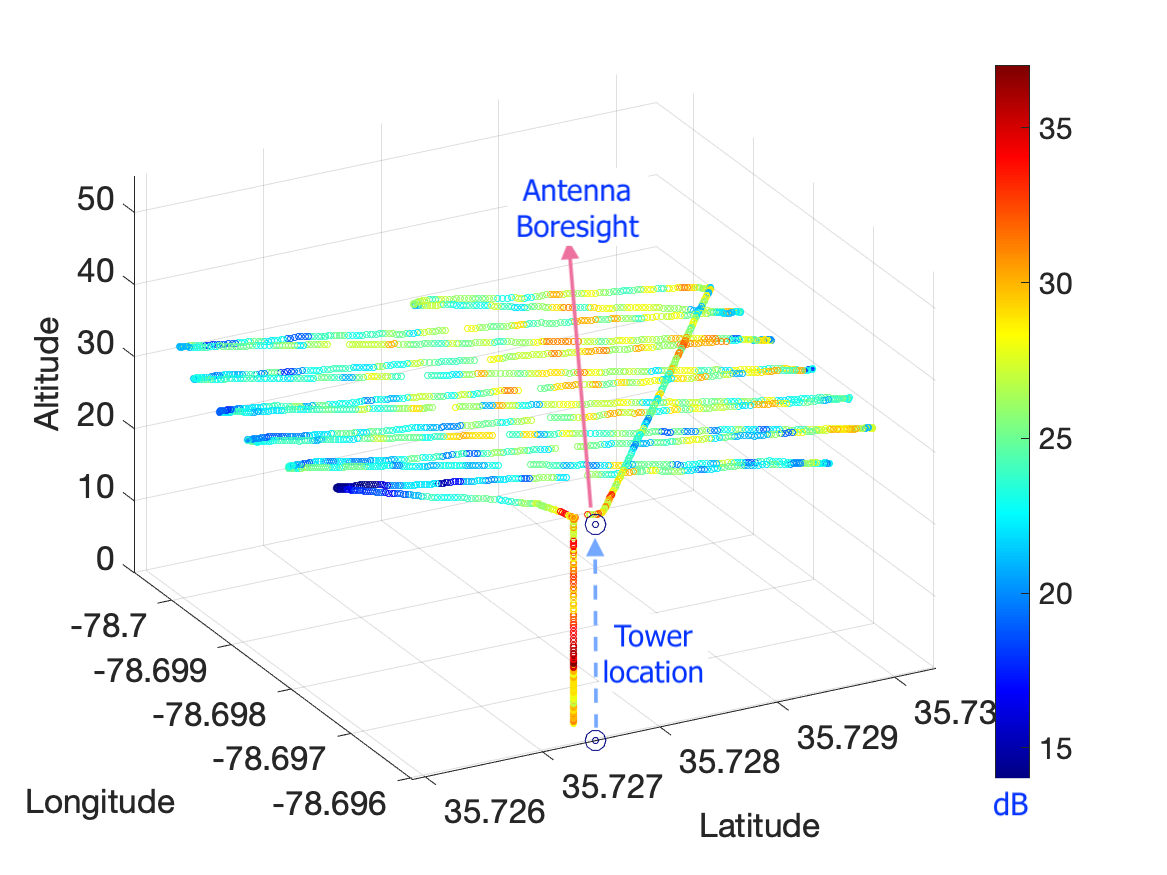}
        \label{fig:sinr3d}
    }

    \caption{\textbf{(a)} RSRP and \textbf{(b)} SINR of LTE and NR carriers vs. distance and time, \textbf{(c)} RSRP and \textbf{(d)} SINR of NR Carrier with UAV geo-location. These figures show that both RSRP and SINR exhibit clear distance- and location-dependent trends, supporting spatially resolved statistical coverage analysis.}
    \label{fig:rsrp_sinr}
    \vspace{-4mm}
\end{figure*}

For Android-based measurements, the SAM UAV carried an SPN payload containing two Android smartphones mounted on custom hardware brackets.
The two Android devices used were: a commercial standard Samsung S21 device with PawPrints, an internal AERPAW-built Android application using open source Android APIs to log radio KPIs, and a modified Samsung S23 containing Keysight Nemo software~\cite{keysightNemoHandy}, with custom firmware, access to internal modem metric, and a wider range of KPIs. A Latte Panda on-board the SPN collected and logged GPS and UAV attitude measurements, obtained from GPS sensors and an Ardupilot, respectively, along with radio KPIs streamed by the Android phone over USB. These Android phones connected to the private 4G/5G Ericsson cell tower. These phones also connected as clients to the Internet Protocol Performance~(iPerf) Server at the AFN, enabling throughput measurements.

\subsection{Dataset Format}

The 5G-NSA datasets are categorized based on the type of portable node used during the measurement campaigns: (a) 5G modem-based SPN and (b) Android device-based SPN (Nemo and PawPrints). The datasets provide synchronized RF/PHY-layer performance metrics, application-layer throughput measurements, and UAV telemetry collected during controlled UAV flight experiments on the AERPAW platform.

The modem-based datasets include LTE and NR KPIs such as RSRP, SINR, Channel Quality Indicator~(CQI), Rank Indicator~(RI), Modulation and Coding Scheme~(MCS), and downlink throughput obtained using a UAV-mounted Quectel 5G modem connected to an Ericsson 5G-NSA network. Complementary Android-based datasets collected using Nemo and PawPrints tools provide LTE and NR KPIs derived from Android APIs, along with throughput measurements when available.

All datasets are released as structured CSV files with synchronized timestamps and UAV geolocation (longitude, latitude, altitude), enabling spatially resolved analysis of aerial cellular performance. Detailed descriptions of file organization, raw and post-processed logs, and parameter-specific data files for both modem-based and Android-based datasets are provided in Appendix~\ref{app:file_structure}.

\subsection{Representative Results}
This section presents a few representative results from the aerial experiments using the above 5G modem and Android devices based portable nodes on the Ericsson 5G-NSA network.

\noindent\textbf{5G modem dataset results:}
As explained above, we used an UAV powered by a 5G Quectel modem to measure and collect data along a zigzag aerial path trajectory. From these experiments the RF/PHY parameters as well as application layer performance parameters were collected for analysis.

\begin{figure*}[!t]
    \centering
    \subfloat[CQI \& RI of NR carrier]{
        \includegraphics[width=0.48\linewidth,trim={0cm 0cm 0cm 0cm},clip]{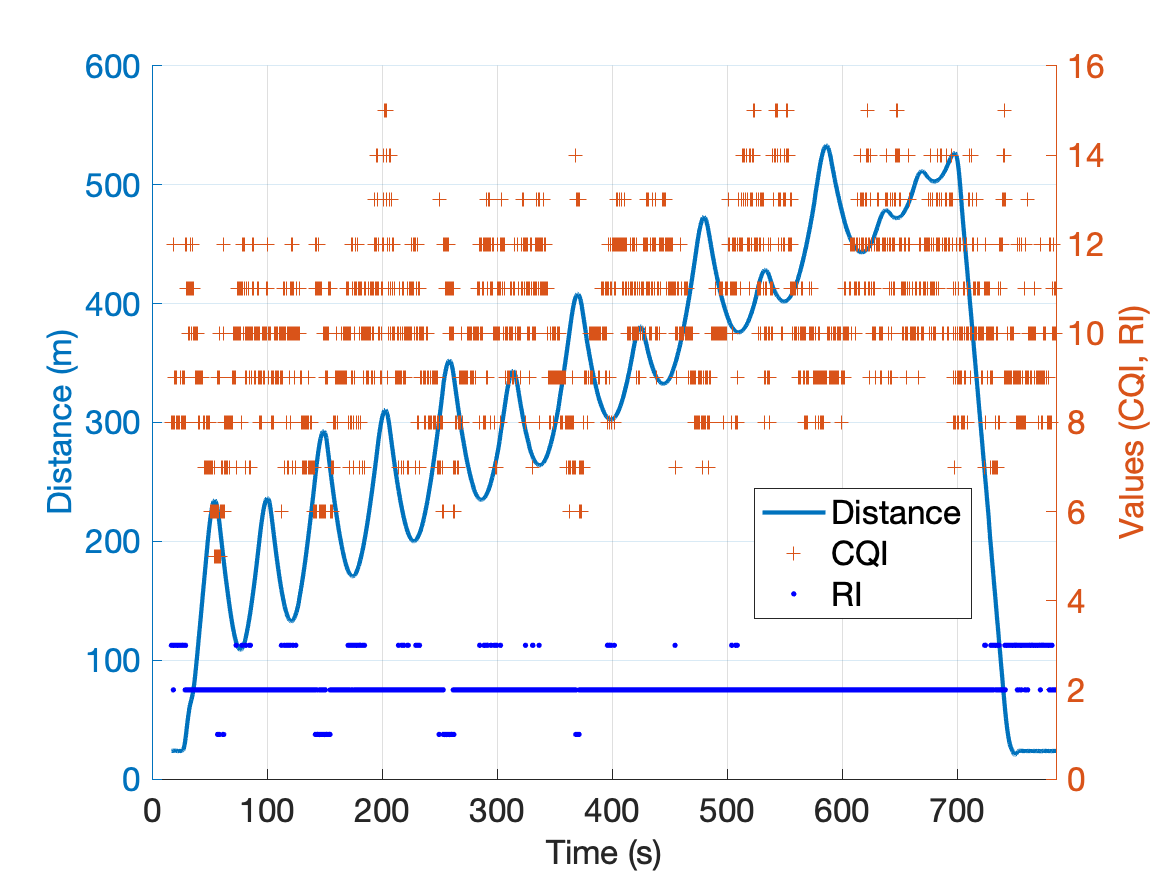}
        \label{fig:cqi}
    }
    \subfloat[CQI of NR carrier]{
        \includegraphics[width=0.48\linewidth,trim={0cm 0cm 0cm 0cm},clip]{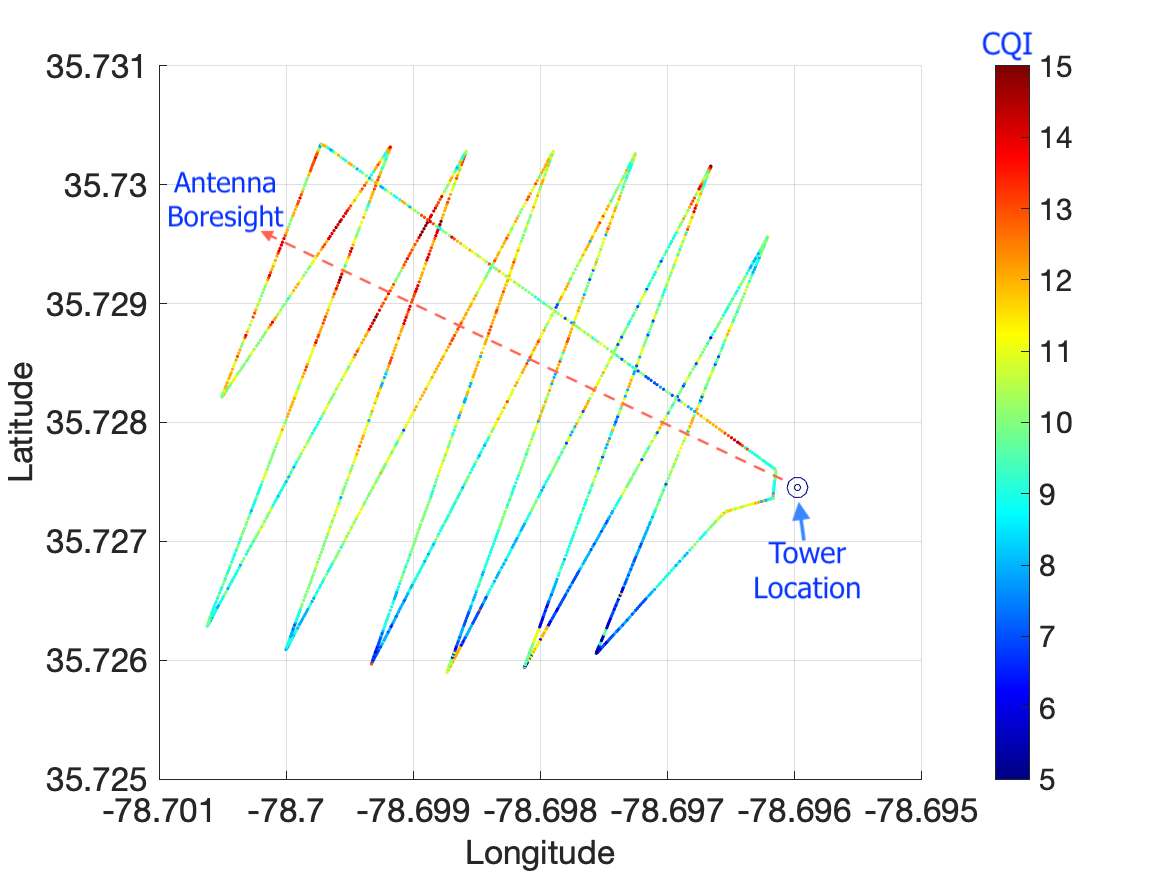}
        \label{fig:cqi2d}
    }
    \vspace{-0.3cm}

    \subfloat[MCS \& RI of NR carrier]{
        \includegraphics[width=0.48\linewidth,trim={0cm 0cm 0cm 0cm},clip]{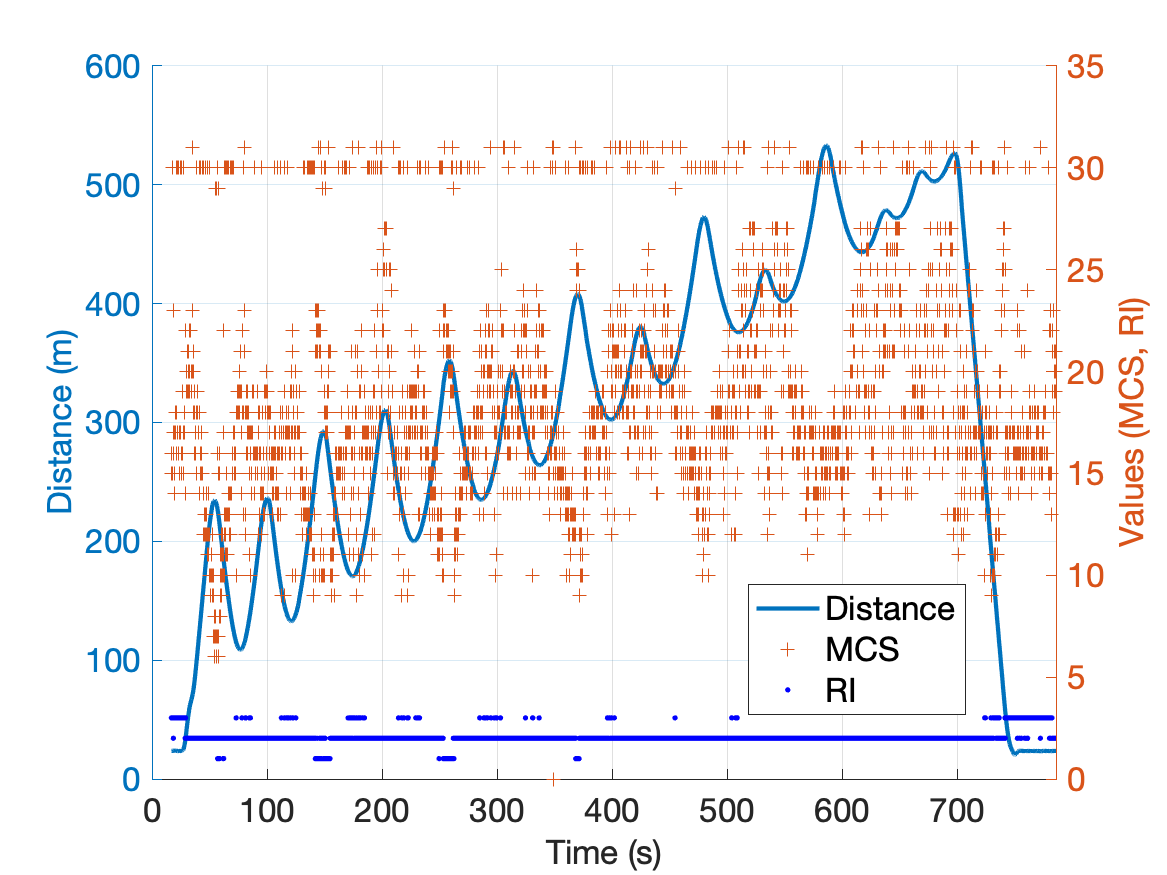}
        \label{fig:mcs}
    }
    \subfloat[MCS of NR carrier]{
        \includegraphics[width=0.48\linewidth,trim={0cm 0cm 0cm 0cm},clip]{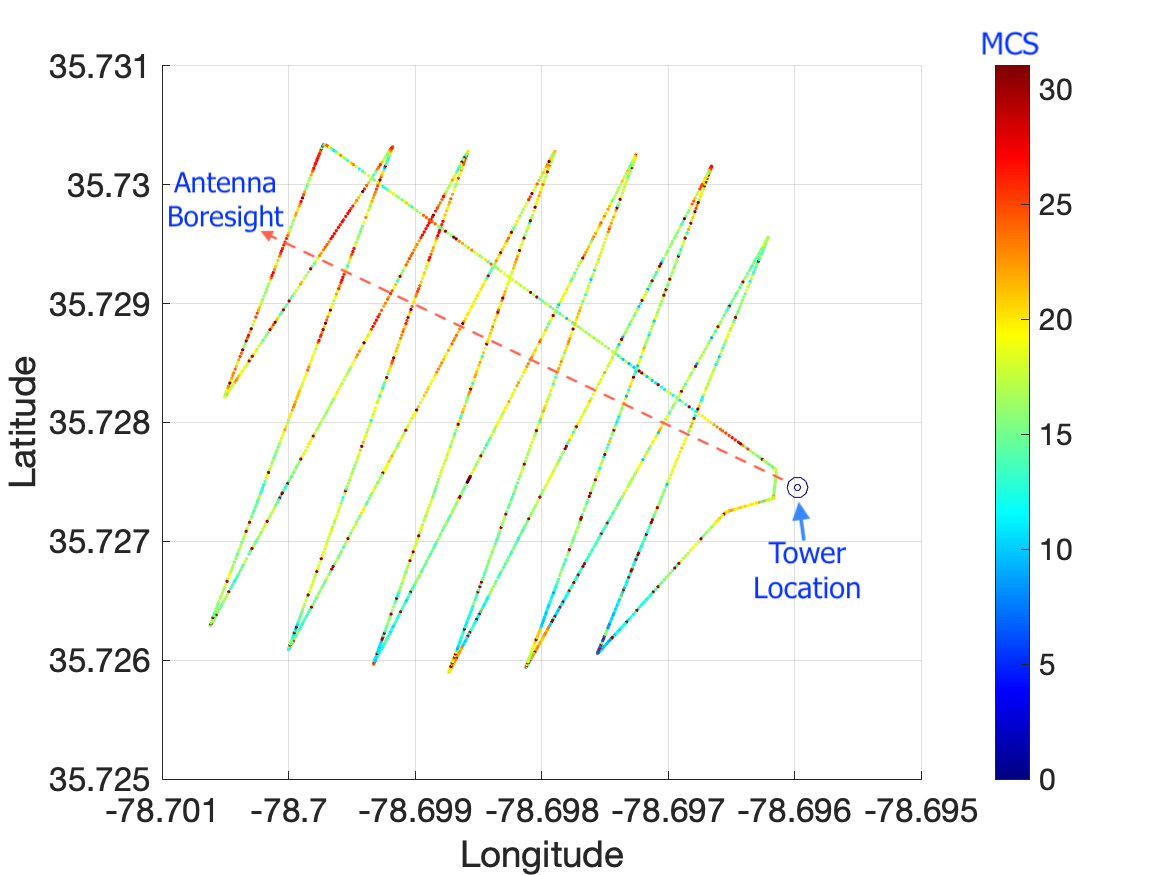}
        \label{fig:mcsd}
    }

    \caption{\textbf{(a)} CQI and \textbf{(c)} MCS of NR carrier with respect to distance and time, \textbf{(b)} CQI and \textbf{(d)} MCS of NR Carrier with UAV geo-location. The adaptation metrics vary consistently with distance and position, indicating that link adaptation responds predictably to aerial channel conditions.}
    \label{fig:CQI_MCS}
    \vspace{-4mm}
\end{figure*}

\begin{figure*}[!t]
    \centering
    \subfloat[Throughput vs distance \& time]{
        \includegraphics[width=0.48\linewidth,trim={0cm 0cm 0cm 0cm},clip]{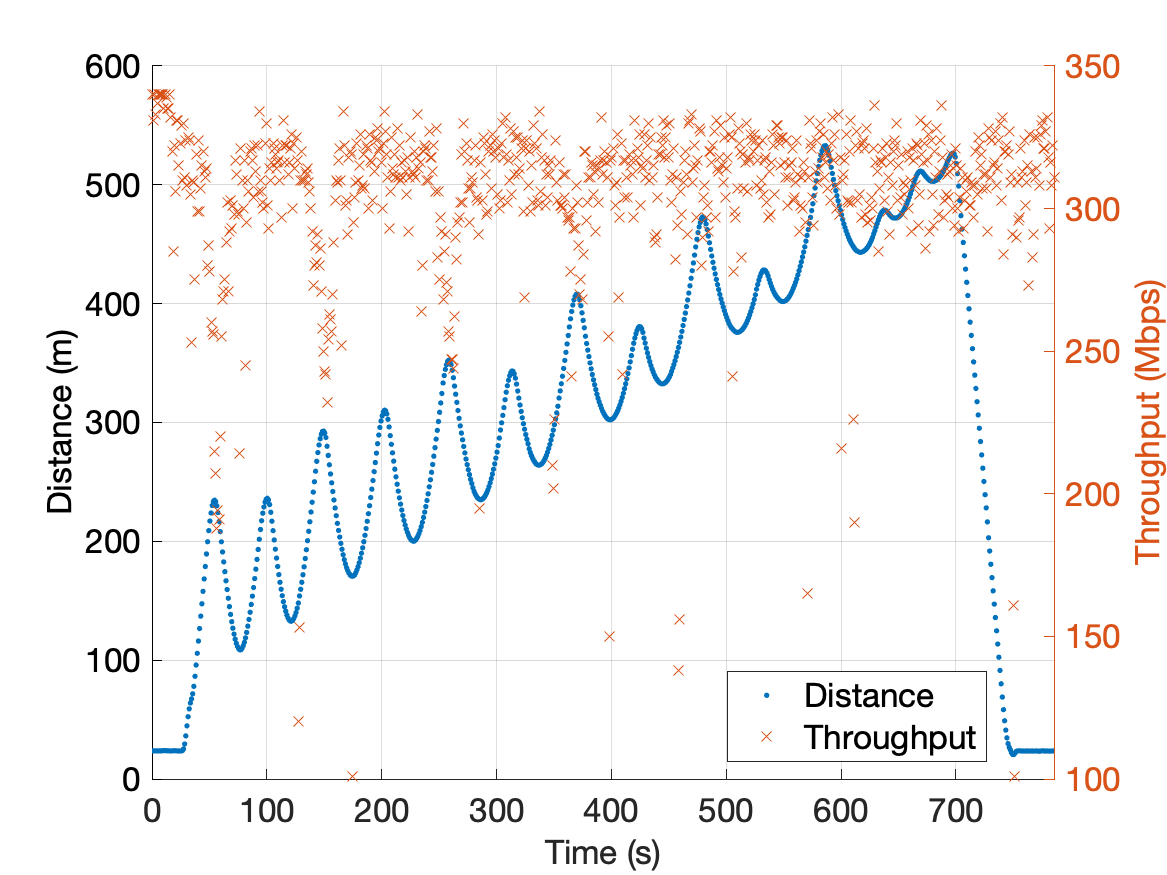}
        \label{fig:tput}
    }
    \subfloat[Throughput vs geo location]{
        \includegraphics[width=0.48\linewidth,trim={0cm 0cm 0cm 0cm},clip]{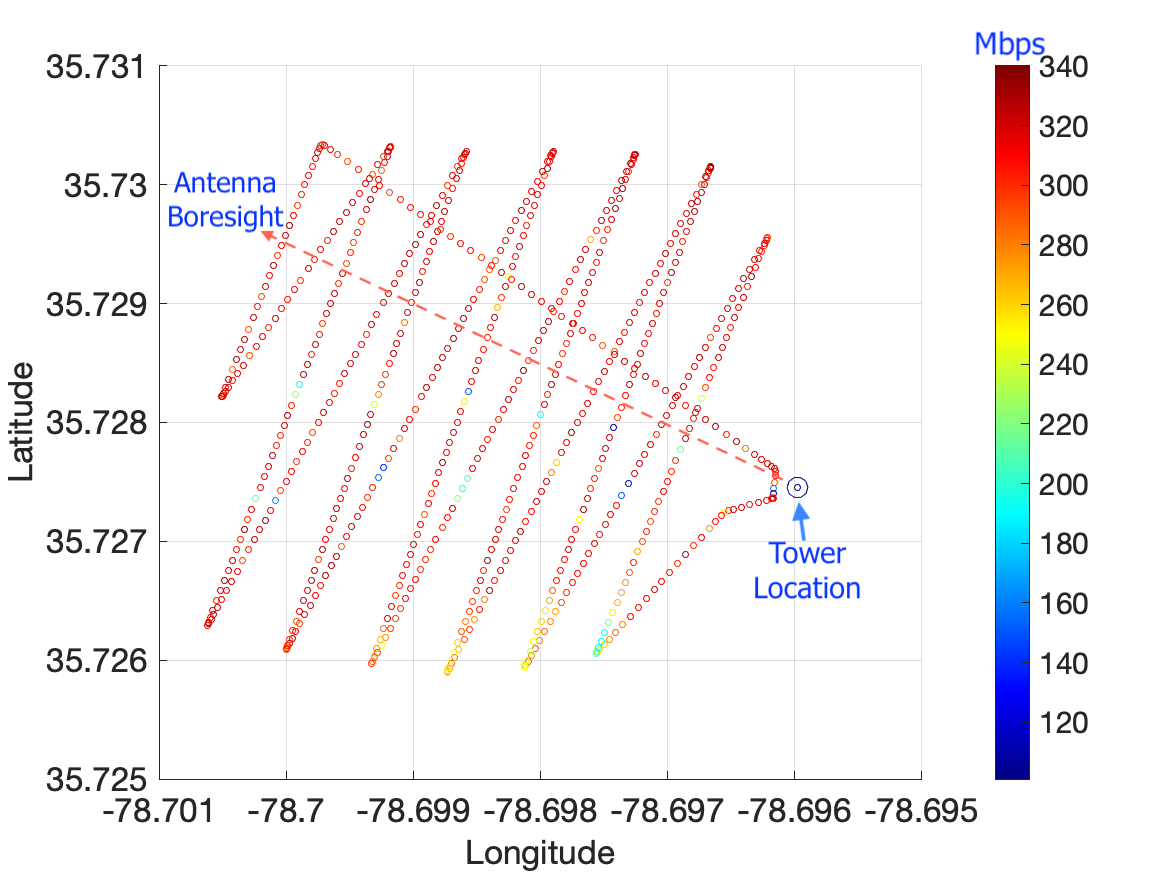}
        \label{fig:tput2d}
    }
    \caption{\textbf{(a)} Throughput with respect to distance and time, \textbf{(b)} Throughput with UAV geo-location. The throughput plots show significant spatial and temporal variability, highlighting that achievable rate statistics are strongly trajectory dependent.}
    \label{fig:Tput}
    \vspace{-4mm}
\end{figure*}

The RSRP and SINR are some of the RF parameters and CQI, RI and MCS are some of the PHY layer parameters measured and logged on the 5G-powered portable node presented here. Fig.~\ref{fig:rsrp} and Fig.~\ref{fig:sinr} show the measured levels of downlink RSRP and SINR on LTE and NR carriers versus distance and time, whereas Fig.~\ref{fig:rsrp3d} and Fig.~\ref{fig:sinr3d} show the same RSRP and SINR of NR carrier along the geo locations of the zigzag UAV flight trajectory. Similarly, Fig.~\ref{fig:CQI_MCS} shows the reported CQI, RI and MCS versus the distance, time and geo location on the LTE and NR downlink carriers. An iPerf client-server app with downlink traffic was used between the portable node and the wired AFN, and Fig.~\ref{fig:Tput} shows the downlink throughput achieved at the application layer in real-time.

\begin{figure*}[t]
    \centering
    \subfloat[Nemo iPerf throughput.]{
        \includegraphics[width=0.3\textwidth, trim={0 0.5cm 0 0}, clip]{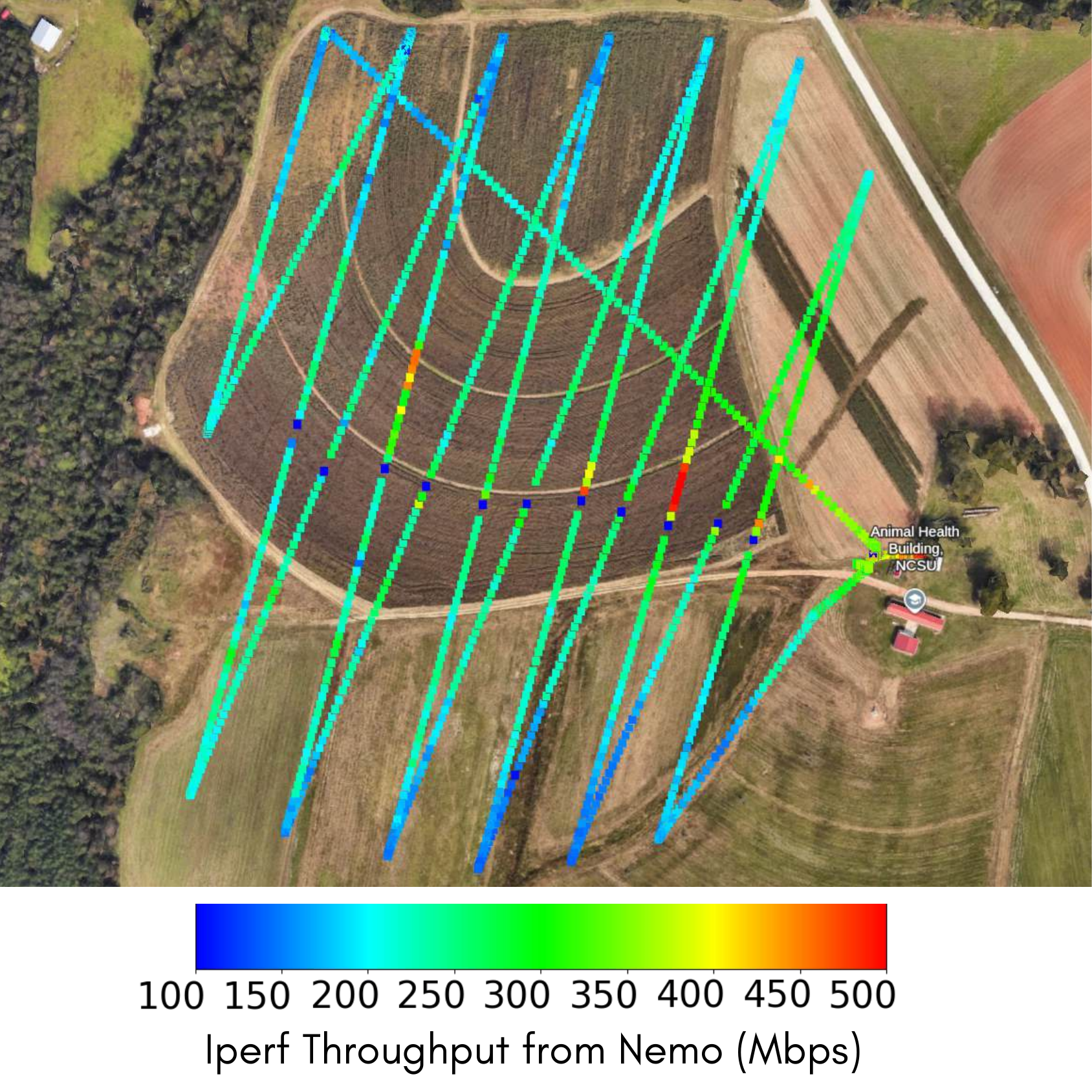}
        \label{fig:nemo_throughput}
    }
    \hspace{1.5mm}
    \subfloat[PawPrints iPerf throughput.]{
        \includegraphics[width=0.3\textwidth, trim={0 0.5cm 0 0}, clip]{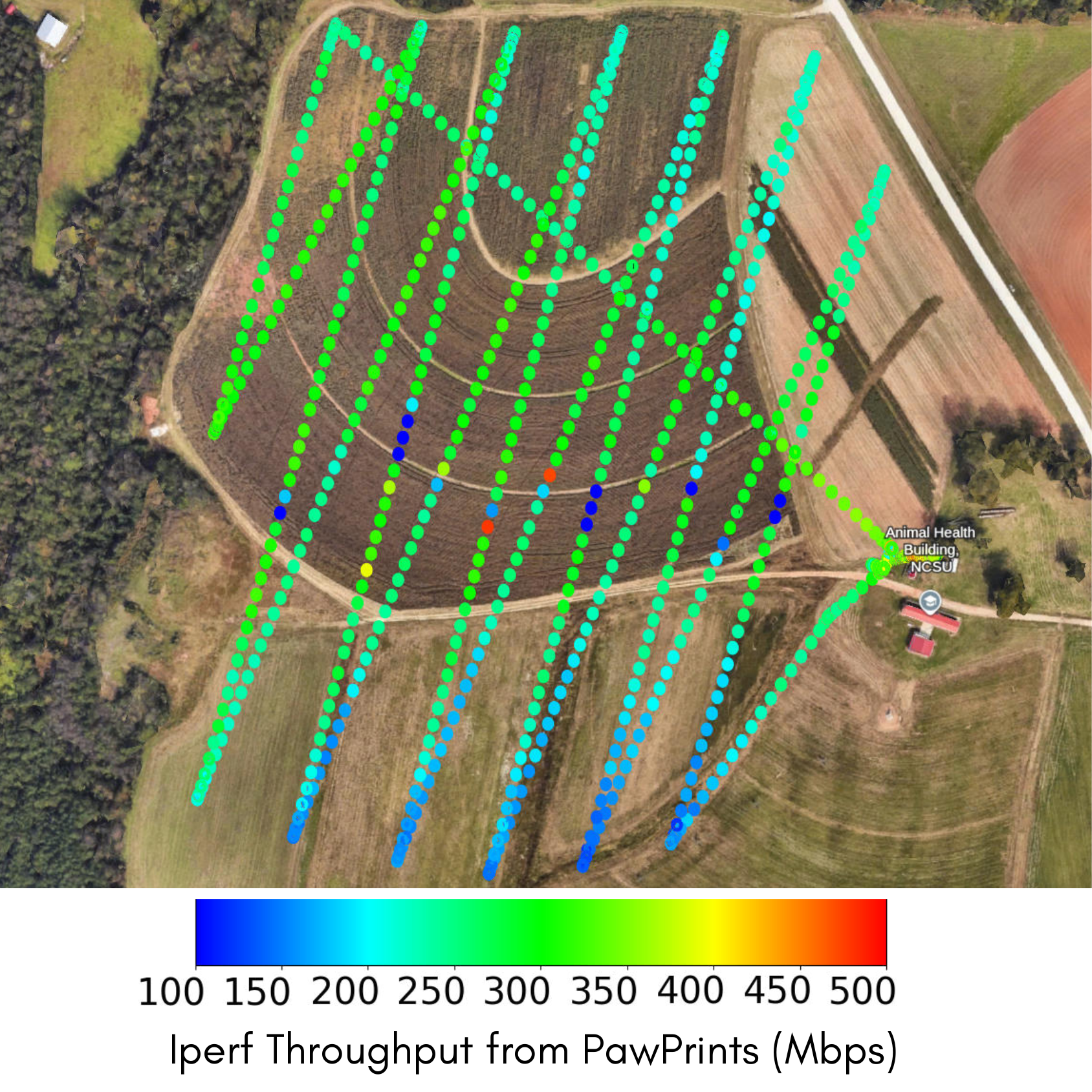}
        \label{fig:pawprints_throughput}
    }
    \hspace{1.5mm}
    \subfloat[5G downlink channel rank (Nemo).]{
        \includegraphics[width=0.3\textwidth, trim={0 0.5cm 0 0}, clip]{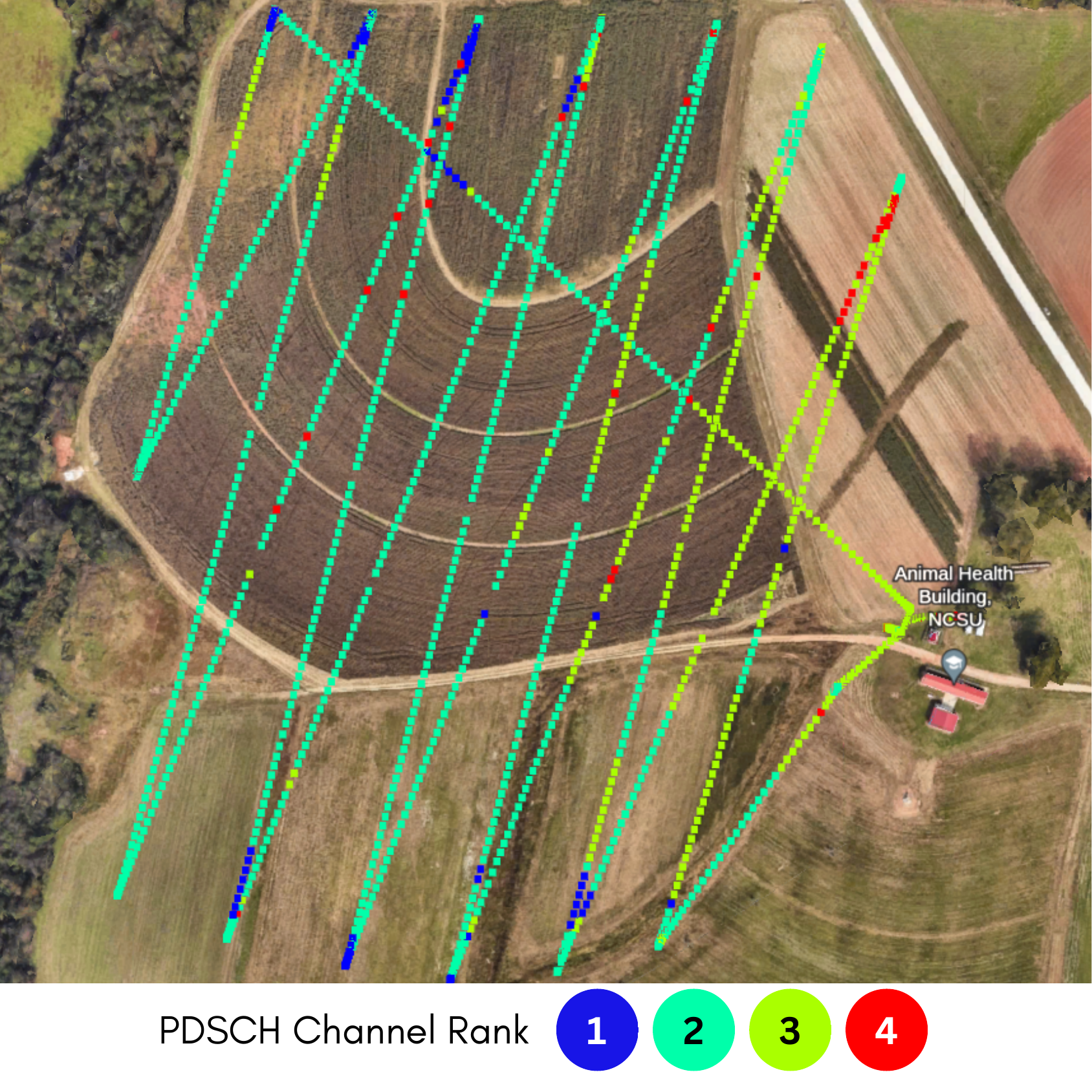}
        \label{fig:nemo_pdsch_rank}
    }

    \caption{Wireless KPI data gathered at AERPAW Lake Wheeler tower 1: \textbf{(a)} and \textbf{(b)} show iPerf throughput observed by Nemo and PawPrints, respectively, while simultaneously sending traffic as iPerf clients; \textbf{(c)} shows the 5G downlink channel rank as recorded by Nemo. The comparison demonstrates that measured throughput statistics depend on the logging tool, which is practically important when benchmarking or cross-validating datasets.}
    \label{fig:PawPrint_KPI_LW1}
\end{figure*}

\noindent\textbf{Nemo and PawPrints dataset results:}
This section presents the representative results from the wireless KPI datasets collected from a UAV in two scenarios: measurements of a private BS at the AERPAW LWRFL with controlled UAV trajectories (using PawPrints and Nemo), and measurements of commercial cell towers from a tethered Helikite during the Packapalooza 2023 festival (using PawPrints). Fig.~\ref{fig:PawPrint_KPI_LW1} depicts some representative results from the first scenario, when the UAV traces sawtooth trajectories in the horizontal plane, at increasing distances from the private Ericsson BS. Fig.~\ref{fig:nemo_throughput} and Fig.~\ref{fig:pawprints_throughput} show a heatmap of the iPerf throughput measured by the Nemo and PawPrints device, respectively, when both were operating as client simultaneously. The heatmaps confirm the reduction in throughput observed near handover regions, particularly at sector boundaries, as evidenced by the blue-shaded areas. Fig.~\ref{fig:nemo_pdsch_rank} shows the channel rank of the physical downlink channel recorded by Nemo during the flight.

\begin{figure}
    \centering
\includegraphics[width=0.95\linewidth]{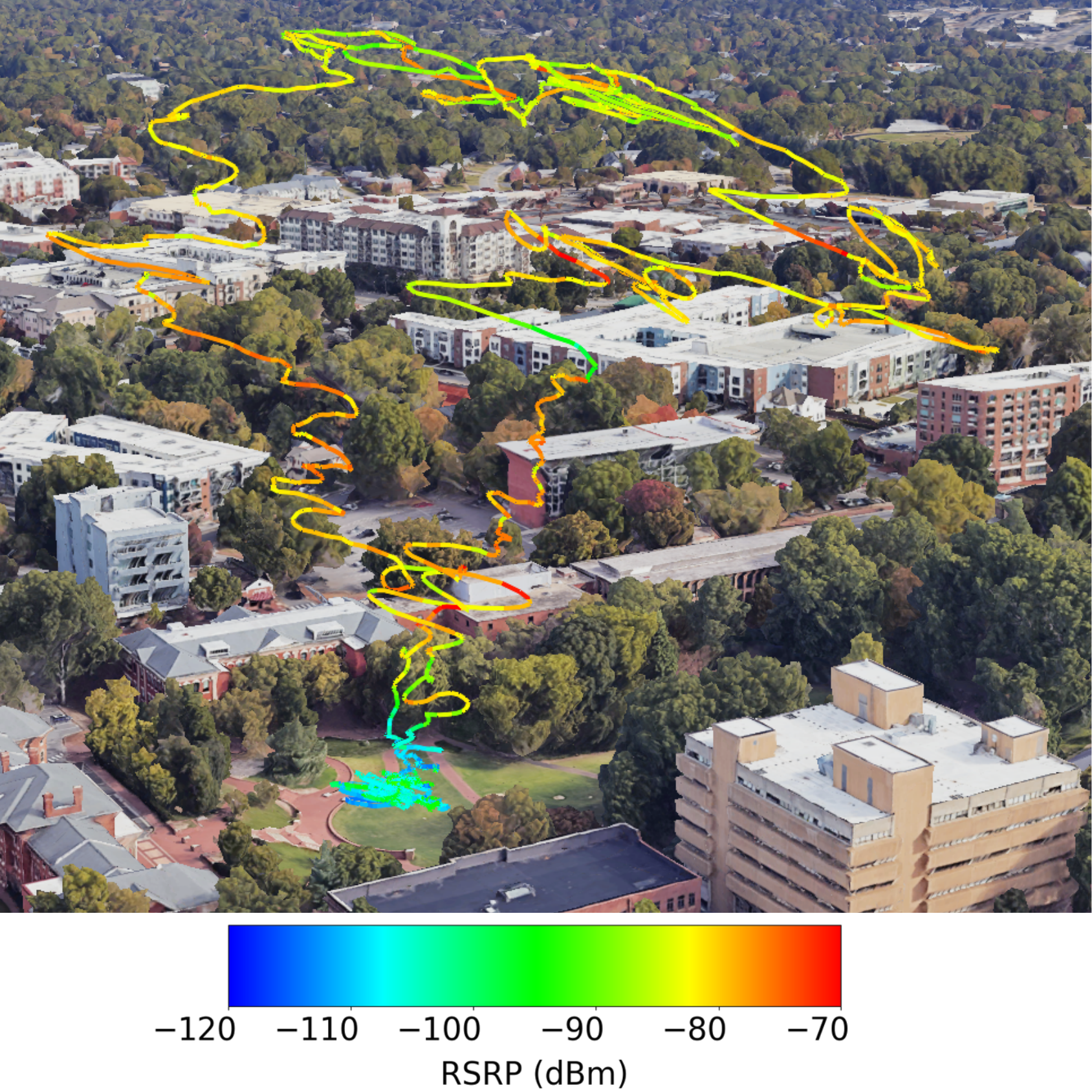}
    \caption{RSRP variation against altitude observed by a PawPrints device on a Helikite at Packapalooza 2023.}
    \label{fig:PawPrint_KPI_Packapalooza}
\end{figure}

Fig.~\ref{fig:PawPrint_KPI_Packapalooza} depicts the changing RSRP of an LTE node with varying Helikite altitude as observed by a PawPrints device during the Packapalooza 2023 event. The PawPrints Packapalooza 2023 dataset also contains extensive records of other nearby commercial LTE cellular towers, operated by AT\&T and their network KPI values near the NC State University campus. Both datasets can be processed using the data processing scripts in the previous sections to analyze network coverage and performance in the aerial dimension. 

The data from these experiments are publicly available at the AERPAW datasets webpage~\cite{aerpaw_datasets} and in Dryad research repository~\cite{asokan2025ericsson5gnsa, singh2023packapaloozadryad, singh2024lakewheelerandroidhorizontalsweeps, singh2024lakewheelerandroidsemicircles, singh2024lakewheelerandroidtwosweeps}.

\subsection{Possible Uses of Dataset} 
The KPIs available in this dataset can be utilized for the following representative research problems.

\noindent\textbf{Study of wireless channel propagation in rural aerial environments:} The empirical data on variations in signal strength, with distance from the BS and elevation and azimuth angles, can be used to validate existing theoretical propagation models and ray-tracing simulators, along with developing new mathematical or machine learning models. Moreover, deeper insights can be obtained by studying the variation in channel rank and channel quality index in aerial scenarios. 

\noindent\textbf{Application layer throughput prediction in rural aerial scenarios:} Throughput prediction models for rural aerial scenarios can be created by studying the relation between physical layer KPIs such as channel rank and received SINR, and iPerf throughput. 

\noindent\textbf{Analysis of commercial cellular network coverage in aerial urban environments:} The Packapalooza dataset contains
received signal strength and quality values at various altitudes, which can be used to study the suitability of commercial cellular networks, traditionally designed to serve ground users, for aerial operations.  

\section{LoRa Propagation Dataset}\label{sec:lora}
The LoRaWAN technology utilizes chirp modulation techniques to support long range, low power communications, exhibiting unique propagation characteristics for measurement. In this section, we describe AERPAW's infrastructure for such measurements and present representative results from LoRaWAN measurement campaigns.

\subsection{Description of Hardware and Software}
\begin{figure}[t!]
    \centering
    \subfloat[LoRaWAN gateway.]{
    \includegraphics[width=0.45\linewidth]{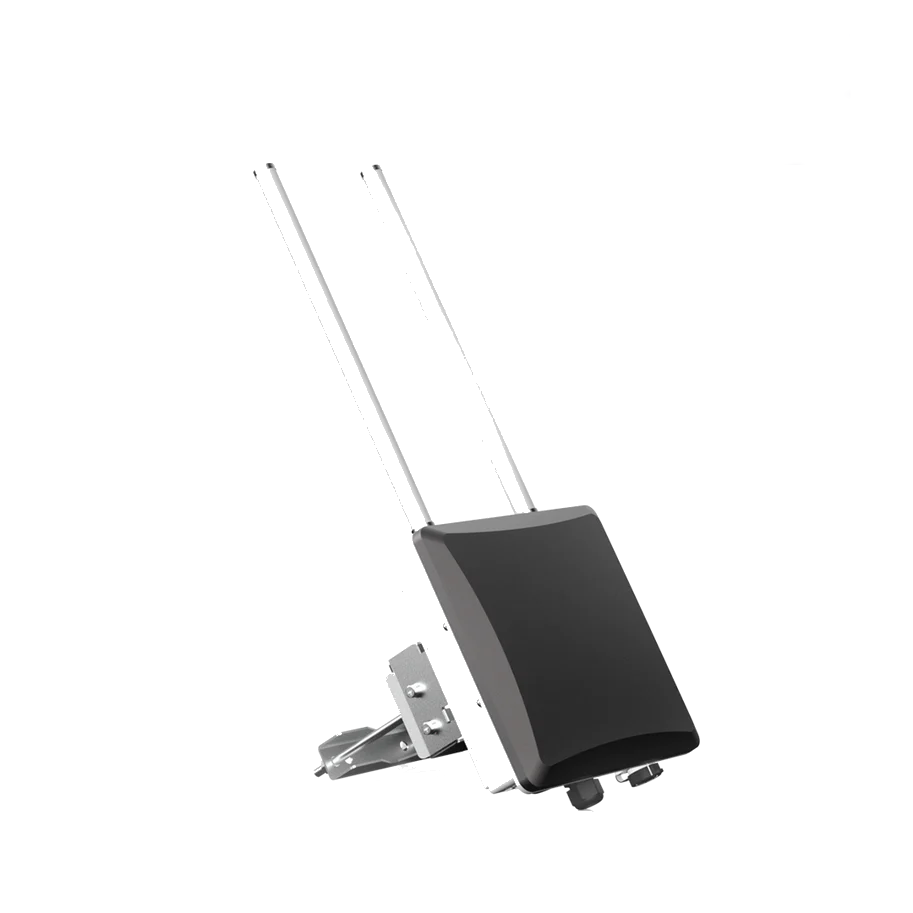}
    \label{fig:lorawanGateway}}
  \subfloat[LoRaWAN dongle.]{    
\includegraphics[width=0.45\linewidth]{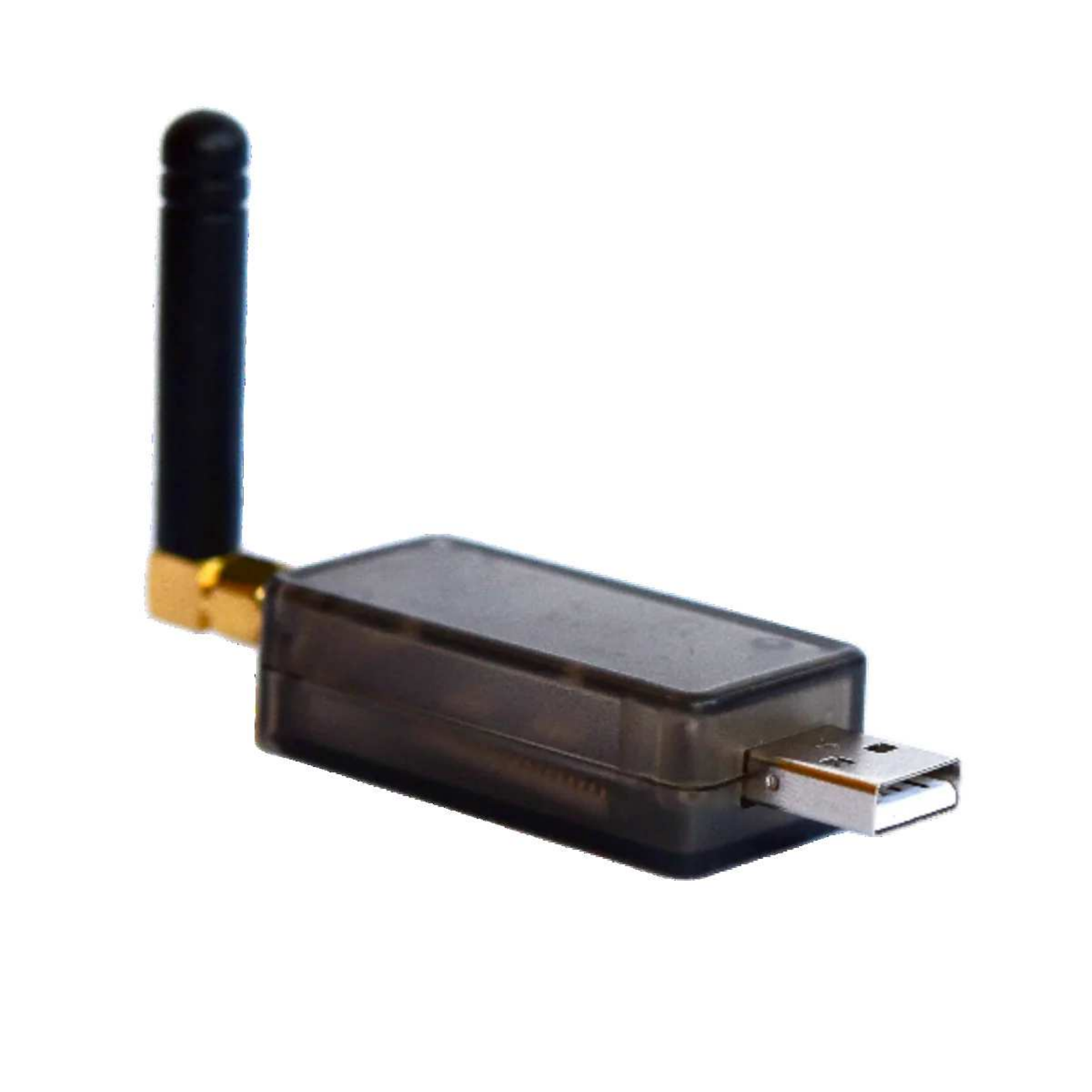}}
    \caption{\textbf{(a)} LoRaWAN gateway and \textbf{(b)} USB-compliant programmable LoRa device, used in the AERPAW system for mobile LoRa experiments.}
    \label{fig:lora_equipment}
\end{figure}

\begin{figure}[t!]
    \centering
    \subfloat[Ground vehicle carrying a LoRaWAN device.
    \label{fig:loraVehicle_ground}]{
        \includegraphics[width=0.7\linewidth]{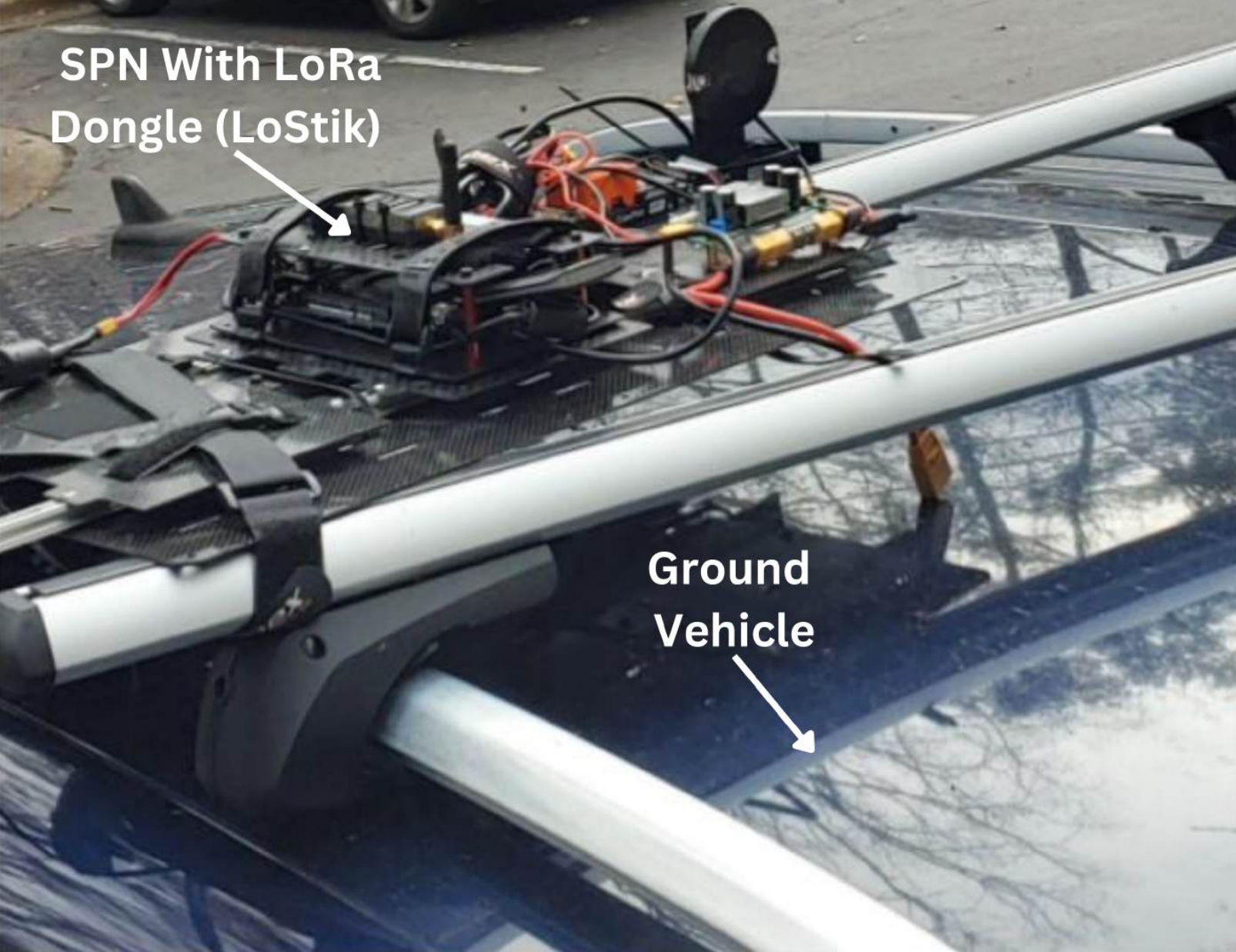}
    }
    \vspace{2mm}

    \subfloat[UAV carrying a LoRaWAN device.
    \label{fig:loraVehicle_uav}]{
        \includegraphics[width=0.75\linewidth]{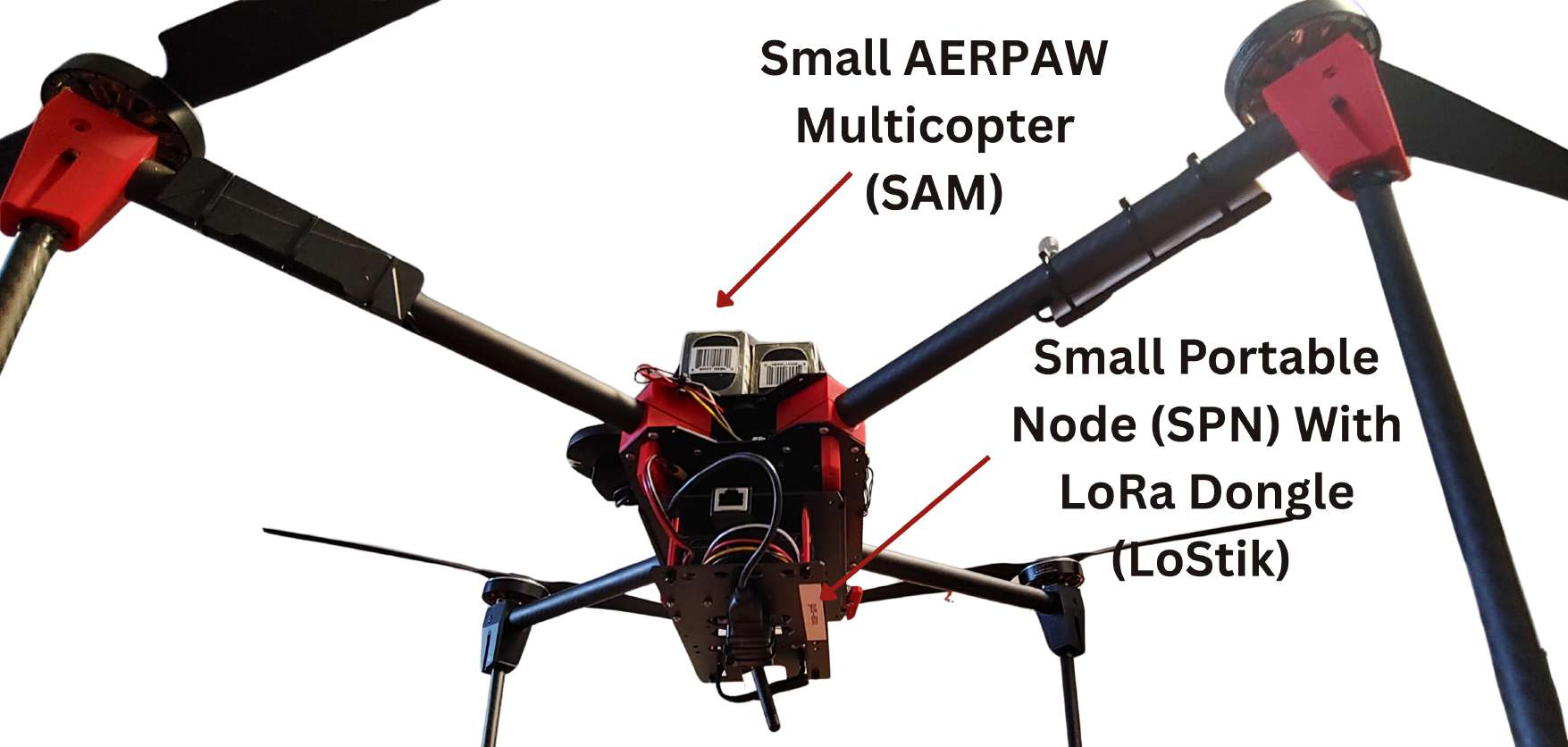}
    }
    \vspace{2mm}

    \subfloat[Helikite carrying a LoRaWAN device.
    \label{fig:loraVehicle_helikite}]{
        \includegraphics[width=0.75\linewidth, trim={0 0.5cm 2.5cm 0}, clip]{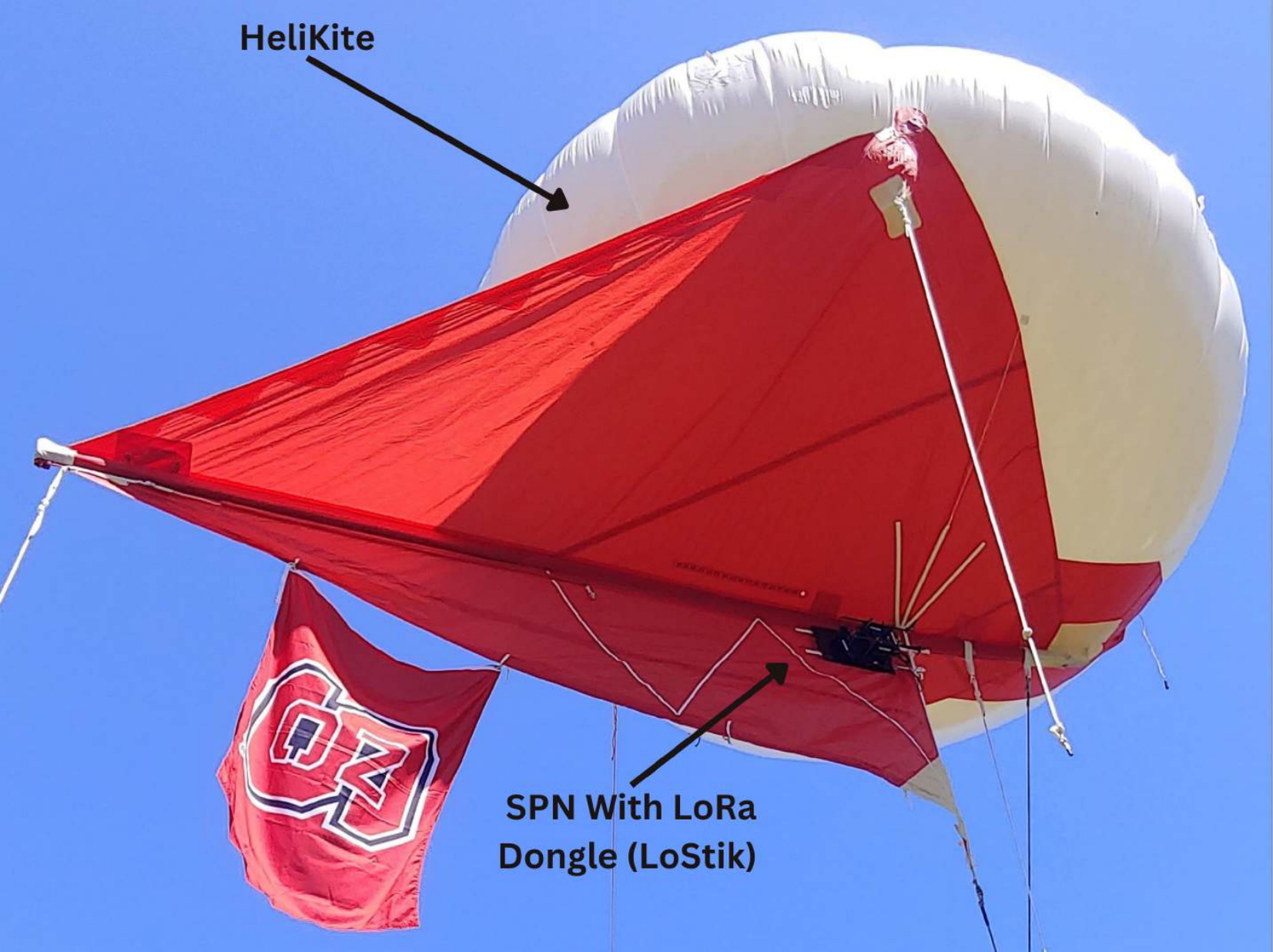}
    }

    \caption{Ground and aerial platforms used in mobile LoRaWAN experiments: \textbf{(a)} ground vehicle, \textbf{(b)} UAV, and \textbf{(c)} helikite, each carrying a LoRaWAN device.}
    \label{fig:loraVehicles}
    \vspace{-3mm}
\end{figure}

The LoRa infrastructure consists of USB-compliant programmable LoRa devices and seven LoRaWAN gateways (see Fig.~\ref{fig:lora_equipment}), which relay LoRa communications over the AERPAW backplane to tenant-dedicated Docker containers for executing application-specific data processing tasks. Standard software stacks are installed on the containers, including Prometheus and PostGRES storage, which feed data for Grafana-based visualization. Fig.~\ref{fig:loraVehicles} shows the mobile platforms used for LoRaWAN measurements, including a ground vehicle (Fig.~\ref{fig:loraVehicle_ground}), a UAV (Fig.~\ref{fig:loraVehicle_uav}), and a helikite (Fig.~\ref{fig:loraVehicle_helikite}). The LoRa devices allow the experiments to configure the spreading factor between 7 to 12, and accordingly set the transmission data rate from DR3 (5.47~kbps) to DR0 (0.25~kbps).

\subsection{Dataset Format}
The LoRa propagation dataset includes measurement logs collected from both the LoRa transmitter and the associated LoRaWAN gateways, enabling end-to-end characterization of aerial and ground LoRaWAN links. The dataset captures detailed packet-level metadata together with synchronized vehicle mobility and positioning information, supporting coverage analysis, signal quality evaluation, and latency assessment in aerial IoT scenarios.

Transmitter-side logs record packet identifiers, transmission parameters (including data rate, bandwidth, code rate, spreading factor, and carrier frequency), transmission timestamps, and vehicle state information such as geographic location, orientation, velocity, and GPS status. Gateway-side logs capture reception metrics for successfully decoded packets, including Received signal strength indicator~(RSSI), signal-to-noise ratio~(SNR), reception timestamps, frequency channels, RF chain identifiers, and gateway location metadata.

In addition to successful transmissions, the dataset includes records of failed packet transmissions and aggregated summaries of gateway-level data rate statistics, enabling reliability and performance analysis across deployment scenarios. Raw signal-level measurements are provided in SigMF format, while processed packet- and gateway-level records are released in structured CSV files suitable for statistical analysis and machine learning applications. Detailed file inventories, directory organization, and conversion script descriptions are provided in Appendix~\ref{app:file_structure}.

\subsection{Representative Results}

\begin{figure}[t!]
    \centering
    \includegraphics[width=0.95\linewidth]{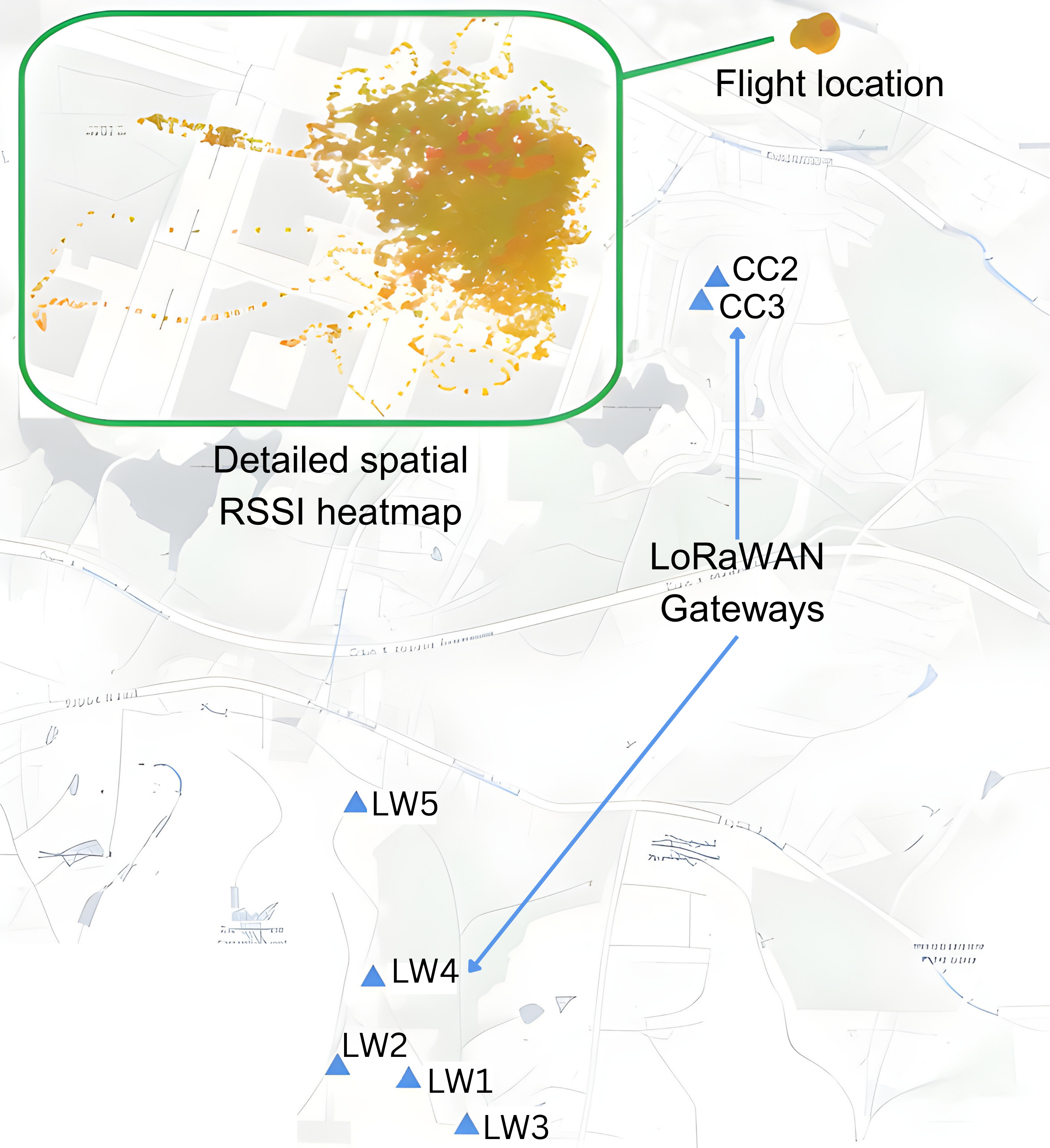}
    \caption{Location of the LoRa tethered Helikite flight, showing location of the six LoRaWAN gateways and the top view of the Helikite locations. The spatial separation of gateways enables statistical evaluation of gateway diversity and coverage overlap.}
    \label{fig:lora_locations}
\end{figure}

\begin{figure}[t]
    \centering
    \subfloat[CDF of received RSSI at various LoRaWAN gateways.]{
        \includegraphics[width=0.4\textwidth]{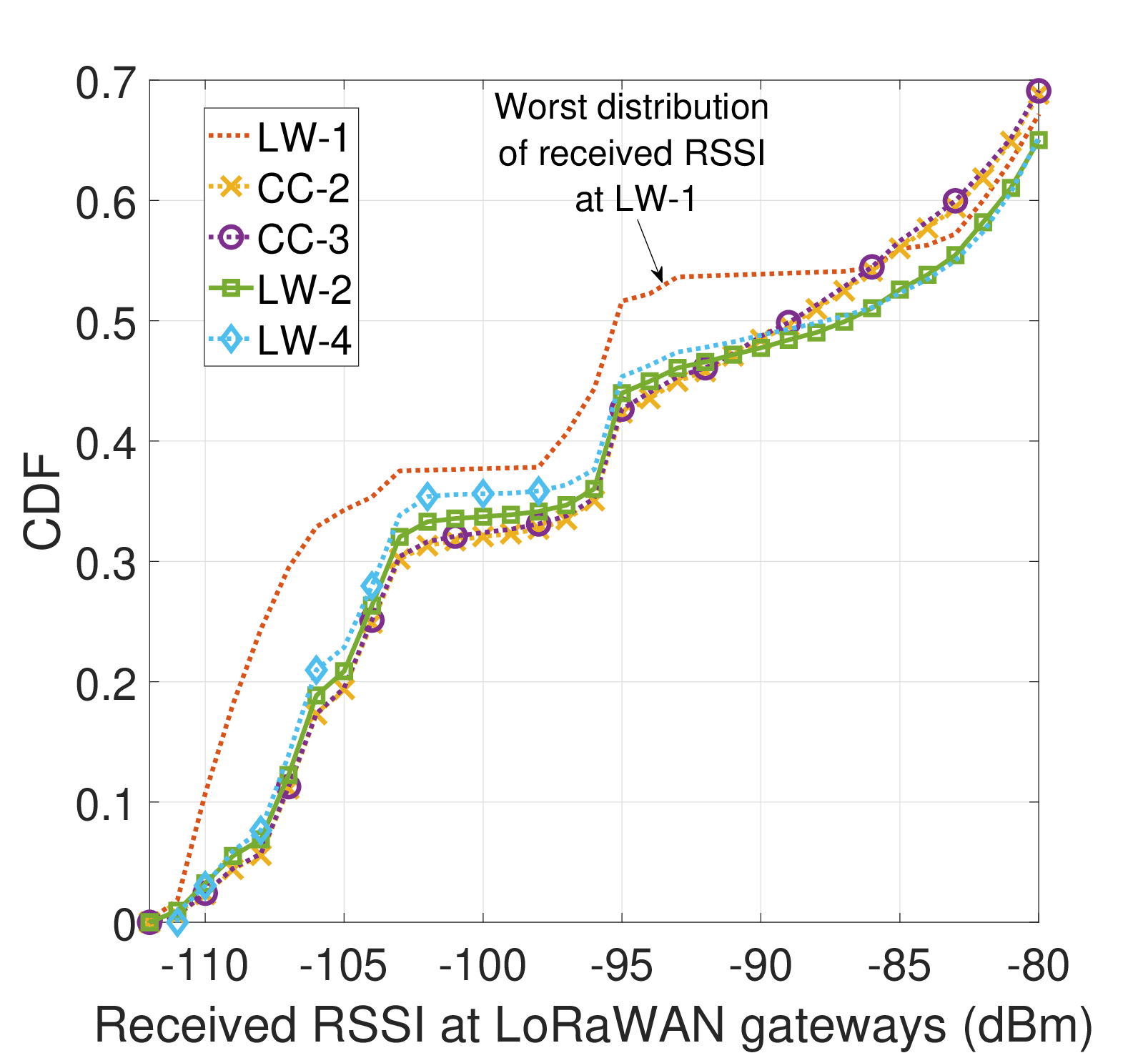}
        \label{fig:lora_rssi_cdf}
    }
\\
    \subfloat[Variation in LoRa packet's received SNR as a function of received RSSI, aggregated over six LoRaWAN gateways.]{
        \includegraphics[width=0.4\textwidth, trim={2cm 0 2cm 0cm}]{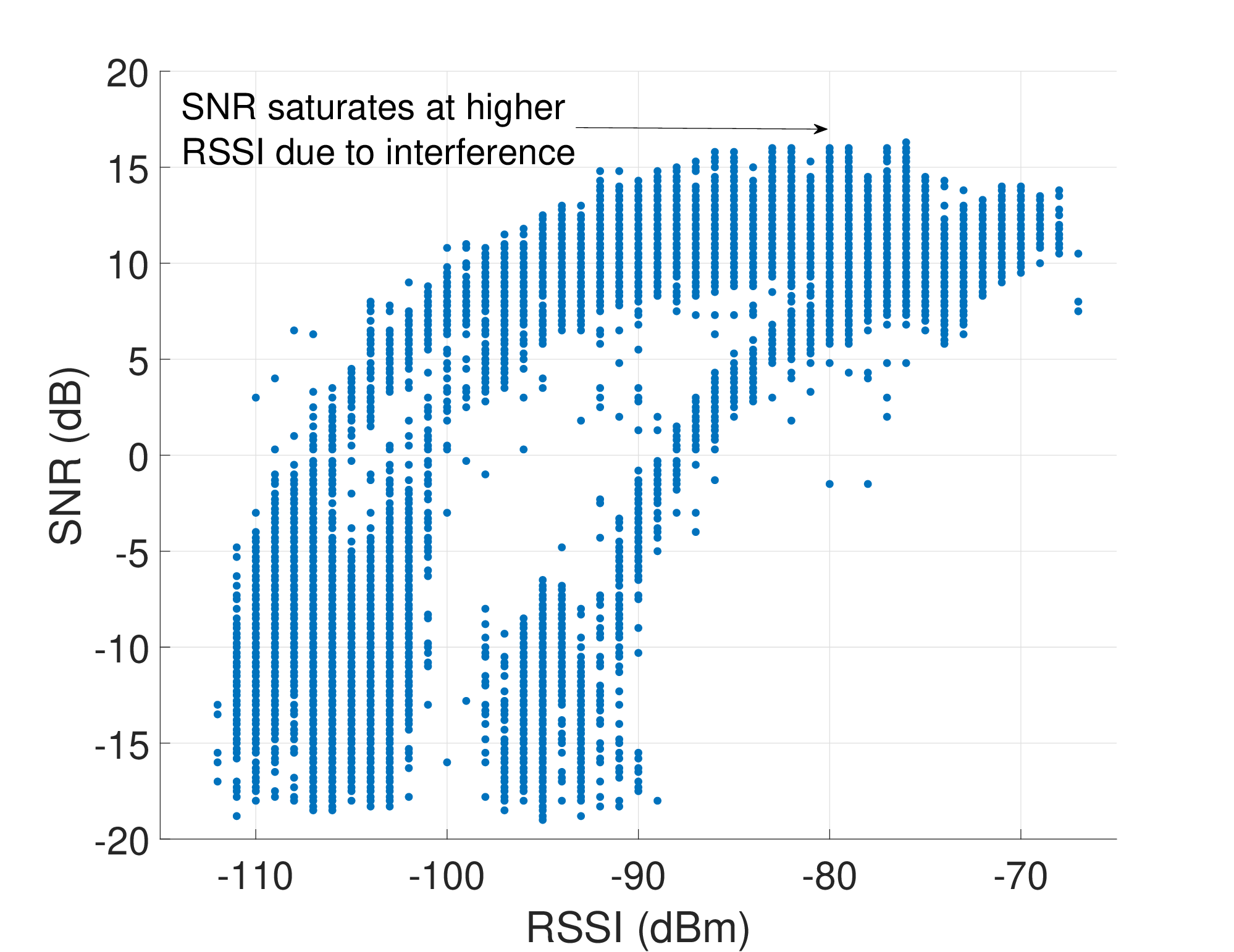}
        \label{fig:lora_rssi_vs_sinr}
    }
    \caption{The KPIs of LoRaWAN links measured during a tethered Helikite experiment, where the LoRa transmitter was elevated, and data were collected across six gateways.
\textbf{(a)} Empirical CDF of received RSSI values observed at multiple gateways.
\textbf{(b)} Relationship between received RSSI and corresponding SNR, highlighting link quality variation.}
    \label{fig:loar_rssi_latency}
\end{figure}

The LoRa transmitter device was carried as a payload by a tethered HeliKite, which traced a free-floating trajectory over the North Campus of NC State University, Raleigh. The LoRa device was controlled by a Latte Panda mini computer, which also logged packet transmission details. The LoRa device transmitted packets, containing a sequence number and timestamp, at intervals of $1.5$~s to six LoRaWAN gateways at multiple locations, as shown in Fig.~\ref{fig:lora_locations}. These gateways recorded the received RSSI, received SNR, and the timestamp of reception. The CDF of received RSSI at the gateways is shown in Fig.~\ref{fig:lora_rssi_cdf}. The variation in received SNR with received RSSI, over all the LoRaWAN gateways, is shown in Fig.~\ref{fig:lora_rssi_vs_sinr}, which indicates that higher RSSI results in a smaller range of SINR values. In contrast, lower RSSI values can cause SIR fluctuations as noise and interference dominate. The dataset not only provides physical layer signal strength parameters but also includes packet-level latency metrics. 

Data from LoRaWAN experiments are publicly available at the AERPAW datasets webpage~\cite{aerpaw_datasets} and in Dryad research repository~\cite{sergio2024lorawanperformance}.

\subsection{Possible Uses of Dataset}
The use cases of this data set include:   

\noindent\textbf{LoRaWAN Optimization for Mobile Aerial Applications:}
The dataset enables evaluation of LoRaWAN performance in both urban and rural environments under mobile and aerial operating conditions. By analyzing received signal strength, packet delivery behavior, and gateway visibility as functions of UAV position and altitude, researchers can assess coverage limitations and inform gateway placement and density strategies. These measurements are particularly relevant for IoT deployments involving mobile or airborne platforms, which remain comparatively underrepresented in existing LoRaWAN measurement studies.

\noindent\textbf{LoRa Physical-Layer Characterization:}
The dataset supports empirical analysis of LoRa physical-layer behavior under varying channel conditions, including changes in noise floor and received signal strength. The availability of altitude- and orientation-tagged measurements enables characterization of air-to-ground LoRa propagation trends and transmission reliability. Such data can be used to evaluate physical-layer robustness and to inform propagation modeling efforts for low-power wide-area aerial communication scenarios.

\section{Multipath Propagation Dataset}\label{sec:multipath}

The behavior of radio signals as they propagate through the environment is a key factor in the design and performance of various wireless systems, ranging from radars to cellular networks.
In this context, multipath propagation datasets provide a crucial pathway to a deeper understanding of wireless communication principles, which ultimately contributes to robust wireless systems.
This section describes the hardware and software components of some of a propagation dataset acquired with an open-source channel sounder \cite{channel_sounder_github, gurses_sichitiu_a2g_uav} from the AERPAW testbed platform.

\subsection{Description of Hardware and Software}

\begin{figure}[!t]
    \centering
    \subfloat[Experiment setup for the USRP~B210-equipped portable node and UAV.]{
        \includegraphics[width=\columnwidth, trim={0 0.5cm 0 0}, clip]{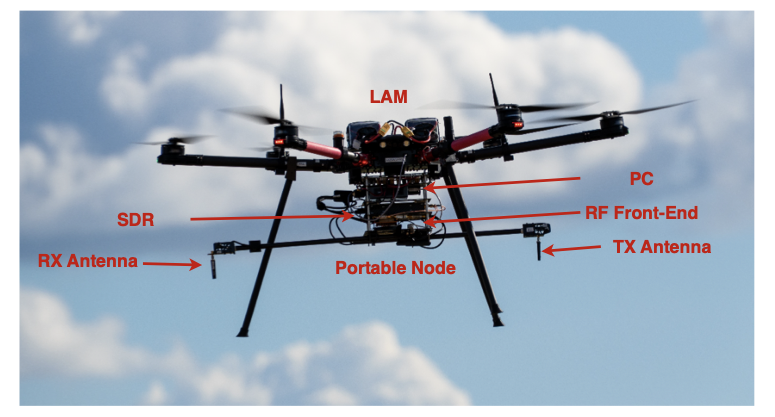}
        \label{fig:cs_a2g_onground}
    }
    \hfill
    \subfloat[Custom-designed GNSSDO system.]{
        \includegraphics[width=\columnwidth, trim={0 0.5cm 0 0}, clip]{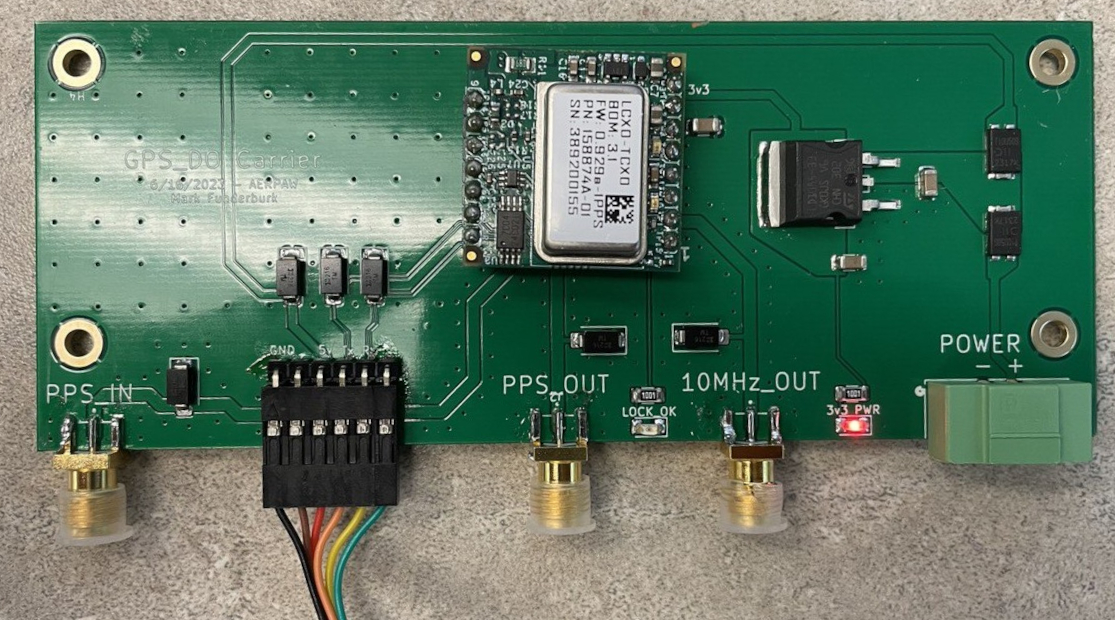}
        \label{fig:AERPAW_GNSSDO}
    }\\[-1.5mm]
    \subfloat[Overall system architecture of the channel sounder.]{
        \includegraphics[width=\columnwidth, trim={0 0.5cm 0 0}, clip]{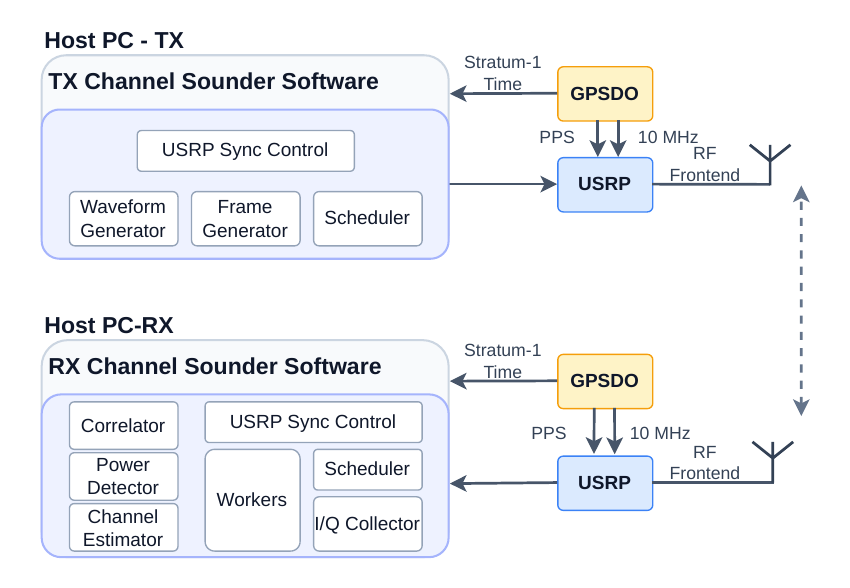}
        \label{fig:arch-channel-sounder}
    }
    \caption{Hardware and system architecture used for A2G multipath channel sounding measurements.
    \textbf{(a)} Field deployment of the portable measurement node based on a USRP~B210, mounted on a UAV platform for aerial data collection.
    \textbf{(b)} Custom-designed GNSSDO providing high-precision timing and synchronization.
    \textbf{(c)} Block diagram of the channel sounder, highlighting key components and signal flow used during the A2G experiments.}
    \label{fig:a2g_GNSSDO_arch}
\end{figure}

The experimental setup involved a UAV and a fixed node located at the AERPAW Lake Wheeler testbed site in Raleigh, North Carolina.
The UAV carries a portable node equipped with a USRP B210, RF
front-end, Intel NUC, and a custom-designed Global Navigation Satellite System disciplined oscillator~(GNSSDO), as shown in Fig.~\ref{fig:cs_a2g_onground}.
The fixed node is configured with identical equipment.

Accurate characterization of A2G multipath wireless channels requires precise temporal alignment between the transmitting and receiving devices.
Multipath propagation inherently introduces time-varying delays and phase shifts, and without stringent synchronization, these effects can be misinterpreted as genuine channel behavior.
The custom-designed GNSSDO system, as shown in Fig.~\ref{fig:AERPAW_GNSSDO}, employed by AERPAW ensures a stable and common timebase for both the UAV and fixed node, minimizing phase ambiguity and enabling reliable quantification of multipath components and time-domain channel impulse responses~(CIRs). 
The GNSSDO system achieves 2.5~ns pulse-per-second~(PPS) accuracy between the nodes, facilitating high-fidelity channel measurements.

Fig.~\ref{fig:arch-channel-sounder} presents the overall architecture of the channel sounder software.
The software, developed in Cython, enables real-time processing of the USRP B210's full bandwidth (56~MHz) and beyond, as described in~\cite{channel_sounder_github}. 
The USRP B210 is synchronized using GNSSDO-generated PPS and 10~MHz reference signals, ensuring accurate timing control and reliable data acquisition. 
The system supports multiple configurable sounding waveforms, including Zadoff-Chu, pseudo-noise~(PN), and chirp sequences.
For the measurements in this study, a Zadoff-Chu sequence of length 401 and root index 200 was used, with each sequence repeated four times.
The sounding was conducted at a measurement frequency of 4~Hz, with a center frequency of 3564, 3620, or 3686~MHz, and a transmit power of 19~dBm.
A total of nine flight experiments were performed at three altitudes (30, 60, and 90 meters) along a 500-meter flight path, with a flight speed of 5~m/s.
This configuration supports a wide range of channel sounding scenarios with high temporal and spatial resolution.

\subsection{Dataset Format}

The A2G channel sounding dataset is released using the SigMF specification~\cite{sigmf_spec}, enabling standardized representation of raw signal measurements together with comprehensive metadata. The dataset captures synchronized complex baseband samples acquired during UAV-based channel sounding experiments and supports detailed analysis of multipath propagation, path loss, and Doppler effects.

The associated metadata describe measurement parameters such as sampling rate, center frequency, capture timing, and waveform configuration, along with UAV state information including geographic location, altitude, and experimental context. Additional fields document the Zadoff--Chu sequence parameters and synchronization settings used during the sounding process, enabling reproducible extraction of channel impulse responses and derived propagation metrics.

\begin{figure*}[!t]
    \centering
    \subfloat[Representative CIR from channel sounding experiment.]{
        \includegraphics[width=\textwidth, trim={1cm 2cm 1cm 2cm}, clip]{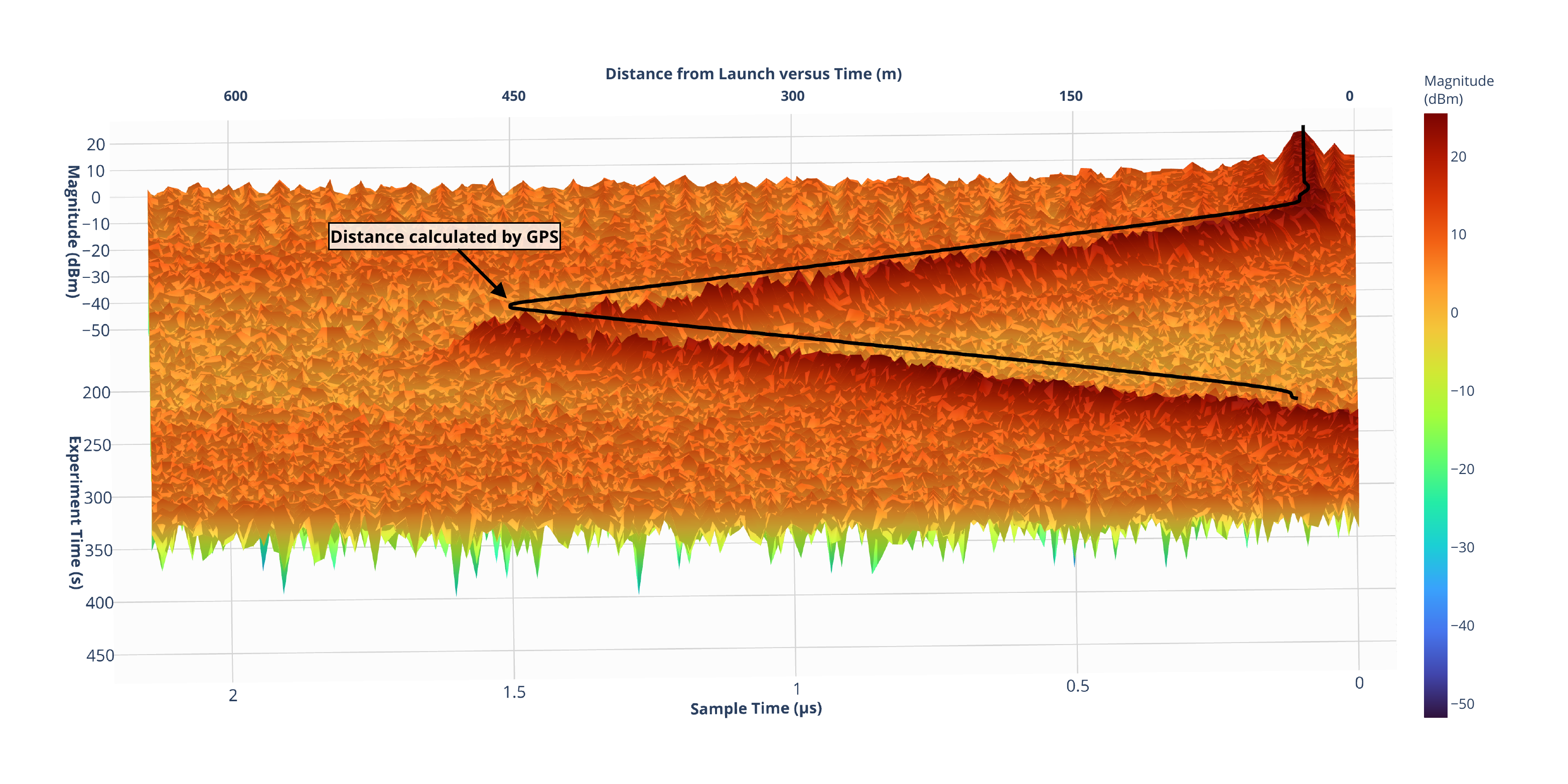}
        \label{fig:cir_a2g_uav}
    }
    \hspace{1.5mm}
    \subfloat[Representative path loss measurement from channel sounding experiment.]{
    \includegraphics[width=\textwidth, trim={0 0 0 2cm}, clip]{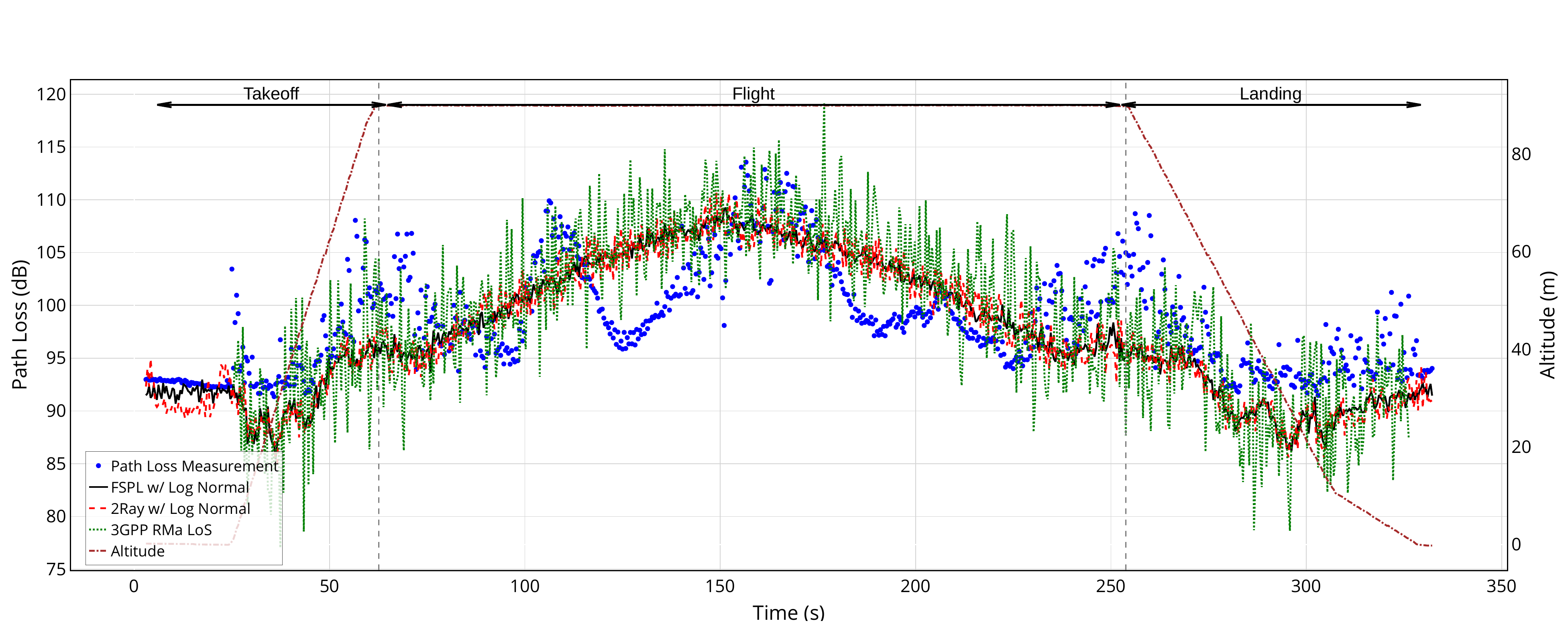}
        \label{fig:pl_a2g_uav}
    }
    \caption{Representative results from the A2G channel sounding campaign using a UAV-mounted transmitter.
\textbf{(a)} Representative CIR illustrating channel impulse response characteristics observed during flight.
\textbf{(b)} Corresponding path loss measurement as a function of time, recorded at an altitude of 90 meters.}
    \label{fig:a2g_rep_results}
\end{figure*}

Representative results obtained from the dataset are illustrated in Fig.~\ref{fig:a2g_rep_results}, including an example channel impulse response and corresponding path loss measurements collected during a UAV flight at 90~m altitude. These examples demonstrate the suitability of the dataset for multipath characterization and propagation modeling.

The dataset is accompanied by post-processing tools that support channel response extraction and visualization using standard signal processing workflows. Detailed file structure, binary format descriptions, and post-processing utilities are documented in Appendix~\ref{app:file_structure}. This organization ensures interoperability with third-party tools while preserving full reproducibility.

\subsection{Representative Results}

This section presents representative results from channel sounding experiments conducted at the AERPAW LWRFL, a rural environment designed for wireless experimentation.
Fig.~\ref{fig:cir_a2g_uav} illustrates a representative CIR obtained using correlation-based processing of the received Zadoff-Chu sequences.
The three-dimensional plot shows the evolution of multipath components over time during a UAV flight.
The horizontal axis represents the delay (in microseconds), the vertical axis indicates the experiment time (indexed per snapshot), and the color scale denotes the received signal magnitude in dB.
The black trace projected onto the back plane indicates the UAV's GPS distance from the fixed transmitter as a function of time. 
The variation in delay spread and path power over time reflects the dynamic nature of the A2G propagation channel, including the impact of UAV motion and altitude variation.
Fig.~\ref{fig:pl_a2g_uav} shows the corresponding path loss measurement over the full flight trajectory. 
The UAV's altitude profile is overlaid to highlight different flight phases, including takeoff, flight, and landing. 
The results demonstrate a clear relationship between the received power and UAV position, consistent with expected large-scale path loss behavior.

\subsection{Possible Uses of Dataset}

The provided dataset enables a wide range of research opportunities in the study and modeling of A2G wireless communication channels.
Given the synchronized high-resolution measurements, as well as the availability of UAV position data, the dataset is well-suited for the following applications:

\noindent\textbf{Air-to-Ground Channel Modeling:}
The dataset enables extraction of key air-to-ground propagation characteristics, including delay spread, Doppler profiles, path loss behavior, and coherence bandwidth. These measurements support the development and evaluation of statistical and geometry-based A2G channel models grounded in empirical observations.

\noindent\textbf{Machine Learning for Wireless Systems:}
The dataset supports training and evaluation of machine learning models for UAV-based wireless applications. The availability of synchronized signal measurements and metadata enables data-driven approaches to tasks such as channel estimation and link quality prediction under realistic aerial operating conditions.

\noindent\textbf{Impact of Altitude and Mobility:}
Measurements collected at multiple altitudes and along controlled flight trajectories enable analysis of how UAV height and mobility influence propagation conditions and coverage. The dataset allows researchers to isolate altitude- and speed-dependent effects on channel behavior.

\noindent\textbf{Waveform and System Design Evaluation:}
The raw I/Q samples and associated metadata support simulation and evaluation of waveform and receiver designs under realistic channel conditions. These data enable assessment of system performance using experimentally captured aerial signals rather than idealized channel assumptions.

\noindent\textbf{Validation of Ray-Tracing and Analytical Models:}
The dataset provides empirical reference data that can be used to validate and calibrate ray-tracing simulations and analytical propagation models. Comparisons between measured and modeled results support assessment of model accuracy in rural air-to-ground environments.

\section{Wireless Localization Dataset}\label{sec:localization_cole}

\begin{figure*}[t]
    \centering
    \subfloat[Keysight N6841A RF sensor deployed on AERPAW tower LW3.]{
        \includegraphics[width=0.3\textwidth]{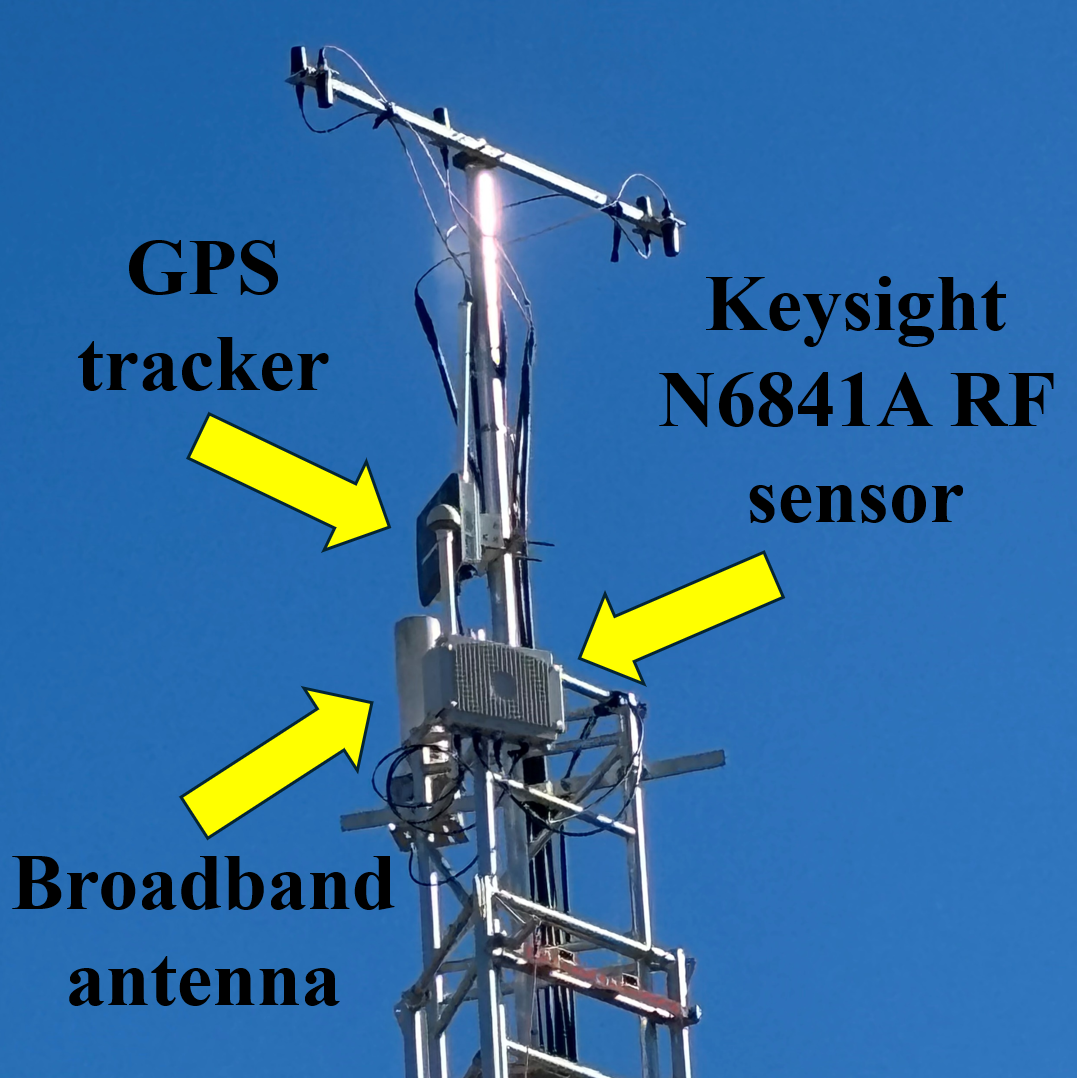}
        \label{fig:rfsensor}
    }
    \hspace{1.5mm}
    \subfloat[AERPAW UAV with SDR portable node.]{
        \includegraphics[width=0.3\textwidth]{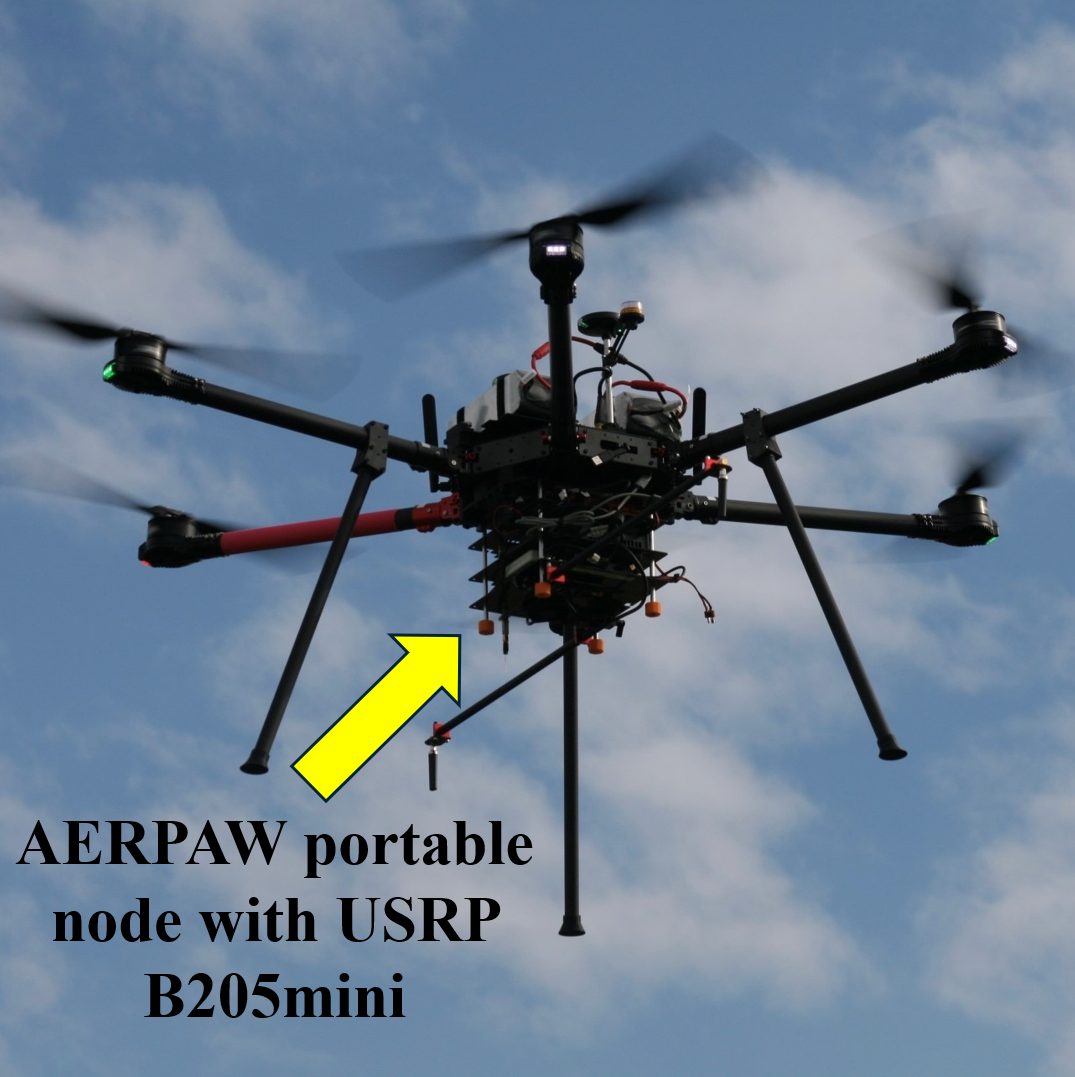}
        \label{fig:uavnode}
    }
    \hspace{1.5mm}
    \subfloat[AERPAW's LWRFL and RF sensor tower locations at LW2, LW3, LW4, LW5.]{
        \includegraphics[width=0.3\textwidth]{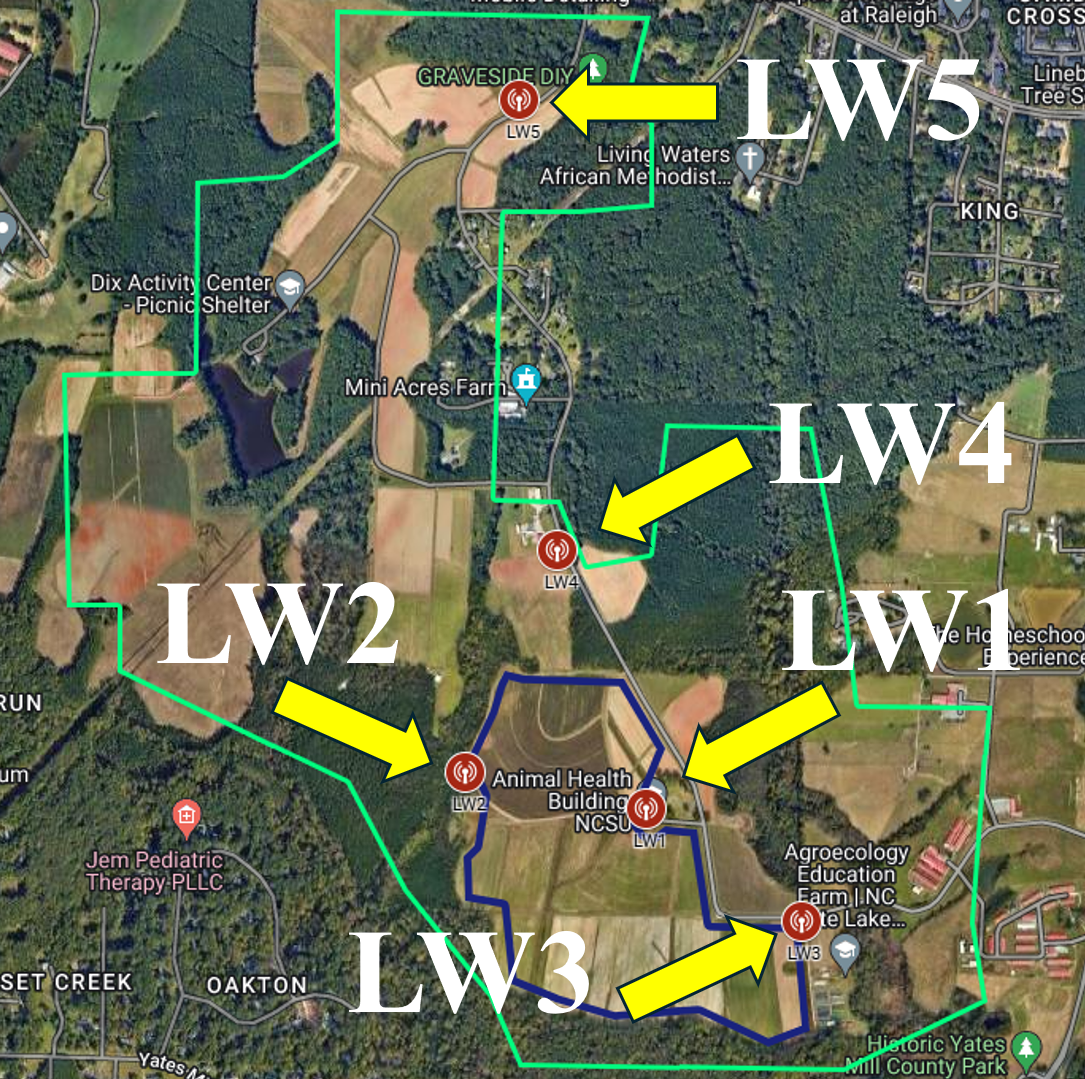}
        \label{fig:lwfrl}
    }
    \caption{Overview of the AERPAW testbed showing the RF sensor, UAV node, and LWRFL site layout. \textbf{(a)} Keysight N6841A RF sensor deployed on AERPAW tower LW3, \textbf{(b)} AERPAW UAV with SDR portable node, and \textbf{(c)} AERPAW's LWRFL and RF sensor tower locations at LW2, LW3, LW4, LW5.}
    \label{fig:localizationoverview}
\end{figure*}

The decreasing cost of UAVs has led to rapid adoption across sectors such as defense, precision agriculture, aerial communications, search-and-rescue, and spectrum monitoring and enforcement. However, their growing presence introduces new challenges for security and airspace management, particularly as UAV activity increases around critical infrastructure. Effective UAV detection and tracking are essential to developing a secure UAV traffic management (UTM) ecosystem \cite{faa2020utm}. A range of studies have explored the use of RF signals for UAV detection, classification, localization, and tracking---including in scenarios involving non-cooperative or potentially malicious drones. Among the various RF-based approaches, Time Difference of Arrival (TDOA)---a multilateration technique that estimates UAV positions by measuring differences in signal arrival times at spatially separated sensors---has been shown to be particularly effective for passive RF sensing and UAV localization \cite{guvenc2018detection, azari2018key}. To support further research, this section introduces two UAV localization and tracking datasets collected using the AERPAW testbed platform, providing researchers with valuable resources for evaluating TDOA-based tracking methods in real-world scenarios.

\subsection{Description of Hardware and Software}

UAV flight experiments were conducted in AERPAW's controlled environment to generate both datasets, using Keysight N6841A RF sensors to collect TDOA measurements. The N6841A (Fig.~\ref{fig:rfsensor}) is capable of detecting, recording, and precisely time-stamping RF signals across a frequency range of 20~MHz to 6~GHz, with a maximum bandwidth of 20~MHz. Equipped with a broadband omnidirectional antenna and GPS-based timestamping, the system uses Keysight's N6854A Geolocation Software and Sensor Management Tool to support TDOA, RSS, and hybrid localization methods for tracking RF sources within approximately a 2 km radius. However, the software is limited to 2D localization and does not estimate altitude.

Fig.~\ref{fig:lwfrl} shows the deployment of these sensors at the LWRFL, where a single N6841A unit is mounted on each of the four towers labeled LW2 through LW5, approximately 10 meters above ground level. The sensors are deployed in a rural environment with mixed LoS conditions due to tree cover and building obstructions. All RF sensors within AERPAW are synchronized using a shared GPS-disciplined clock infrastructure and operate on the same local network, with individual IP addresses assigned within a common subnet. Centralized management via the Keysight Geolocation Server ensures network-level time synchronization, which is essential for accurate TDOA-based localization. During experiments, the N6841A units capture I/Q data from RF signals, which are subsequently processed to estimate UAV positions. However, the specific algorithms used for TDOA extraction and position estimation are proprietary to Keysight and not publicly available.

In our previous work \cite{dickerson2025tdoa}, a 3.32~GHz channel sounding waveform was transmitted from a UAV-mounted SDR (Fig.~\ref{fig:lwfrl}), and the RF sensors localized the UAV along multiple repeated trajectories. The dataset associated with this study is publicly available at~\cite{dickerson2025aerpaw}. In a separate study~\cite{bhattacherjee2022experimental}, TDOA-based localization was performed using downlink control signals from a DJI Inspire 2 UAV operating in the 2.400--2.483~GHz ISM band with a 20~MHz bandwidth. The dataset corresponding to this experiment is also publicly available at~\cite{uditalocalization}.


\subsection{Dataset Format}

The TDOA-based UAV localization datasets associated with~\cite{dickerson2025tdoa,dickerson2025aerpaw} provide time-synchronized localization estimates and performance metrics collected during multiple UAV flight experiments on the AERPAW platform. Each dataset corresponds to an individual flight conducted at a specified altitude and signal bandwidth, enabling controlled evaluation of TDOA localization accuracy under varying propagation conditions.

For each flight, the dataset records estimated UAV positions together with ground-truth coordinates, center frequency information, timestamps, and localization performance indicators. These indicators include correlation-based metrics and error statistics that quantify localization accuracy relative to known ground truth. In addition, the datasets include binary indicators describing LoS or NLoS conditions between the UAV transmitter and multiple AERPAW RF sensor towers, enabling analysis of the impact of propagation conditions on localization performance.

A related set of TDOA localization datasets reported in~\cite{bhattacherjee2022experimental,uditalocalization} follows a similar structure but differs in the representation of LoS information and the separation of measurement and ground-truth data. Together, these datasets support benchmarking of TDOA localization algorithms, evaluation of LoS/NLoS effects, and comparative studies across experimental configurations.

The datasets are released in structured CSV format, along with auxiliary processing tools to support visualization and performance evaluation. Detailed file inventories, CSV field descriptions, and helper script information are provided in Appendix~\ref{app:file_structure}.

\subsection{Representative Results}

Representative examples from the collected datasets are presented in this section to demonstrate typical localization behavior and characteristics.

\begin{figure}[!t]
	\centering
        \subfloat[UAV trajectory at 40 meter altitude color-coded by LoS conditions to towers LW2-4.]{\includegraphics[width=0.95\linewidth]{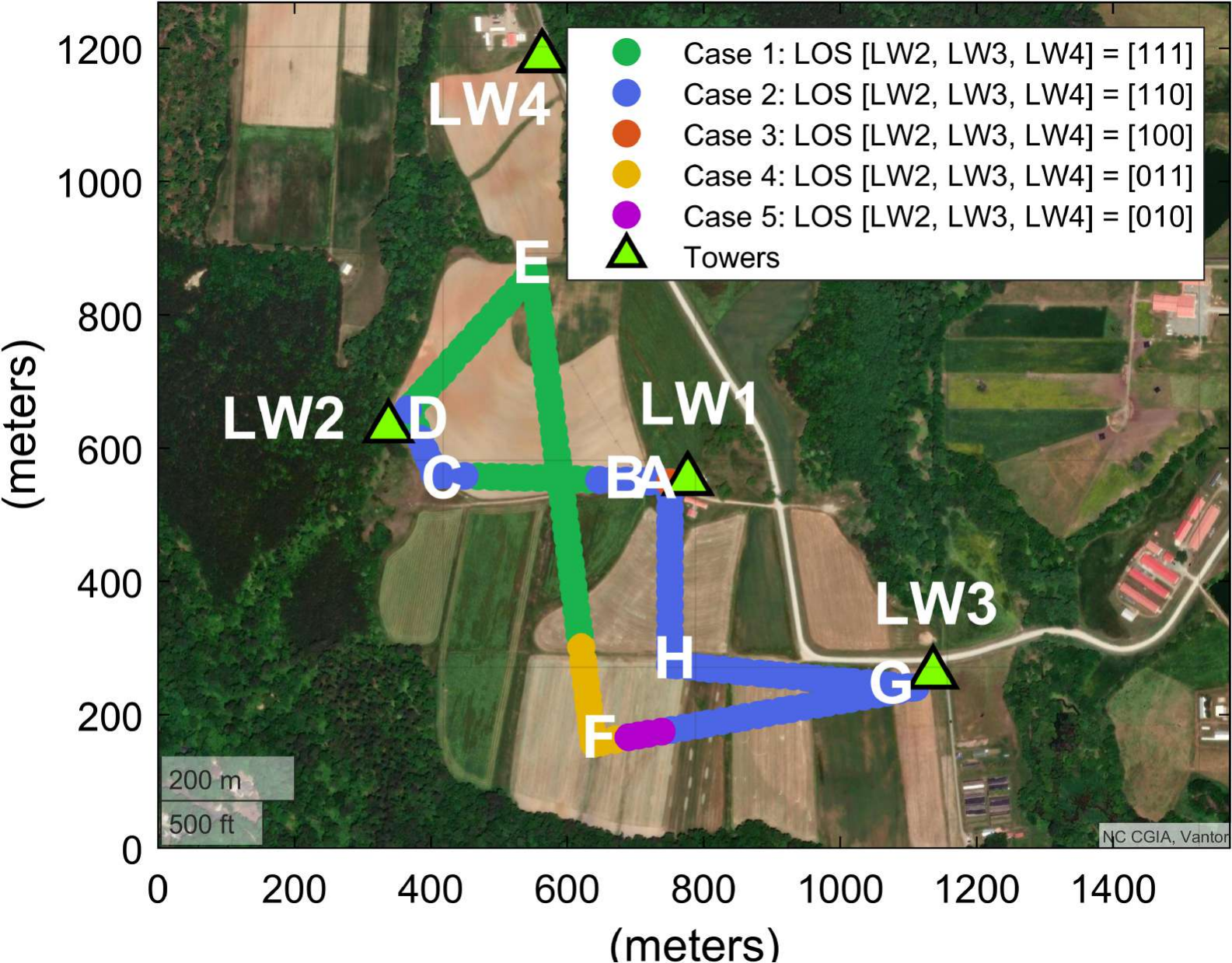}\label{fig:40mtraj}}
        \vspace{-0.02in}
        \subfloat[Ground truth UAV trajectory color-coded by localization error.]{\includegraphics[width=\linewidth]{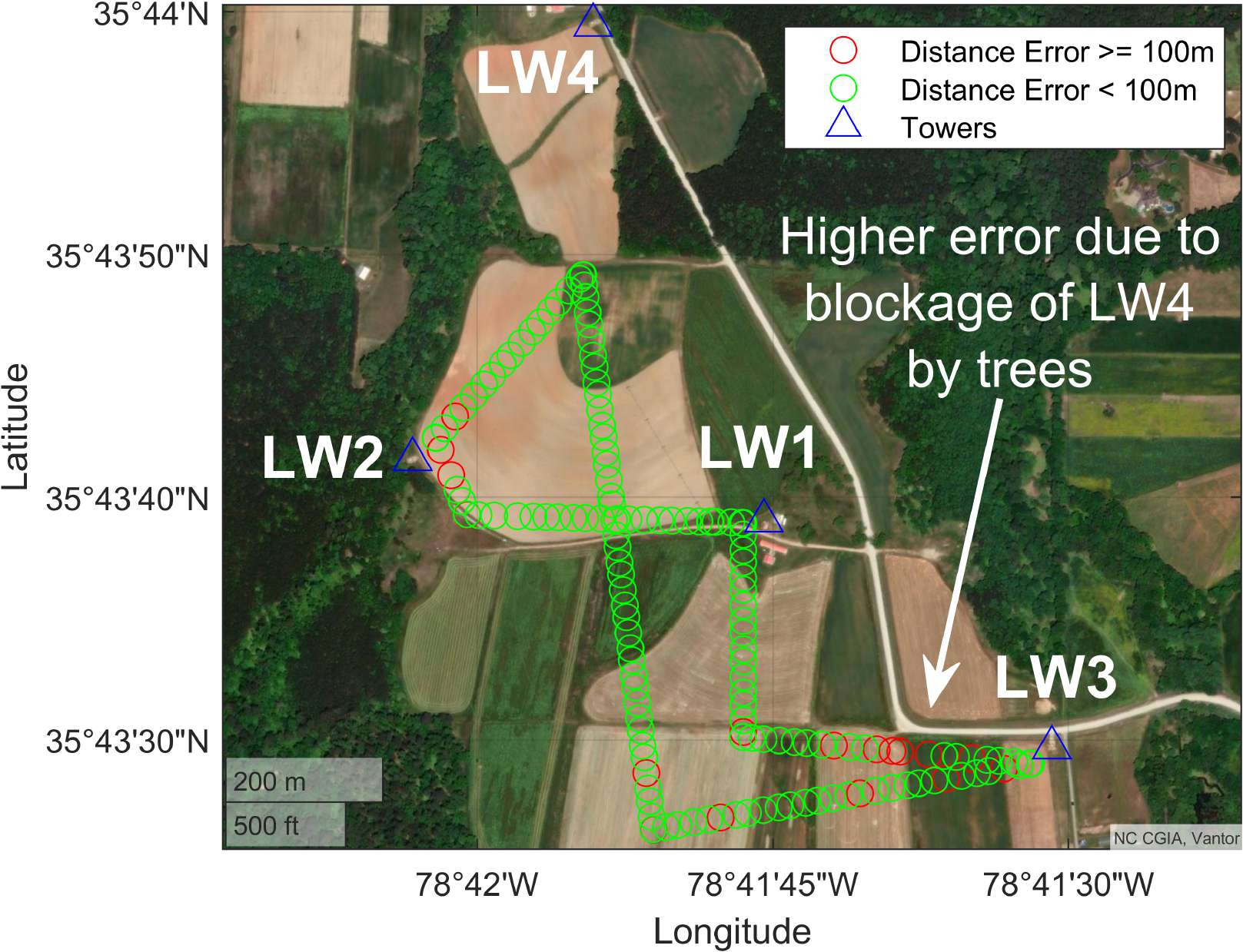}\label{fig:error_map}}
        \vspace{-0.01in}
	\caption{UAV trajectory and localization performance at 40~m altitude.
\textbf{(a)} UAV flight trajectory color-coded by LoS visibility to cellular towers LW2 through LW4, based on geometry and terrain data.
\textbf{(b)} Ground truth trajectory of the UAV color-coded by the magnitude of localization error, highlighting spatial variation in positioning accuracy.}
\label{fig:40mtraj_error_map}\vspace{-0.1in}
\end{figure}

Fig.~\ref{fig:40mtraj} illustrates the UAV's 40-meter altitude trajectory overlaid on the LWRFL site map, showing LoS conditions relative to three RF sensor towers (LW2, LW3, and LW4), which are indicated by green triangles. The trajectory is color-coded based on five distinct LoS scenarios, where each case indicates LoS or NLoS status relative to the three towers. For example, Case 1 (green) represents segments where the UAV maintained LoS to all three towers, while Case 5 (purple) corresponds to areas where only LW3 maintained LoS. The color-coding of the trajectory clearly highlights frequent transitions between LoS conditions as the UAV moves through the field, illustrating the mixed LoS/NLoS environment characteristic of the LWRFL site. A key takeaway from this figure is that LoS availability varies substantially along the trajectory, even at a fixed altitude, leading to position-dependent measurement quality for TDOA-based localization.

Fig.~\ref{fig:error_map} presents a color-coded visualization of localization error along the UAV's trajectory. Each point represents a ground truth UAV position, with circles indicating the corresponding localization error relative to the nearest estimated coordinate. Green circles mark locations where the position error is less than 100 meters, while red circles identify points where the error exceeds 100 meters. Blue triangles indicate the locations of the RF sensor towers used for TDOA-based localization.

As shown in the figure, the majority of the trajectory is associated with low localization error (green), suggesting consistent and accurate TDOA-based position estimates along most of the UAV's flight path. Higher error regions (red) are concentrated in specific trajectory segments that coincide with reduced LoS availability (due to foliage obstructions) or unfavorable sensor geometry. Together, these results highlight two key observations: (i) reliable localization is achievable across most of the trajectory when favorable LoS conditions are available, and (ii) localization performance degrades in predictable regions where LoS conditions or geometric dilution of precision (GDOP) are poor. This emphasizes the strong coupling between propagation conditions, sensor geometry, and localization accuracy in real-world environments.

\subsection{Possible Uses of Dataset}

The UAV TDOA localization datasets serve as a valuable resource for advancing research in RF-based localization, UAV tracking, and passive sensing systems. Its real-world measurements, mixed LoS/NLoS conditions, and ground-truth references enable exploration of the following research directions:

\noindent\textbf{Tracking Filter Development and Evaluation:}
The dataset supports the development and evaluation of tracking filters, such as Kalman and particle filters, by providing real-world UAV trajectory and localization measurement sequences. These data enable state estimation studies under mixed line-of-sight and non-line-of-sight conditions.

\noindent\textbf{TDOA Localization Algorithm Benchmarking:}
The dataset enables performance evaluation and benchmarking of TDOA-based localization algorithms using real-world RF measurements. It supports direct comparison against industry-standard implementations, including Keysight-based TDOA localization, under mixed LoS/NLoS conditions.

\noindent\textbf{Multi-Sensor Fusion:}
The dataset supports the development and evaluation of multi-sensor fusion algorithms by combining TDOA-based RF localization data with complementary sensing modalities. Such integration enables improved UAV tracking and state estimation in complex propagation environments.

\noindent\textbf{Non-Line-of-Sight Modeling and Analysis:}
The dataset enables characterization and modeling of NLoS effects by providing real-world localization errors alongside explicit LoS/NLoS labels. These data support studies on NLoS detection, mitigation strategies, and bias-aware localization techniques.

\noindent\textbf{CRLB Modeling and Validation:}
The dataset supports theoretical performance analysis through CRLB modeling and enables empirical validation by comparing theoretical bounds against measured TDOA-based localization errors under varying sensor geometries and environmental conditions.

Together, these research directions highlight the dataset's value as a resource for advancing localization, tracking, and sensing technologies in real-world UAV applications.

\section{UAV-collected RSS Measurements for RF Source Localization}\label{sec:localization_saad}
\begin{figure}
\includegraphics[width=0.9\linewidth]{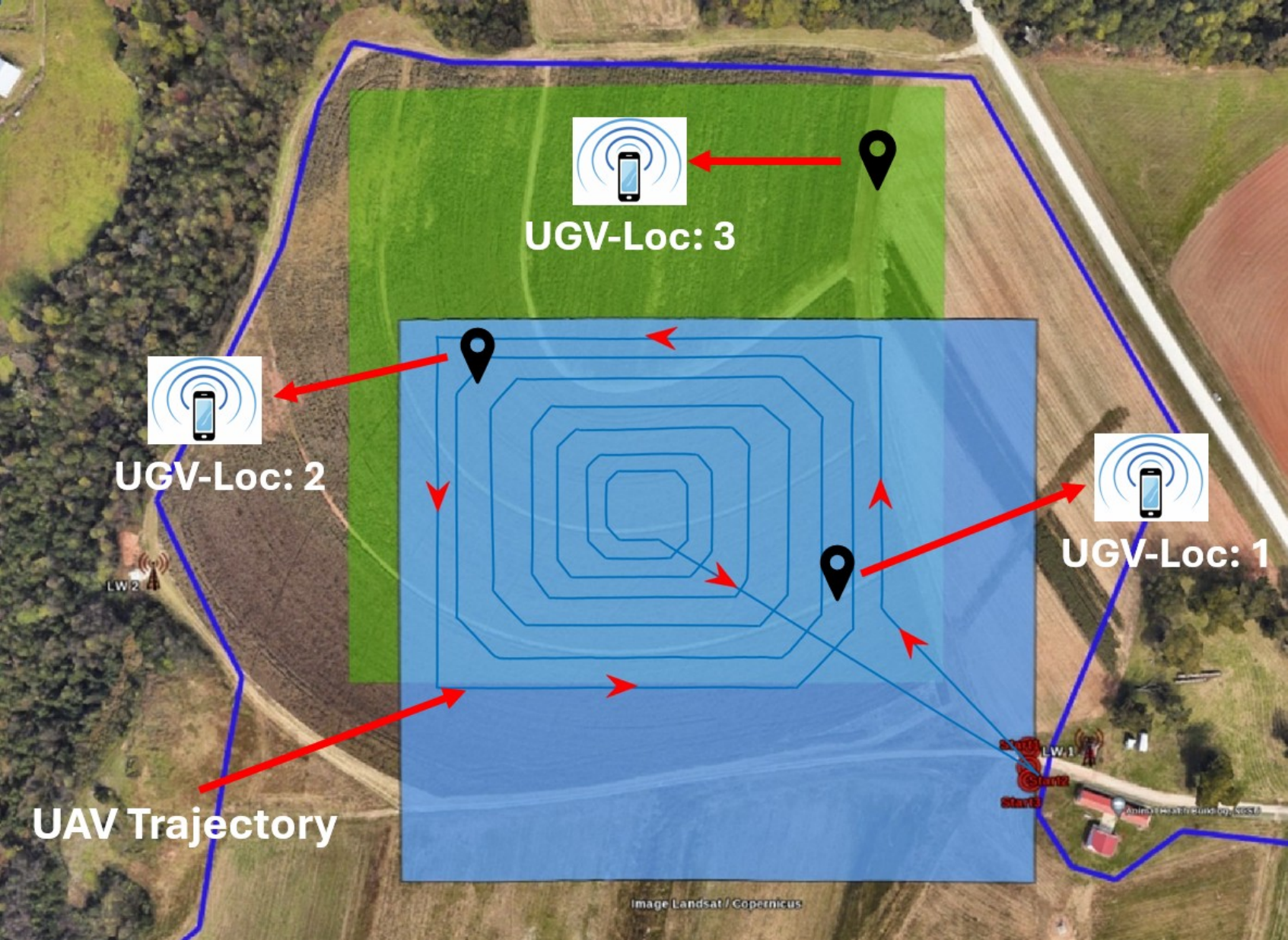}
	\centering
	\caption{AFAR Challenge setup highlighting the UAV flight zone (blue) and the designated area where the UGV could be hidden (green). The UAV used during the experiment is shown in Fig.~\ref{fig:localizationoverview}.}
	\label{fig1:AFAR}
\vspace{-0.3cm}
\end{figure}

The AERPAW Find A Rover~(AFAR) Challenge~\cite{aerpaw_afar_challenge} was a national-level competition designed to promote research in UAV-assisted RF localization. Organized under the AERPAW testbed, the challenge aimed to accelerate innovation by providing a standardized experimental environment for evaluating RF localization algorithms using UAVs.  In this competition, UAVs were deployed to locate an RF-emitting unmanned ground vehicle~(UGV) based solely on signal measurements. In the AFAR Challenge, the UGV could be placed anywhere within a designated search area  (marked in green in Fig.~\ref{fig1:AFAR}), while the UAV was restricted to fly in the flight zone (marked in blue in Fig.~\ref{fig1:AFAR}). Teams were free to design either autonomous or fixed waypoint-based UAV trajectories to locate the UGV, with flight constraints of 20-110 meters altitude and speeds up to 10 m/s.
The challenge consisted of two phases: development in a DT environment, and deployment in a real-world testbed at LWRFL, NC, as detailed in \cite{masrur2025collection}. Each of the five finalist university teams independently devised UAV flight trajectories and localization algorithms as part of the competition.

\subsection{Description of Hardware and Software}
The AFAR dataset includes data collected from both a DT simulation environment and a real-world wireless testbed, both deployed via the AERPAW platform. Each experimental run involved a UAV serving as the receiver and a UGV operating as the RF signal transmitter. The UAV and UGV were equipped with portable SDR nodes based on the USRP B205mini.

Each SDR was connected to an Intel NUC 10 mini-PC equipped with an i7-10710U processor, 64~GB of RAM, and a 1~TB SSD, enabling real-time onboard signal processing. Transmissions employed a GNU Radio-based channel sounder that generated a pseudo-random bit sequence~(PRBS) of length 4095 using a degree-12 Galois LFSR. The sequence was pulse-shaped using a root-raised cosine filter and transmitted at a 2 MHz sampling rate over 3.0–4.2 GHz using a wideband antenna. At the receiver, frequency offset correction and correlation with the original PRBS enabled extraction of the CIR, from which RSS and received signal quality~(RSQ) values were derived.

The DT environment mirrored the physical setup using containerized software emulation. A virtual USRP (V-USRP) and a channel emulator VM (CHEM-VM) simulated RF propagation based on real-time UAV-UGV position updates. Experiment logic ran in Experiment VMs (E-VMs), and UAV mobility was emulated using Software-In-The-Loop (SITL) vehicles, all orchestrated through AERPAW's geofencing and control interfaces.

\subsection{Dataset Format}

The AFAR challenge dataset is organized to reflect the structure of the UAV-based RF source localization experiments conducted during the competition. Data were collected from five finalist teams across three distinct UGV transmitter placements and two environments, namely a DT simulation and the real-world AERPAW testbed. This results in a total of 30 experimental scenarios, enabling systematic evaluation of localization performance across teams, environments, and deployment configurations.

For each experiment, the dataset includes time-synchronized received signal measurements, signal quality indicators, and UAV navigation data, allowing joint analysis of RF observables and UAV motion. The measurements capture both RSS and RSQ, together with precise timestamps and UAV state information, supporting trajectory-aware localization and algorithm benchmarking.

The dataset structure supports comparative analysis between simulated and real-world environments, as well as cross-team and cross-location evaluation. All data are released in structured, machine-readable formats suitable for statistical analysis, visualization, and machine learning workflows. Detailed directory organization, file inventories, log formats, and example data snippets are provided in Appendix~\ref{app:file_structure}.

\subsection{Representative Results}
\begin{figure}[!t]
    \centering
    \subfloat[Autonomous trajectory.]
    {\includegraphics[width=0.48\linewidth]{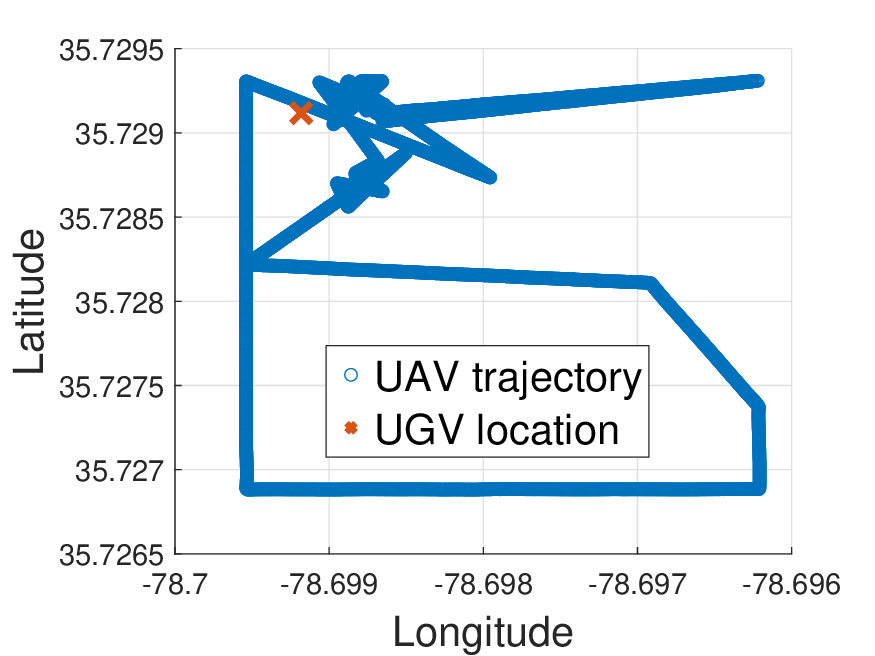}}
    \vspace{-0.02in}
    \subfloat[Fixed waypoint trajectory.]
    {\includegraphics[width=0.48\linewidth]{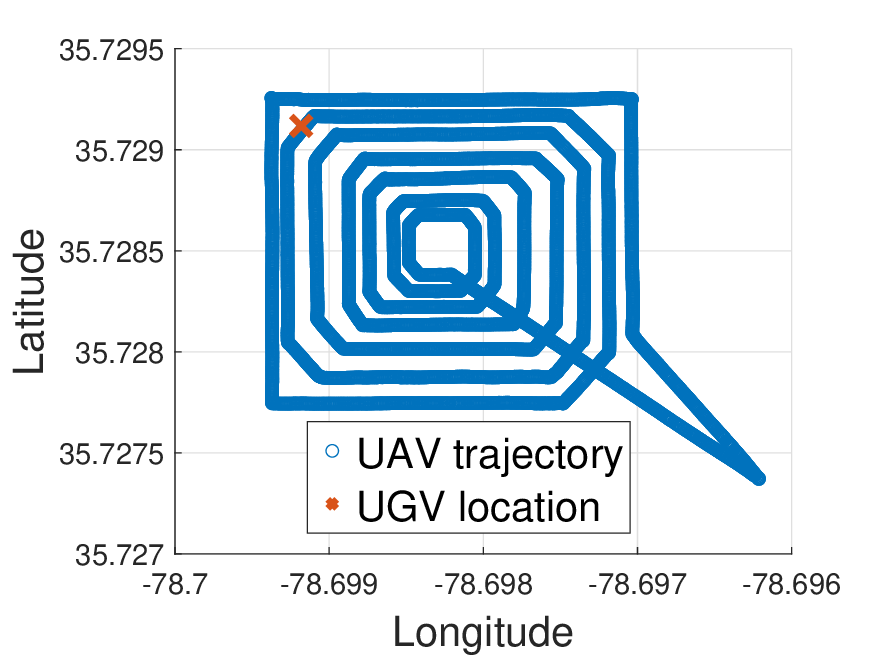}}
    \vspace{-0.01in}
    \caption{Comparison of UAV trajectories: autonomous (Team-300) vs. fixed waypoint (Team-309).}
\label{fig:Autovsfix}
\end{figure}

The AFAR dataset reflects diverse RF and mobility dynamics, with teams employing distinct UAV trajectories-three teams using autonomous trajectories and two adopting fixed waypoint trajectories. Example trajectories for Team-300 (autonomous) and Team-309 (fixed) are shown for comparison in Fig.~\ref{fig:Autovsfix}.

\begin{figure*}[th]
    \centering

    \subfloat[UAV altitude and speed over time.]{
        \includegraphics[width=0.3\textwidth]{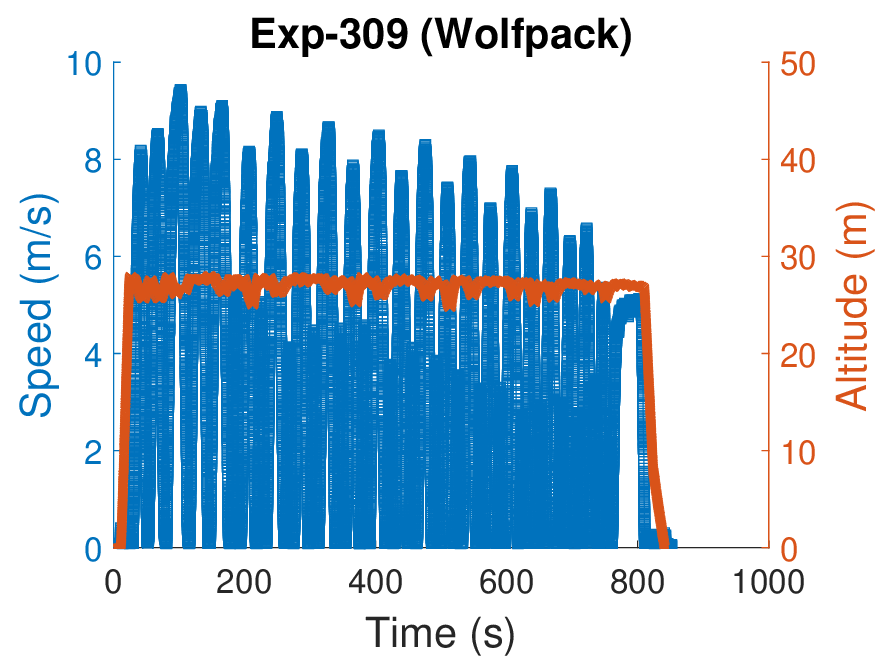}
        \label{Alt_and_Speed_309L2}
    }
    \hspace{1.5mm}
    \subfloat[RSS over the UAV trajectory.]{
        \includegraphics[width=0.3\textwidth]{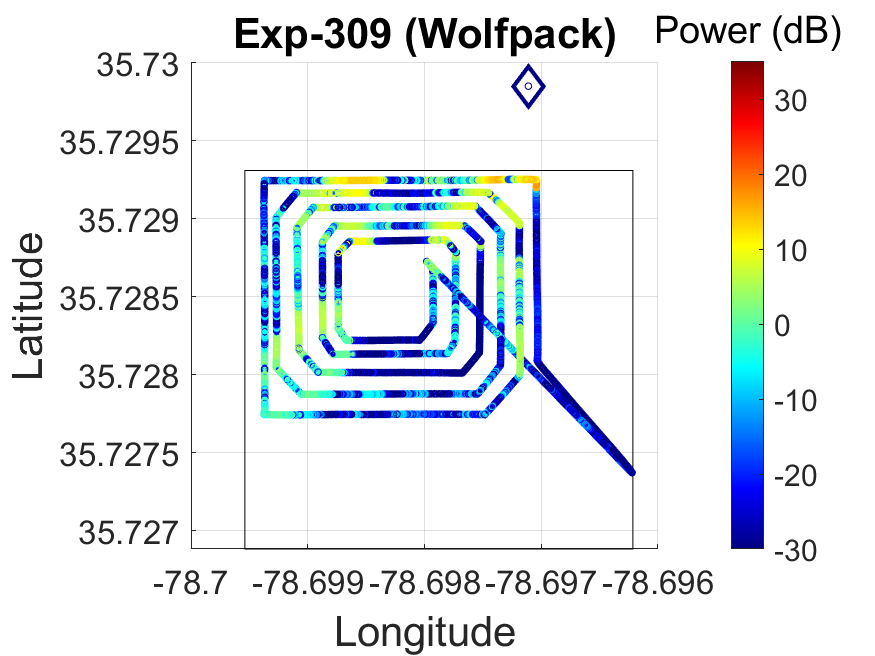}
        \label{Heatmap_309_L2}
    }
    \hspace{1.5mm}
    \subfloat[DT vs. real-world RSS and distance versus time.]{
        \includegraphics[width=0.3\textwidth]{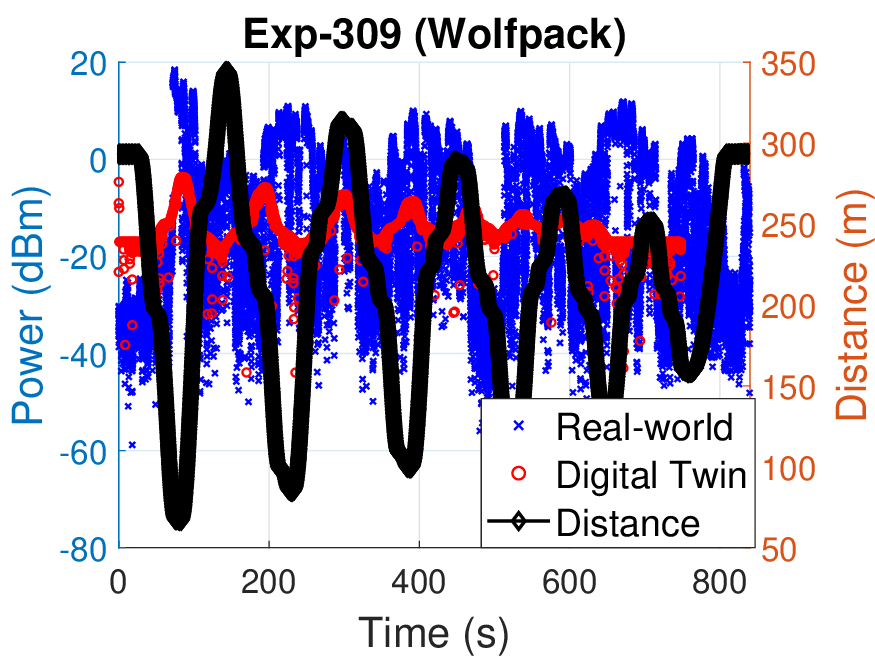}
        \label{Real_vs_DT_309}
    }
    \caption{Representative signal and mobility characteristics from the AFAR dataset for the team Wolfpack. \textbf{(a)} UAV altitude and speed over time, \textbf{(b)} RSS over the UAV trajectory, and \textbf{(c)} DT vs. real-world RSS and distance versus time.}
\label{fig:Team309}
\end{figure*}

The measurements in the AFAR dataset exhibit considerable variability resulting from differences in UAV trajectories, UGV placements, and environment-specific propagation effects. KPIs such as RSS, RSQ, and UAV motion parameters (e.g., speed and altitude) vary significantly across locations. For instance, the UAV speed and altitude profile for Team-309, who employed a fixed waypoint trajectory, are shown in Fig.~\ref{Alt_and_Speed_309L2}. As the UAV approaches each waypoint, its speed increases, then decreases upon arrival, before accelerating again toward the next target. This cyclical speed pattern is characteristic of waypoint-based navigation. Additionally, the UAV maintains a relatively constant altitude of approximately 30 meters throughout the mission.

Fig.~\ref{Heatmap_309_L2} presents RSS heatmaps overlaid on the UAV flight paths. The diamond marker indicates the RF source location. As expected, RSS is strongest when the UAV is in close proximity to the source and weakens with increasing distance. However, the spatial distribution of received power is not uniform, reflecting the influence of multipath and shadowing effects.

To highlight the contrast between simulated and real-world signal behavior, Fig.~\ref{Real_vs_DT_309} plots RSS and distance against time for both DT and real-world environments. In the DT environment, RSS remains relatively smooth and predictable. In contrast, the real-world data exhibits significant fluctuations, even at similar distances, due to dynamic factors such as fading, body blockage, and environmental clutter. This discrepancy underscores the importance of accounting for real-world propagation effects when developing and validating RF localization algorithms.

\subsection{Possible Uses of Dataset}

The AFAR dataset serves as a comprehensive resource for advancing research in wireless communications, RF-based localization, and UAV-enabled signal intelligence. Its rich content and dual-environment structure (DT and real-world) enable the following research directions:

\noindent\textbf{A2G Channel Propagation Modeling:} The dataset enables realistic modeling of A2G wireless channels, accounting for UAV-specific factors such as altitude, elevation/azimuth angles, velocity, and orientation (roll, pitch, yaw). These measurements help characterize the propagation environment under mobility and elevation diversity \cite{masrur2025bridging}.

\noindent\textbf{Antenna Gain and Shadowing Analysis:} The dataset facilitates the evaluation of directional antenna performance and gain variations due to UAV-body shadowing. This is particularly relevant for understanding signal attenuation in NLoS conditions and UAV maneuvers.

\noindent\textbf{Performance Benchmarking:} Researchers can perform comparative analysis of UAV-assisted RF localization algorithms. The dataset allows assessment of trajectory efficiency, signal quality, and overall localization accuracy under controlled and real-world constraints.

\noindent\textbf{Data-Driven Localization Algorithms:} The time-synchronized RF and positional data provide a robust basis for training and evaluating machine learning models for RF source localization, including regression, classification, or hybrid approaches.

\noindent\textbf{Signal Strength and Fading Prediction:} Deep learning models can be trained to forecast RSS/RSQ values under varying mobility conditions. This supports proactive planning in UAV-assisted sensing and communication tasks.

\noindent\textbf{Flight Path Optimization:} Using the dataset, UAV trajectories can be optimized for better link reliability, minimal energy use, or improved localization precision, either via reinforcement learning or optimization-based techniques.

\noindent\textbf{Simulation-to-Reality Transfer Learning:} The paired DT and real-world measurements enable transfer learning strategies that improve model robustness across synthetic and physical environments.

In summary, the AFAR dataset bridges multiple disciplines, offering a reproducible platform to study wireless localization, adaptive mobility strategies, and signal-aware autonomy in UAV networks.

\section{UAV Signal Classification Dataset}\label{sec:UAV_detection}

In recent times, malicious UAVs have become a global threat to society. Even in modern warfare, the use of UAVs has altered the dynamics of traditional military operations, providing strategic advantages to state actors as well as established military forces. Besides, low cost, low altitude and low speed consumer UAVs (or micro-UAVs) pose a unique threat to both military assets and civilians. As a result, researchers have been investigating different techniques for UAV identification.  Some of these techniques include RF, radars, computer vision (optical and infra-red cameras), high-energy lasers, and acoustic techniques~\cite{Wahab_survey, UAV_LowGrazing_angle}. Each of these techniques has its own advantages and challenges. 

To achieve long range detection, identification, and improved localization of a wide-range of UAVs, RF-based techniques are commonly preferred. In addition, RF-based techniques for detecting and identifying UAVs can operate in all weather conditions. As a result, UAV detection, identification (classification), and neutralization using RF-based electronic warfare~(EW) and signals intelligence~(SIGINT) systems are becoming popular. These techniques exploit electromagnetic spectrum or directed energy to detect, identify, and interdict an incoming drone. Consequently, these systems are comparatively more effective than alternative detection approaches (e.g. camera and acoustics) because they can operate in all weather conditions and achieve long detection ranges~\cite{Wahab_survey}. However, due to the ubiquitousness of electronics and communication systems, especially in the ISM band in urban centers, it could be difficult to accurately detect and identify specific UAV signals in the presence of intentional jammers and non-intentional EM radiators/interference.  To mitigate this issue, researchers at AERPAW recognized the need to develop a dataset of unique UAV RF signals extracted from popular commercial UAVs. This dataset could be used to develop signal processing algorithms and machine learning models that can improve the detection and identification of specific consumer UAVs in the presence of interference.

\begin{figure}
\includegraphics[width=0.8\linewidth]{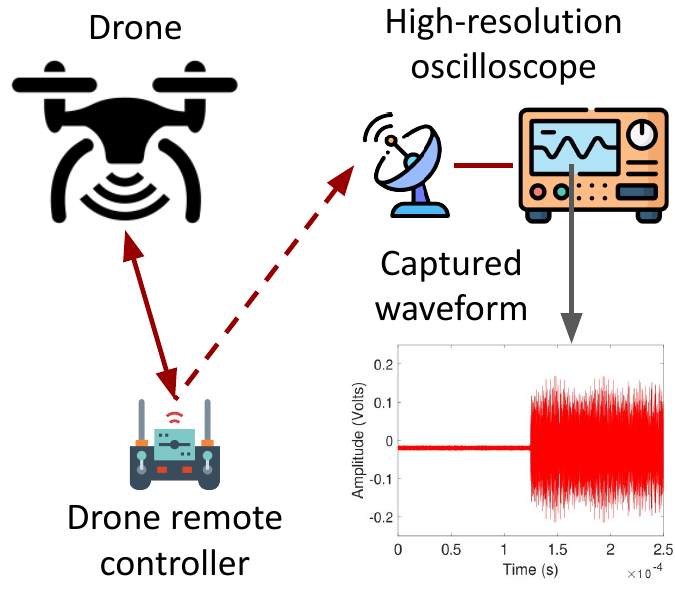}
	\centering
	\caption{The experimental setup for detecting and capturing RF signals from UAV controllers. The input of the receiver has an LAN, a bandpass filter and a parabolic antenna receiver.}
	\label{fig1:SETUP_1}
\vspace{-0.3cm}
\end{figure}

\begin{figure*}[!t]
    \centering
    \subfloat[Graupner MC-32]{
        \includegraphics[width=0.23\linewidth,trim={0cm 0cm 0cm 0cm},clip]{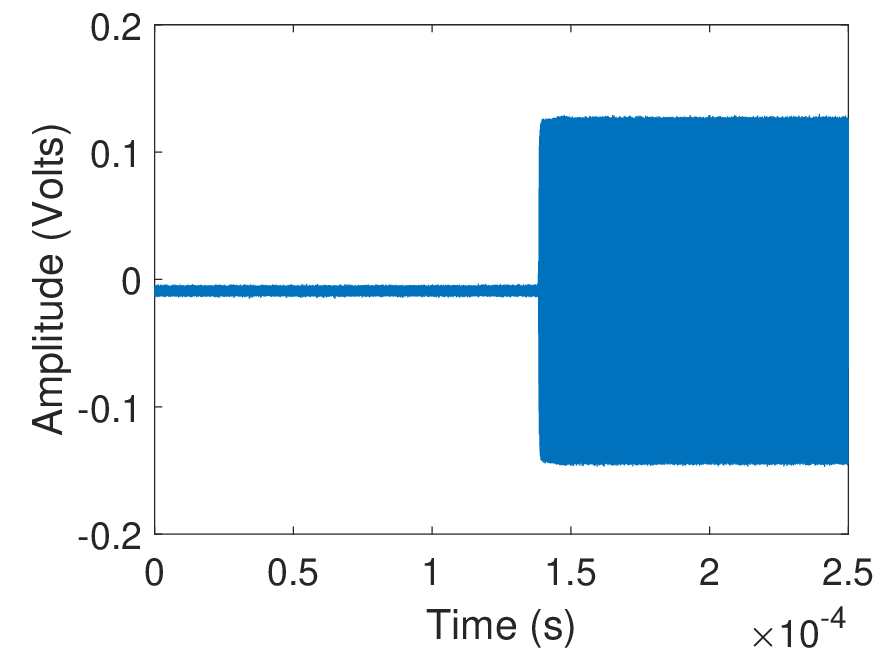}\label{fig:Graupner_MC-32}}
    \subfloat[Spektrum DX6e]{
        \includegraphics[width=0.23\linewidth,trim={0cm 0cm 0cm 0cm},clip]{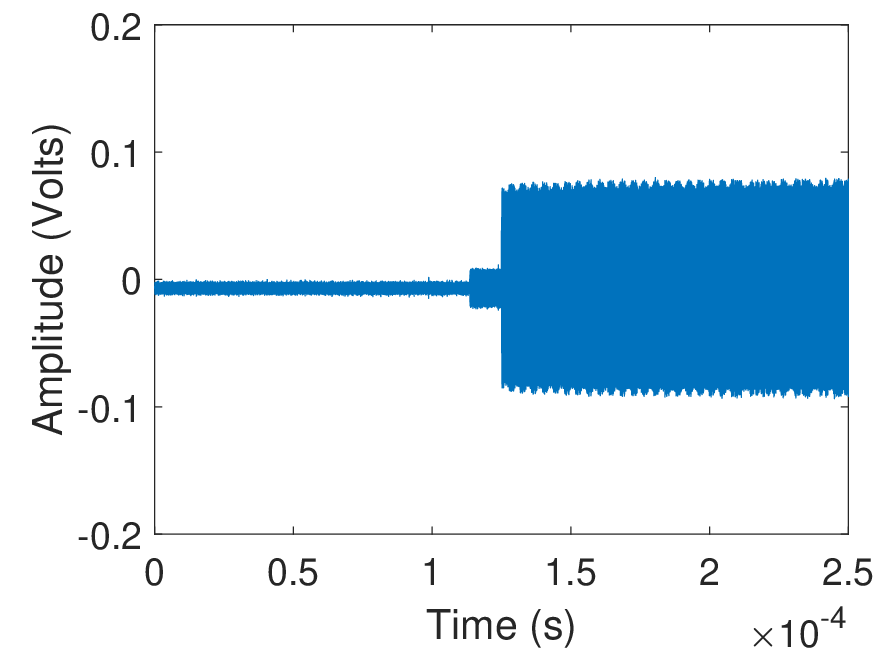}\label{fig:Raw_data_DX6e}}
    \subfloat[Futaba T8FG]{
        \includegraphics[width=0.23\linewidth,trim={0cm 0cm 0cm 0cm},clip]{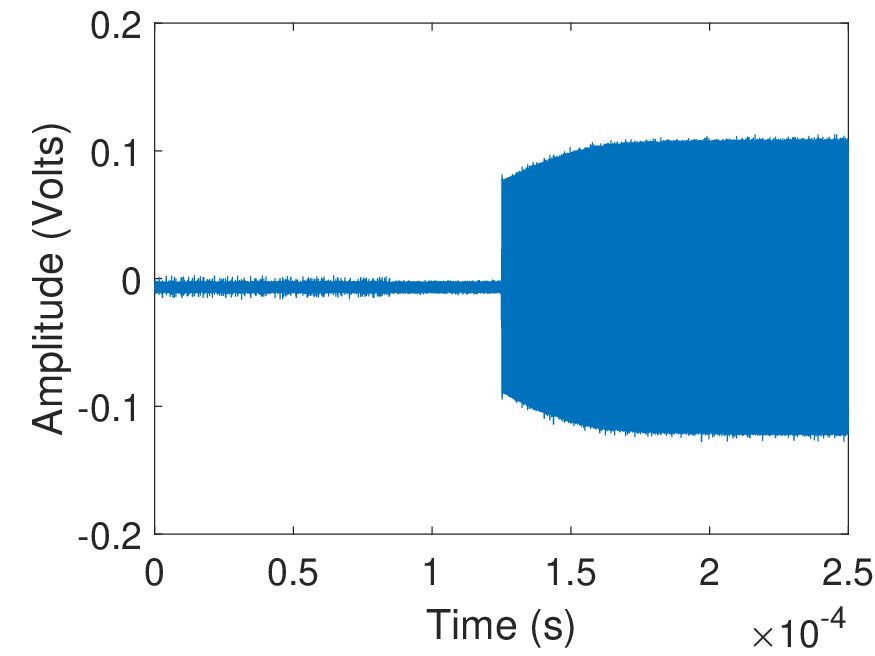}\label{fig:T14SG}}
    \subfloat[DJI Phantom 4 Pro]{
        \includegraphics[width=0.23\linewidth,trim={0cm 0cm 0cm 0cm},clip]{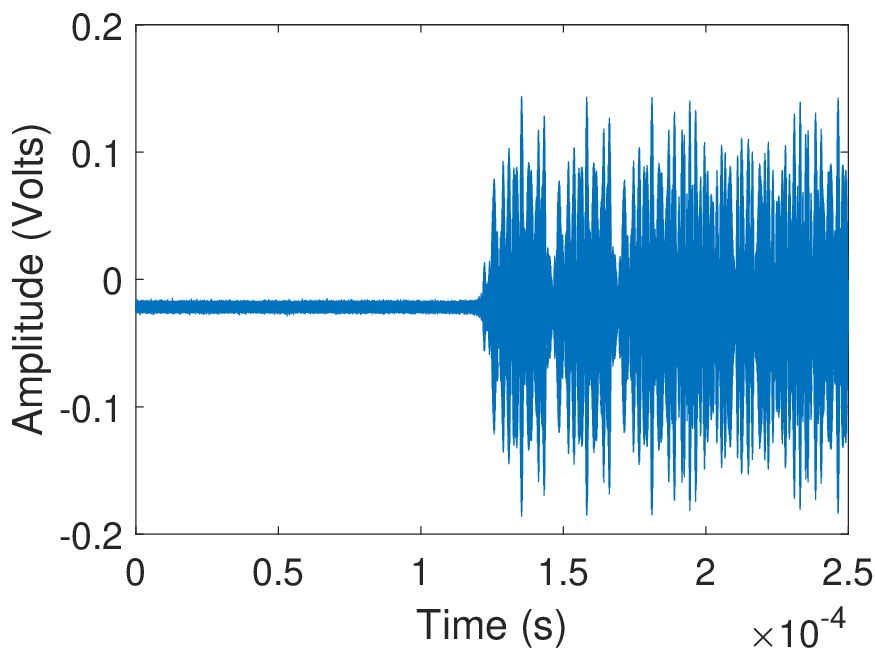}\label{fig:DJI_Phantom4Pro}}
    \vspace{-2mm}\\
    
    \subfloat[DJI Inspire 1 Pro]{
        \includegraphics[width=0.23\linewidth,trim={0cm 0cm 0cm 0cm},clip]{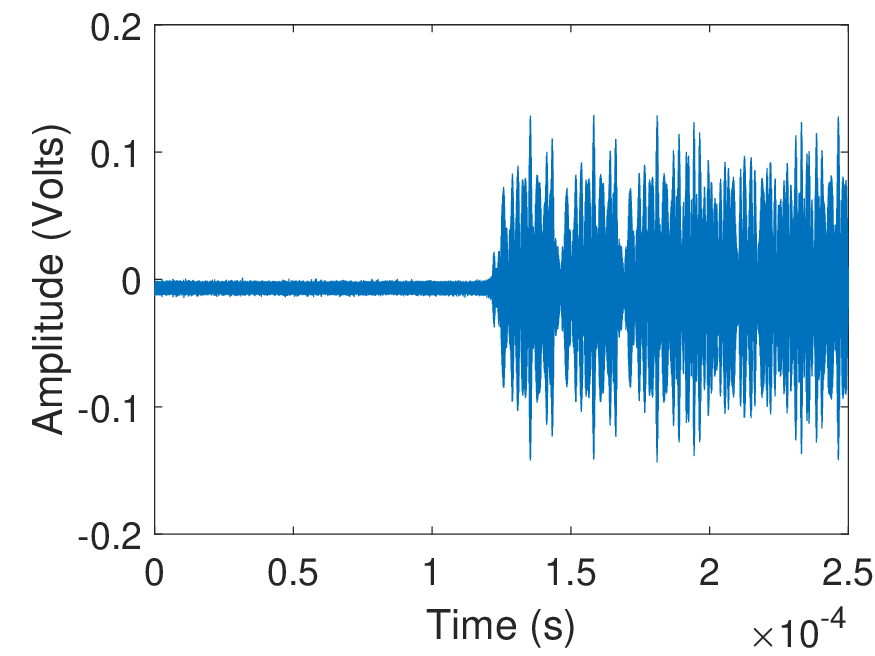}\label{fig:DJI_Inspire_1}}
    \subfloat[JR X9303]{
        \includegraphics[width=0.23\linewidth,trim={0cm 0cm 0cm 0cm},clip]{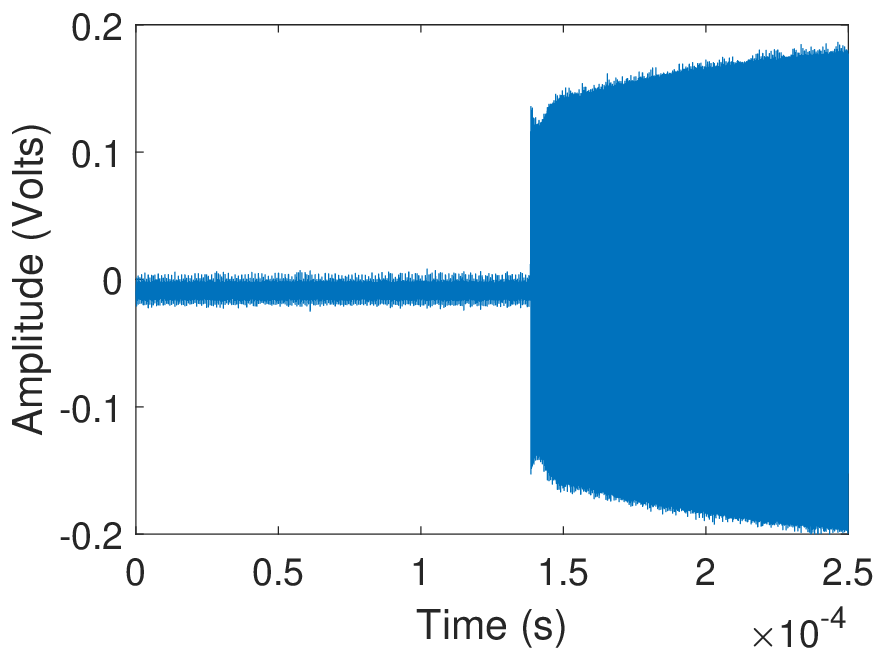}\label{fig:Raw_data_JR_x930324GHz}}
    \subfloat[Jeti Duplex DC-16]{
        \includegraphics[width=0.23\linewidth,trim={0cm 0cm 0cm 0cm},clip]{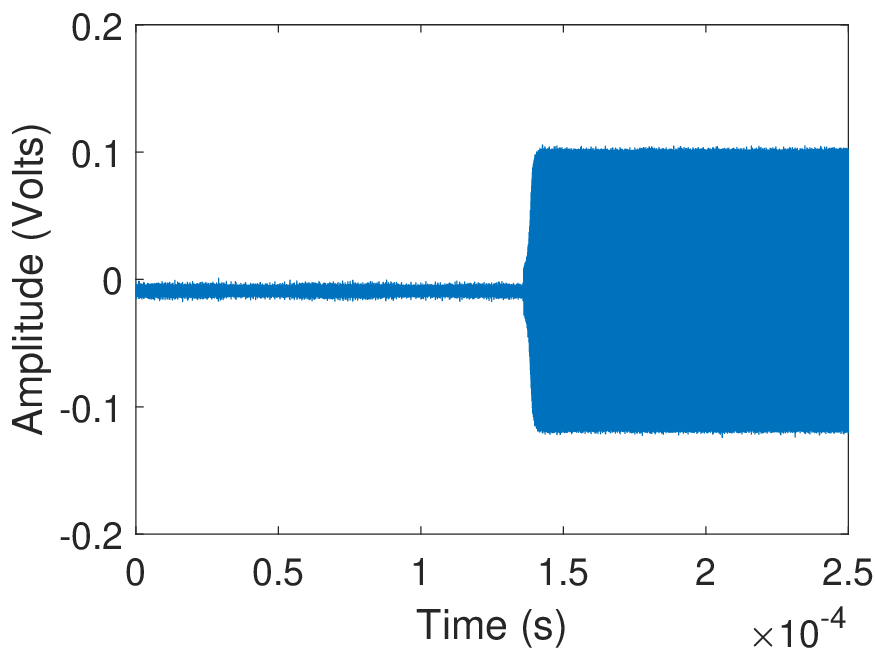}\label{fig:Raw_data_dc-16}}
    \subfloat[FlySky FS-T6]{
        \includegraphics[width=0.23\linewidth,trim={0cm 0cm 0cm 0cm},clip]{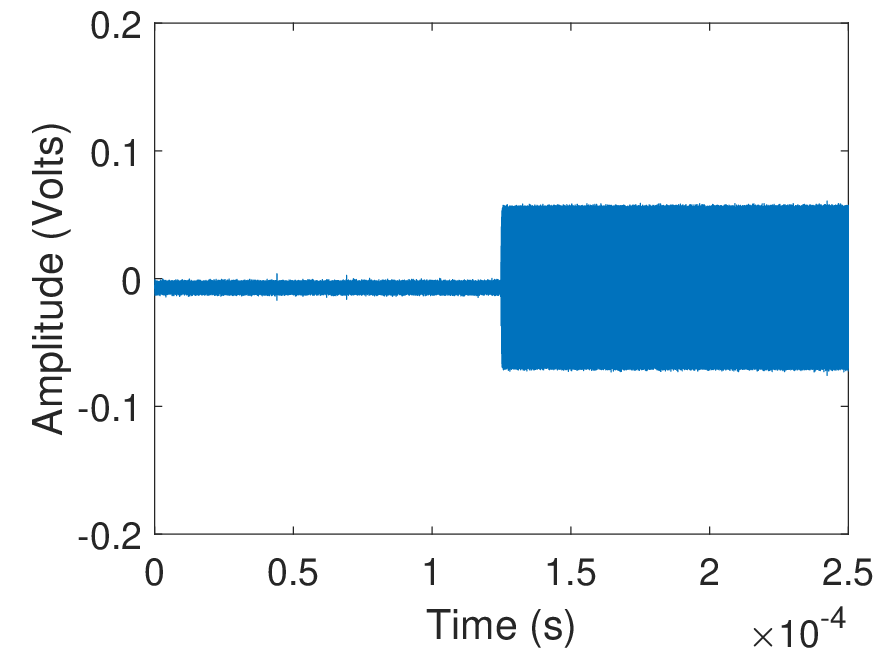}\label{fig:FLY_SKY_FST6}}
    \vspace{-2mm}\\
    
    \subfloat[DJI Matrice 600 UAV]{
        \includegraphics[width=0.23\linewidth,trim={0cm 0cm 0cm 0cm},clip]{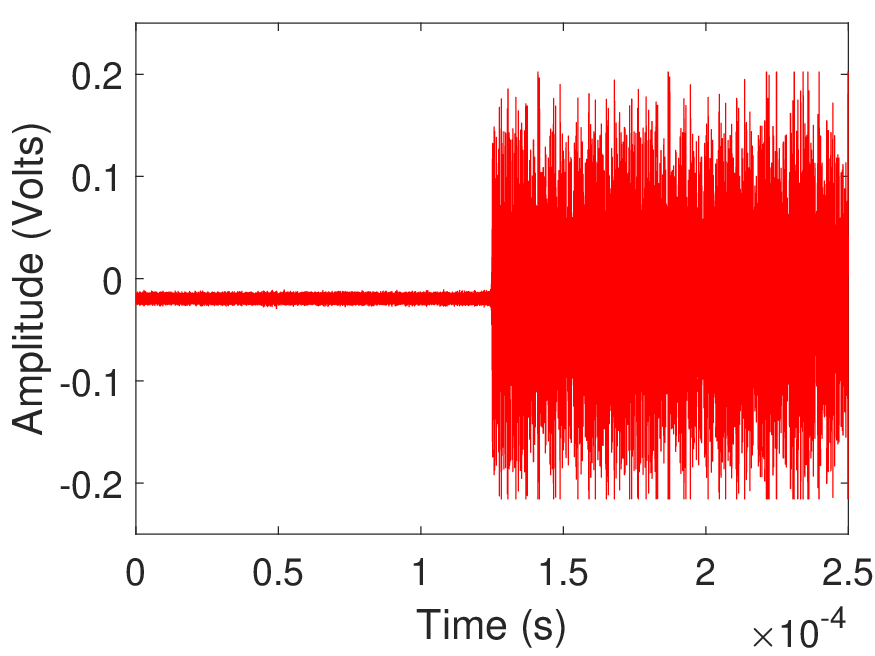}\label{fig:M600_Outdoor_UAV_10m}}
    \subfloat[DJI Phantom 4 Pro UAV]{
        \includegraphics[width=0.23\linewidth,trim={0cm 0cm 0cm 0cm},clip]{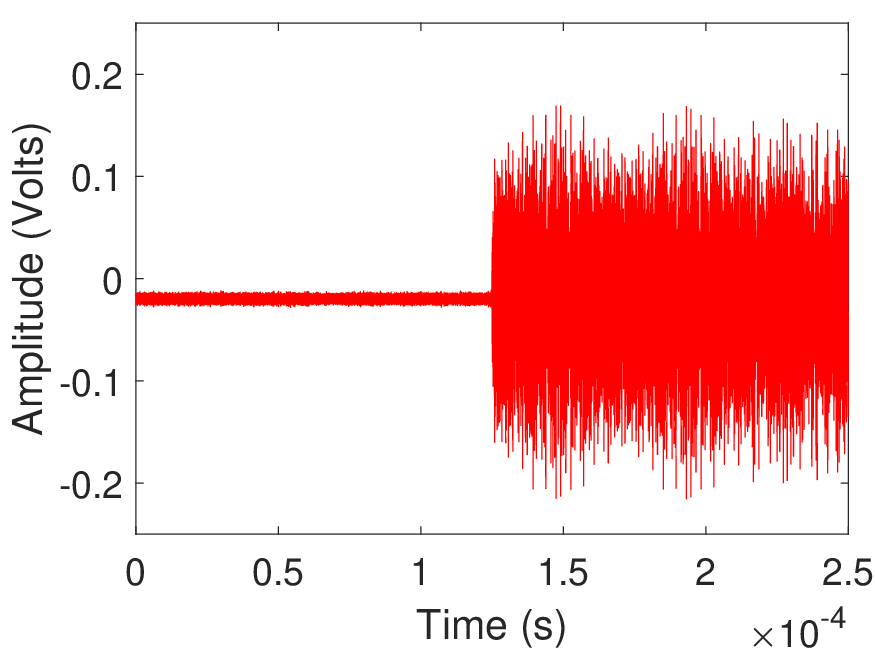}\label{fig:Phantom4PRO_Outdoor_UAV_10m}}
    \subfloat[DJI Inspire 1 Pro UAV]{
        \includegraphics[width=0.23\linewidth,trim={0cm 0cm 0cm 0cm},clip]{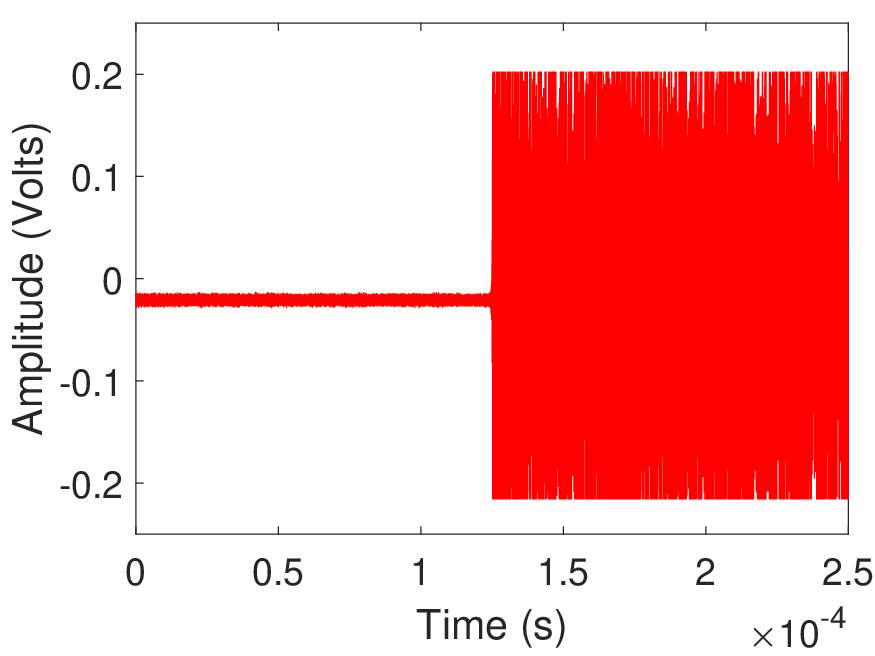}\label{fig:DJI_Inspire_10m}}
    \subfloat[DJI Mavic Pro UAV]{
        \includegraphics[width=0.23\linewidth,trim={0cm 0cm 0cm 0cm},clip]{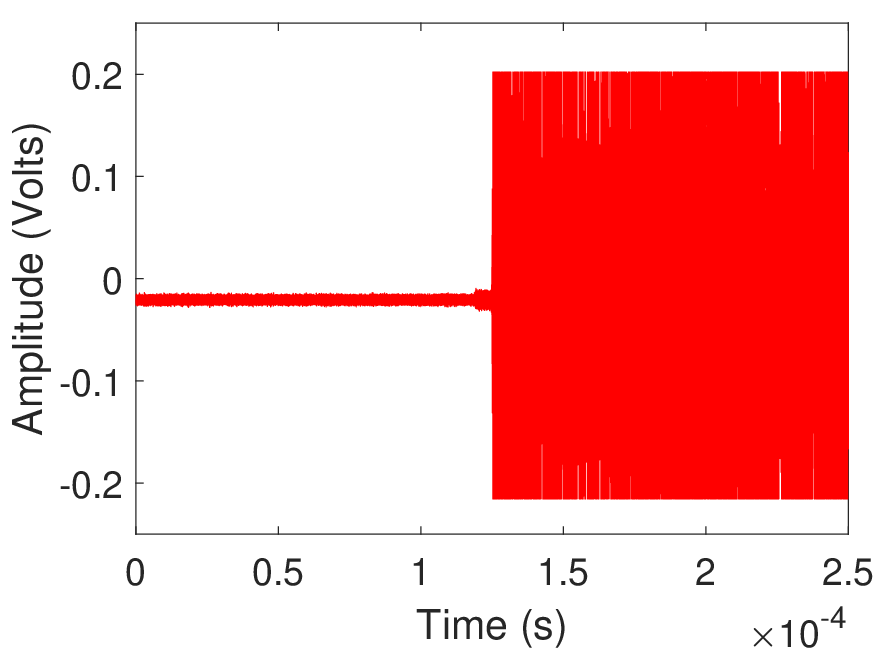}\label{fig:mavicpro_Outdoor_UAV_10m}}
    
  	\caption{This shows a sample of RF signal captured from eight different UAV controllers and four different UAVs while on flight: \textbf{(a)} Graupner MC-32, \textbf{(b)} Spektrum DX6e, \textbf{(c)} Futaba T8FG, \textbf{(d)} DJI Phantom 4 Pro, \textbf{(e)} DJI Inspire 1 Pro, \textbf{(f)} JR X9303, \textbf{(g)} Jeti Duplex DC-16, \textbf{(h)} FlySky FS-T6, \textbf{(i)} DJI Matrice 600 UAV, \textbf{(j)} DJI Phantom 4 Pro UAV, \textbf{(k)} DJI Inspire 1 Pro UAV, \textbf{(l)} DJI Mavic Pro. Reproduced from~\cite{UAV_RF_detection_classification}, licensed under CC BY 4.0.}
	\label{fig1:AFAR_1}
    \vspace{-4mm}
\end{figure*}

\begin{figure*}[th]
    \centering

    \subfloat[]{
        \includegraphics[width=0.5\textwidth]{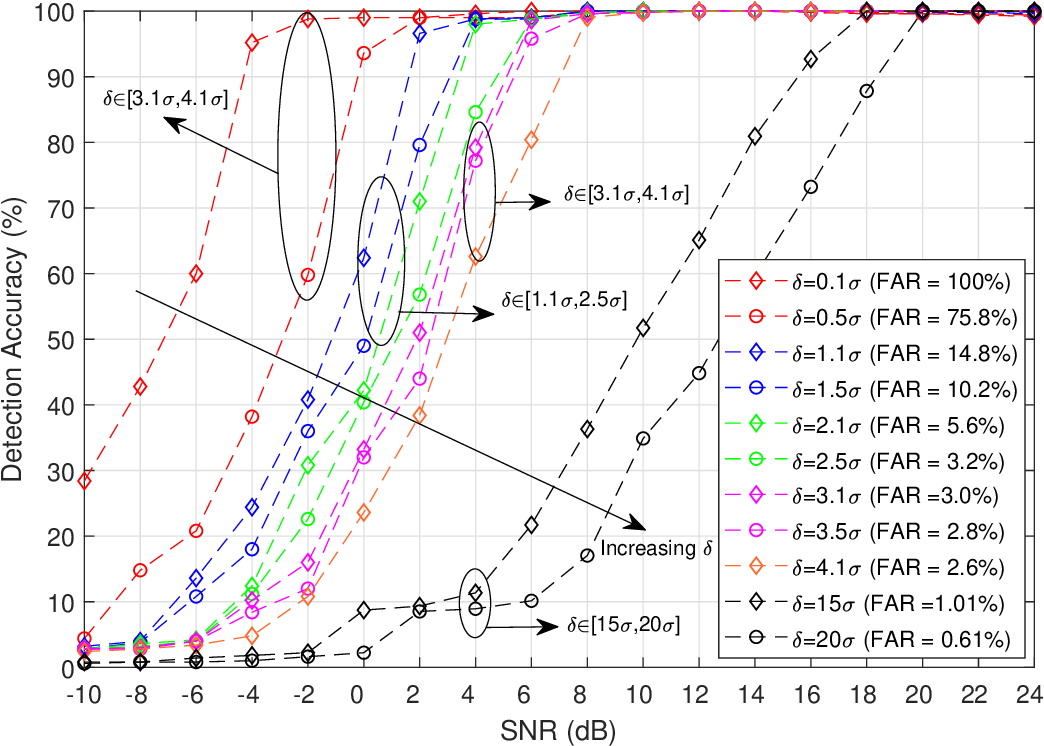}
        \label{detection_1}
    }
    \subfloat[]{
        \includegraphics[width=0.435\textwidth]{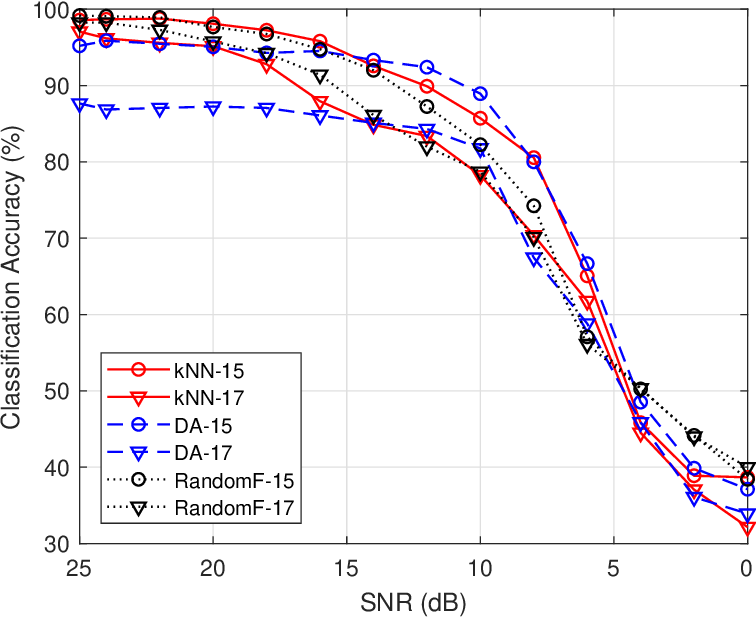}
        \label{classification_1}
    }

    \caption{Representation Result for UAV Detection and Classification Using the UAS signal classification dataset: \textbf{(a)} The performance of the proposed detection systems as a function of the SNR, the detection threshold ($\delta$), and the fixed FAR, \textbf{(b)} The performance of the ML-based model as a function of the SNR, the number of UAV RF controller signal category extracted from the UAS signal classification dataset. Reproduced from~\cite{UAV_RF_detection_classification}, licensed under CC BY 4.0.}  
    \label{fig:UAV_Detect_classification}
\end{figure*}

\subsection{Description of Hardware and Software}
For the data collection, we designed a simple passive RF surveillance receiver that continuously listens to RF signals in the environment and saves the data for further processing. The RF signals captured are time-varying modulated signals from popular commercial UAV controllers. 

The experimental setup is shown in Fig.~\ref{fig1:SETUP_1}. In the figure, a 24~dBi parabolic antenna operating in the 2.4~GHz ISM band listens to RF signals from UAV controllers, which are used to send control and navigation commands to an incoming UAV. 
A high-gain parabolic antenna is used to extend the receiver coverage range.
The output signal from the antenna is fed to the receiver through an RF chain that combines a LNA and an RF bandpass filter. The LNA amplifies weak signals, while the bandpass filter ensures the input signal is band-limited to prevent receiver saturation and nonlinear distortion.
For the experiment, the receiver is a high-resolution mixed signal Keysight oscilloscope (6~GHz Keysight MSOS604A) capable of sampling the captured signal at 20~GSa/s. This high sampling rate ensures the detection system captures all the transient features or fingerprints of the captured UAV RF controller signals. Also, within the receiver in Fig.~\ref{fig1:SETUP_1}, there is a custom MATLAB script for signal detection, data preprocessing and transformation of the raw signal.

Fig.~\ref{fig1:AFAR_1} shows samples of UAV RF remote controller (RC) signals and UAV emitted signals captured using the detection systems in Fig.~\ref{fig1:SETUP_1}~\cite{UAV_RF_detection_classification}. 

\subsection{Dataset Format}

The UAV signal classification dataset comprises RF recordings from 17 commercial UAV controllers (15 of which are unique) manufactured by eight different vendors, including DJI, Futaba, Hobby King, and Turnigy~\cite{UAV_RF_Dataset_IEEEPortal}. The dataset captures controller-specific RF signatures transmitted in the 2.4~GHz ISM band and is designed to support detection, identification, and classification of UAV control signals under varying SNR conditions.

Each RF signal category in the dataset is associated with metadata describing the UAV controller type, manufacturer, and acquisition parameters. This metadata enables systematic benchmarking of RF-based UAV detection and classification algorithms, as well as comparative studies across controller types and interference conditions. Representative examples of dataset usage for detection and classification tasks are shown in Fig.~\ref{fig:UAV_Detect_classification}.

To facilitate reproducible research, the dataset is released together with supporting software utilities that enable extraction of RF samples and associated metadata, as well as the construction of custom datasets for algorithm evaluation. These tools allow researchers to assemble controller-specific or mixed datasets tailored to supervised learning, feature extraction, and classification experiments. Detailed descriptions of dataset organization, metadata fields, and software utilities are provided in Appendix~\ref{app:file_structure}.

\subsection{Representative Results}
The UAS signal classification dataset has been used to validate several UAV detection and classification algorithms. The results from these works have been published in the literature~\cite{microUAV_RF_Detection, UAV_RF_detection_classification, UAV_RF_Fatih_arxiv}. In~\cite{microUAV_RF_Detection}, the authors described a process for detecting and identifying Micro-UAVs using data extracted from the UAS signal classification dataset. The detection phase is based on the naïve Bayes approach using Markov models. Once the UAV signals have been detected, the identity of the UAV is determined (or classified) using classical machine learning algorithms. The authors showed that using features selection techniques such as neighborhood component analysis (NCA), the kNN machine learning algorithm achieved the highest classification accuracy of 96.3\% at an SNR of 25~dB. In addition, the study showed that as the SNR reduces, simulating a drone moving farther away from the detection system, the detection and classification of the machine learning algorithms reduces. The limitation of this study is the absence of interference signal from the environment. In \cite{UAV_RF_detection_classification}, the authors extended the study in~\cite{microUAV_RF_Detection} to include the presence of wireless interference signals from Wi-Fi and Bluetooth enabled devices. Fig.~\ref{fig:UAV_Detect_classification} shows the performance of the detection and classification system as a function of the SNR. 

From Fig.~\ref{fig:Graupner_MC-32}, we see that for a fixed false alarm rate (FAR), increasing the detection threshold ($\delta$) will reduce the detection accuracy. Also, from Fig.~\ref{fig:Raw_data_DX6e}, we see that the accuracy of the classification system depends on the machine learning model deployed, the SNR, and the number of UAV classes in the database. Once again, the study shows that it is possible to achieve an accuracy of 98.13\% in classifying 15 different UAV controllers using classical machine learning models like kNN and random forest.  In~\cite{UAV_RF_Fatih_arxiv}, the authors investigated the impact of using the convolutional neural network (CNN) to classify/identify UAV RF controller signals. The CNN models are trained using spectrogram images representation of the raw UAV controller RF signals. The CNN model achieved an accuracy of about 92\%.

\subsection{Possible Uses of Dataset}
Our UAS signal classification dataset can be used for many practical applications. They include the following:

\noindent\textbf{Benchmarking Detection and Classification Algorithms:}
The dataset supports benchmarking of UAV RF detection and classification algorithms against classical and learning-based approaches under consistent measurement conditions. Because the RF signals are captured in a controlled environment, the dataset enables comparative evaluation of algorithm performance as functions of noise level and antenna temperature.

\noindent\textbf{Impact of Intentional and Unintentional Interference:}
The dataset enables investigation of the impact of both intentional interference, such as jamming signals, and unintentional radiators on UAV RF detection and classification performance. As demonstrated in~\cite{UAV_RF_detection_classification}, the data can be used to study in-band and out-of-band interference from technologies such as Wi-Fi, BLE, Zigbee, and other common emitters, supporting electromagnetic compatibility analysis in shared unlicensed spectrum.

\noindent\textbf{Synthetic UAV RF Dataset Generation:}
The dataset supports generation of synthetic UAV RF signals using data-driven techniques such as generative adversarial networks and variational autoencoders. It can also serve as input for data augmentation, statistical modeling, and Monte Carlo simulation methods, enabling the creation of expanded training datasets to improve model robustness and generalization.

\section{UAV Trajectory, RSRP, and Throughput Dataset in Emulated and Simulated Environments}\label{sec:uav_trajectory}
UAV flight path, including throughput and RSRP measurements, is crucial to developing, experimenting with, and evaluating next-generation wireless networks and DT solutions. The provided datasets capture critical A2G propagation and link characteristics as a function of trajectory, as well as interactions with emulated and simulated environments. The advantage of the emulation environment is that it executes the same software and protocol stack used in real-world testbeds, running on SDR-based radios and UAV platforms. However, it can be relatively slow and computationally intensive for development purposes. In contrast, simulation abstractions accelerate the design and testing of UAS and radio algorithms that can subsequently be deployed in emulation and real-world testbeds, though this comes at the expense of reduced realism.

\subsection{Description of Hardware and Software}
The measurements were obtained through a combination of MATLAB simulation and DT emulation~\cite{Hossen2025}, utilizing AERPAW's cutting-edge experimental facilities at Lake Wheeler field. The UAV testbed includes an LTE SISO radio-equipped UAV, GPS receivers, and flight controllers that ensure accurate trajectory tracking. Field experiments were conducted in the Lake Wheeler area, where high-fidelity RSRP measurements were collected from four BSs (LW1-LW4). 

A MATLAB-based simulation environment was developed to model UAV flights in virtual settings, with field measurements used for validation.
The AERPAW DT further supports emulation of UAV-to-base-station communications, offering parameterized control over mobility patterns, radio environments, and network conditions.

Custom MATLAB and Python scripts are provided for data processing, analysis, and visualization. 
The dataset and associated processing and simulation code are publicly available at~\cite{UAVSimFramework2025, Hossen2025Dryad}.

\subsection{Dataset Format}

This dataset provides trajectory-aware radio performance measurements for UAV-assisted cellular systems, combining emulated and simulated data to enable systematic comparison of link behavior under controlled mobility patterns. It includes RSRP, SNR, throughput, and UAV state information collected along both fixed and autonomous UAV trajectories.

Three complementary categories of measurements are included: emulated RSRP measurements along predefined UAV trajectories, simulated RSRP measurements along the same fixed trajectories as used in the emulation environment, and the RSRP measurements for autonomous UAV trajectories while maintaining the geofencing constraints. For all categories, the dataset records time-synchronized UAV position (latitude, longitude, altitude) together with the RSRP measurements.

The dataset is released in structured CSV format suitable for statistical analysis, visualization, and machine learning applications. It supports comparative evaluation of emulation and simulation fidelity, as well as investigation of trajectory design for coverage analysis and throughput evaluation. Detailed directory organization, CSV field definitions, and post-processing scripts used to generate the reported results are documented in Appendix~\ref{app:file_structure}.

\begin{figure*}[t!]
    \centering
    \subfloat[
    ]{\includegraphics[trim={0cm, 7.5cm, 0cm, 7cm},clip,width=.24\linewidth]{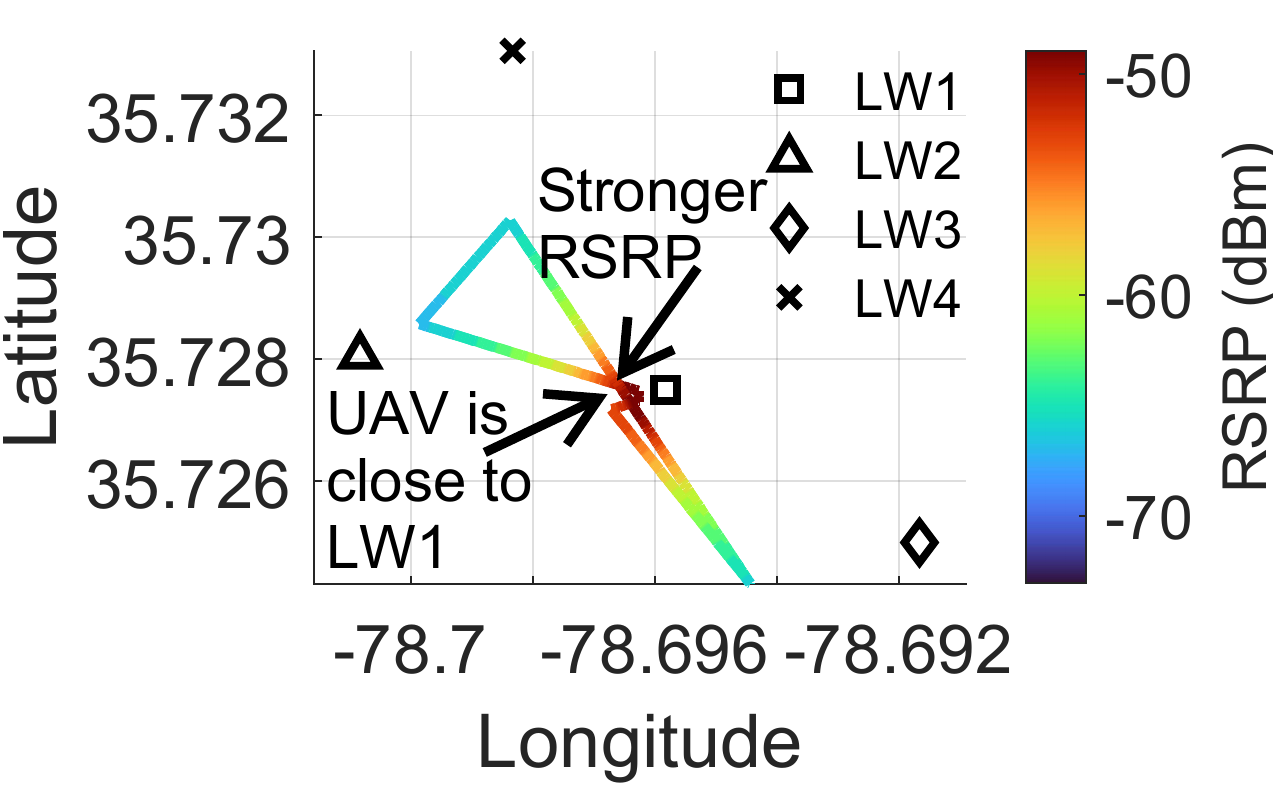}      \label{fig:fixed_trajectory_emumulation_rsrp_LW1}
    }
    \subfloat[    
    ]{\includegraphics[trim={0cm, 7.5cm, 0cm, 7cm},clip,width=.24\linewidth]{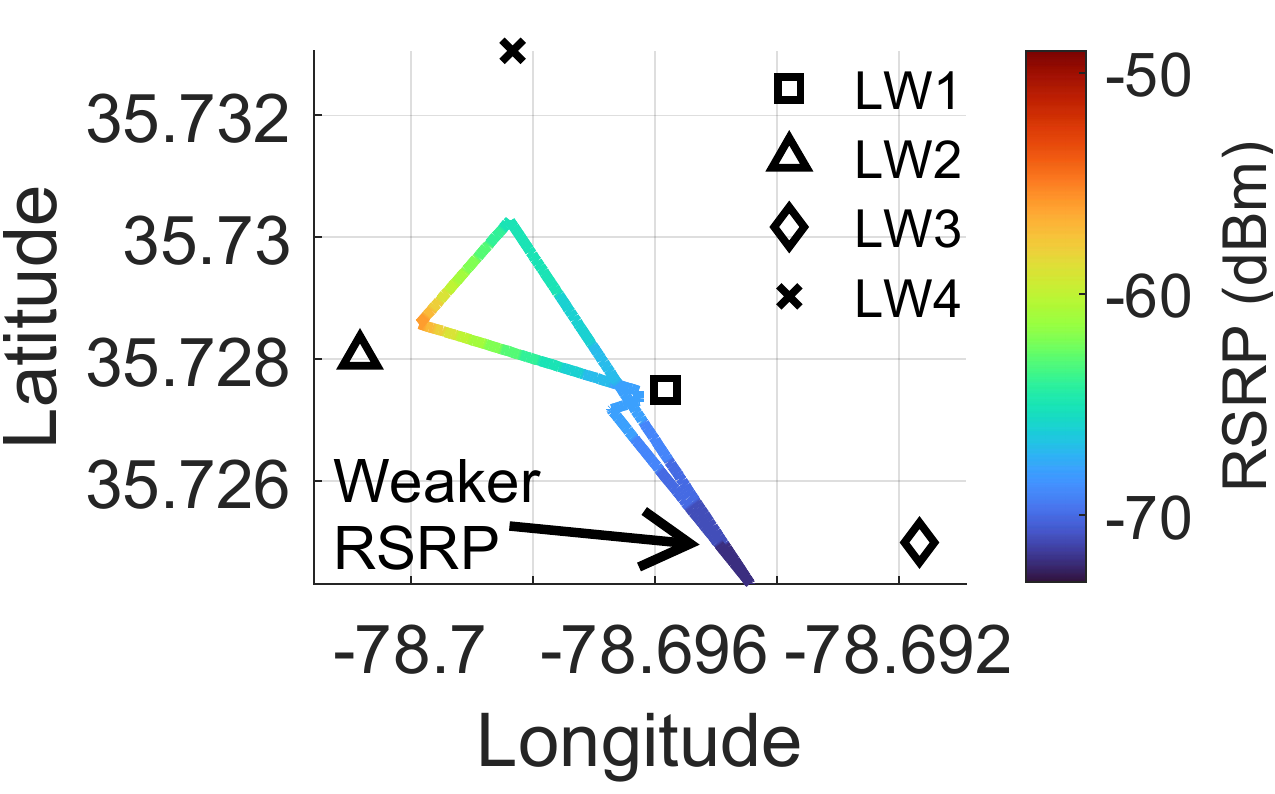}        \label{fig:fixed_trajectory_emumulation_rsrp_LW2}
    }
    \subfloat[    
    ]{\includegraphics[trim={0cm, 7.5cm, 0cm, 7cm},clip,width=.24\linewidth]{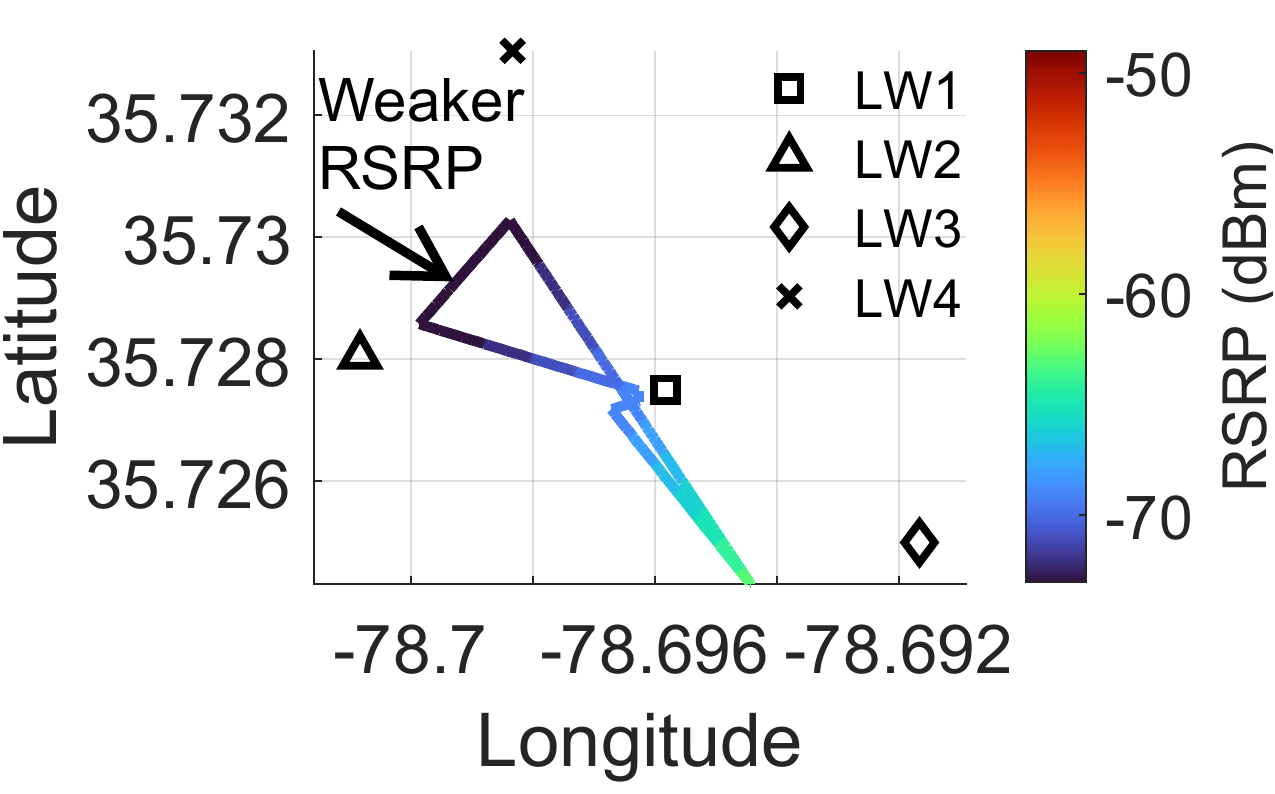}        \label{fig:fixed_trajectory_emumulation_rsrp_LW3}
    }
    \subfloat[    
    ]{\includegraphics[trim={0cm, 7.5cm, 0cm, 7cm},clip,width=.24\linewidth]{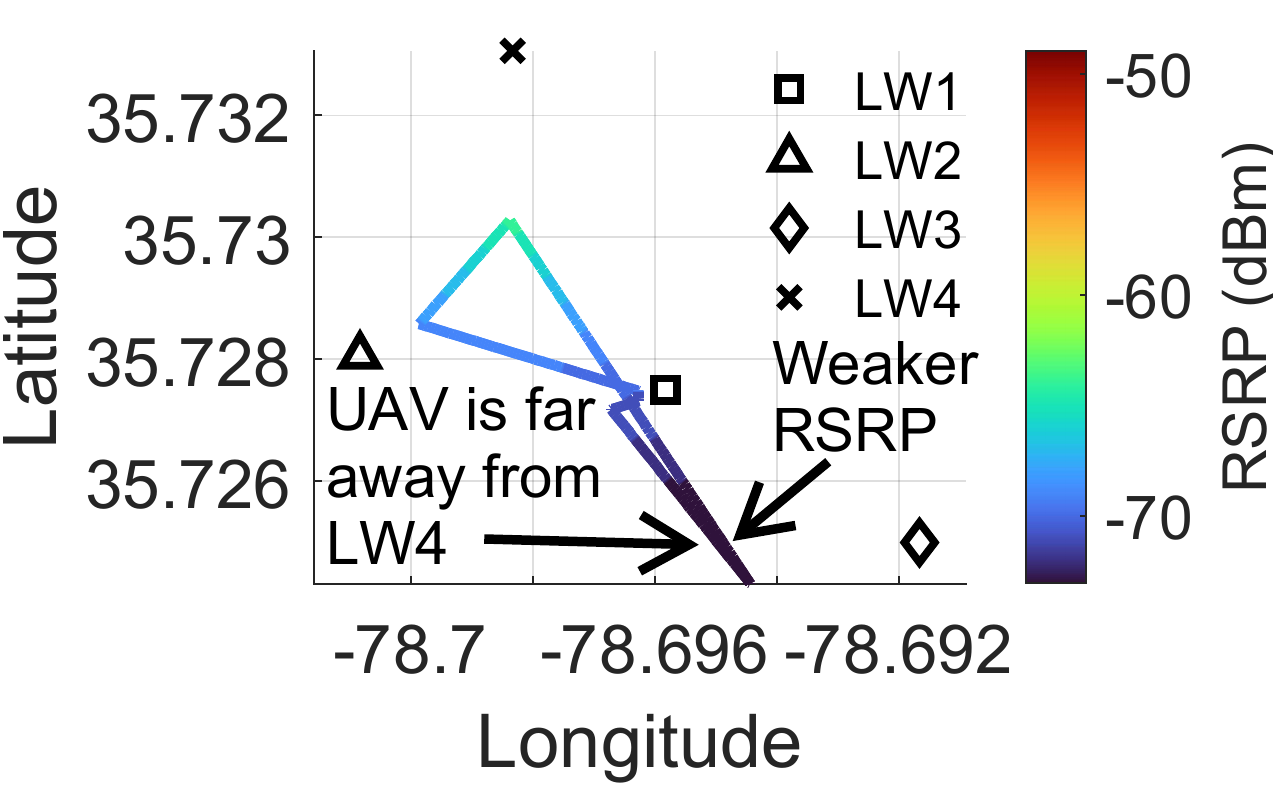}        \label{fig:fixed_trajectory_emumulation_rsrp_LW4}
    }
    \caption{RSRP measurement with respect to \textbf{(a)} LW1, \textbf{(b)} LW2, \textbf{(c)} LW3, and \textbf{(d)} LW4 for fixed trajectory in emulation.}    
    \label{fig:fixed_rsrp_emulation}
\end{figure*}

\begin{figure*}[t!]
    \centering
    \subfloat[
    ]{\includegraphics[trim={0cm, 7.5cm, 0cm, 7cm},clip,width=.24\linewidth]{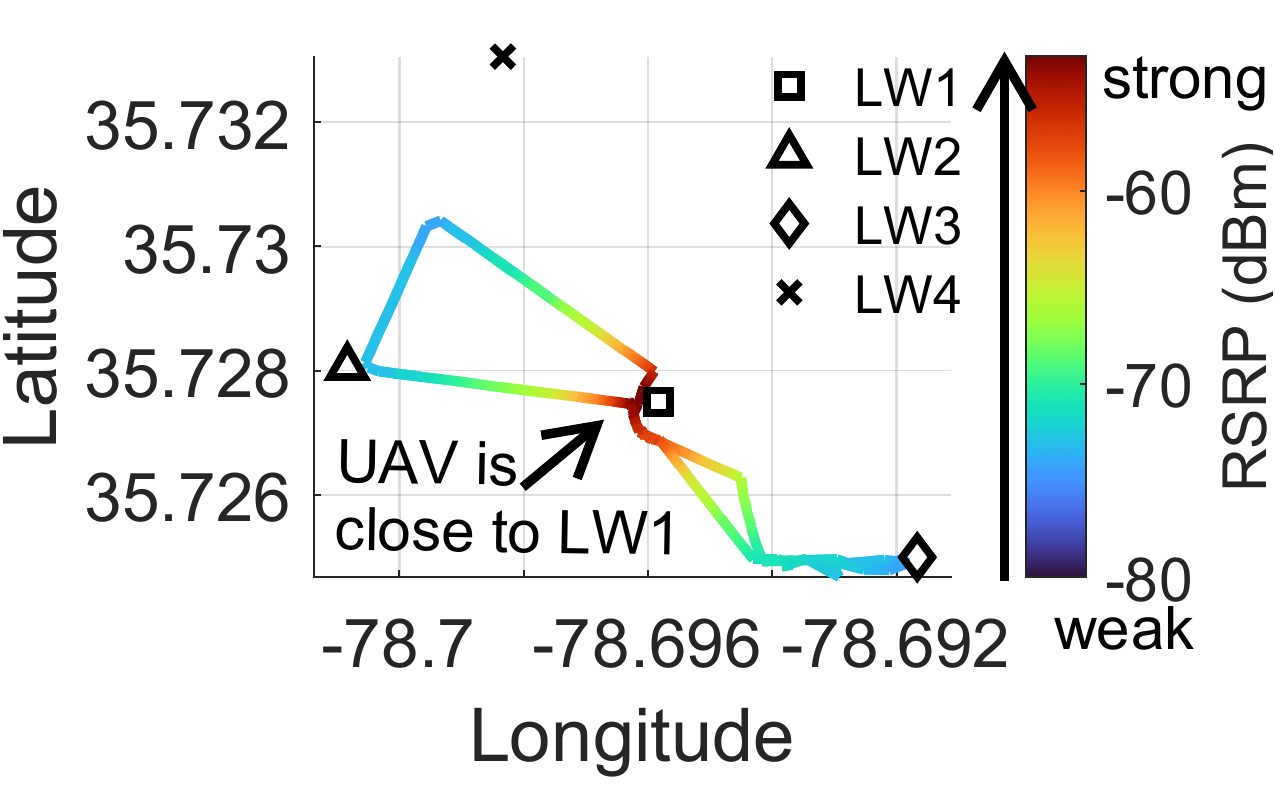}        \label{fig:autonomous_trajectory_simulation_rsrp_LW1}
    }
    \subfloat[    
    ]{\includegraphics[trim={0cm, 7.5cm, 0cm, 7cm},clip,width=.24\linewidth]{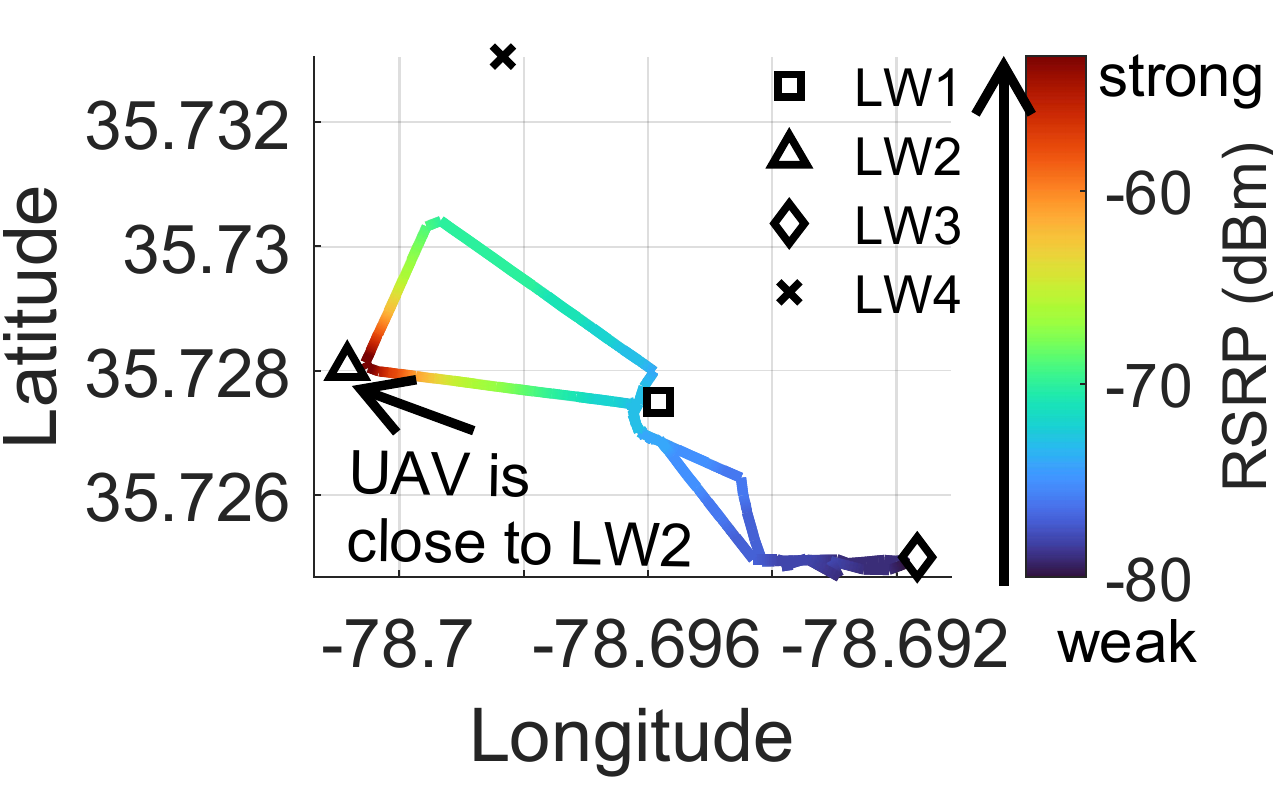}        \label{fig:autonomous_trajectory_simulation_rsrp_LW2}
    }
    \subfloat[    
    ]{\includegraphics[trim={0cm, 7.5cm, 0cm, 7cm},clip,width=.24\linewidth]{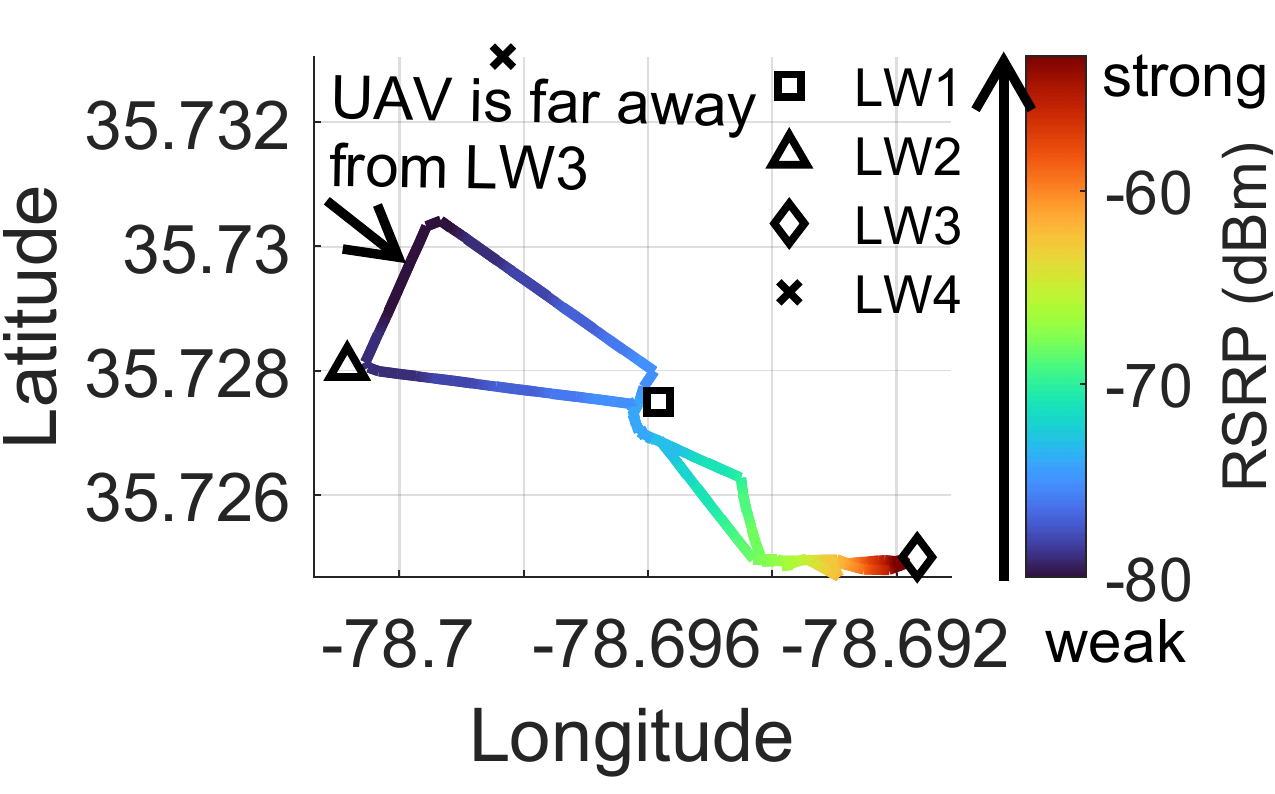}        \label{fig:autonomous_trajectory_simulation_rsrp_LW3}
    }
    \subfloat[    
    ]{\includegraphics[trim={0cm, 7.5cm, 0cm, 7cm},clip,width=.24\linewidth]{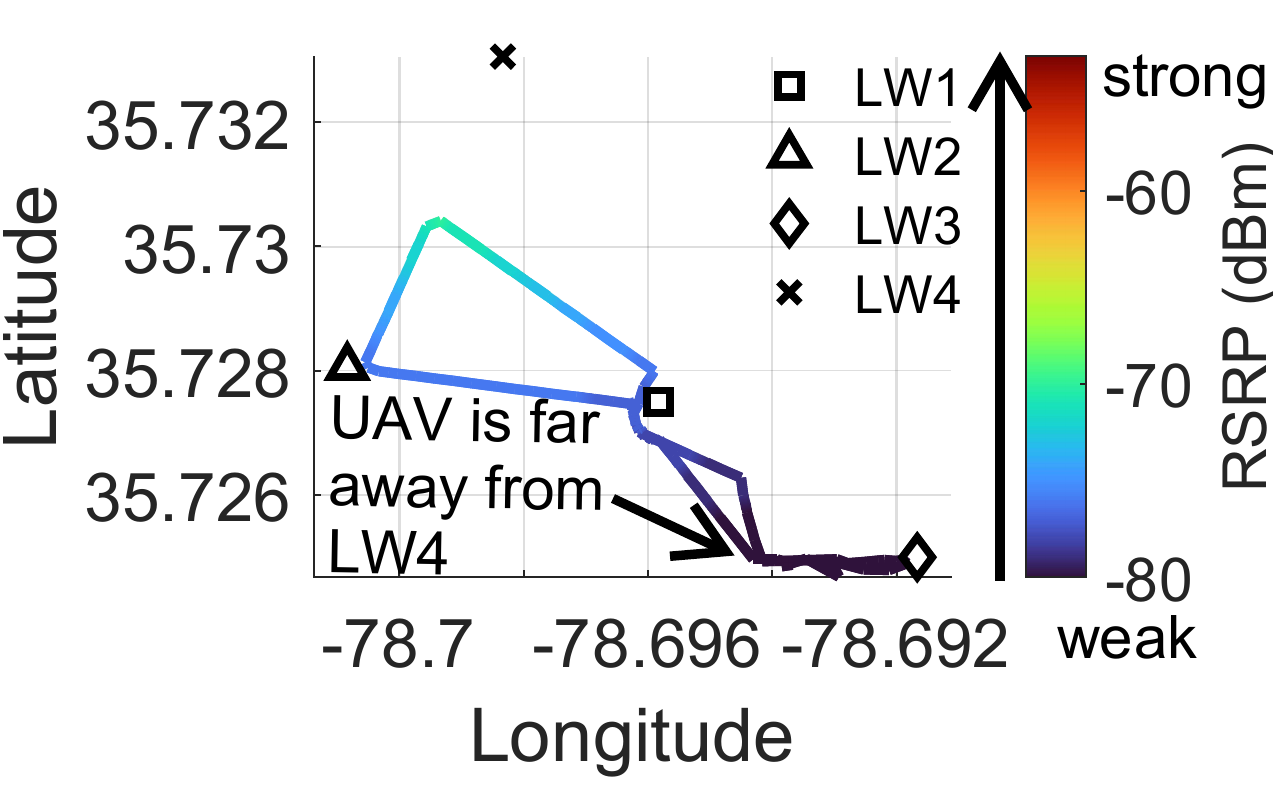}        \label{fig:autonomous_trajectory_simulation_rsrp_LW4}
    }
    \caption{RSRP measurement with respect to \textbf{(a)} LW1, \textbf{(b)} LW2, \textbf{(c)} LW3, and \textbf{(d)} LW4 for autonomous trajectory in simulation.}    
    \label{fig:autonomous_rsrp_simulation}
\end{figure*}

\subsection{Representative Results}
The representative results for emulation and simulation measurements and the impact of trajectory design are discussed in Figs.~\ref{fig:fixed_rsrp_emulation} and \ref{fig:autonomous_rsrp_simulation}, respectively. Fig.~\ref{fig:fixed_rsrp_emulation} presents emulation-based RSRP measurements along the fixed UAV trajectory with respect to four base stations (LW1–LW4). As expected, stronger signal levels are observed when the UAV is closer to a BS. For example, Fig.~\ref{fig:fixed_trajectory_emumulation_rsrp_LW1} shows higher RSRP values near LW1, which can be attributed to the short distance and the absence of significant obstructions such as trees, compared to other BSs. In \cite{Hossen2025}, we demonstrated a high similarity between the simulation and emulation RSRP measurements. 

Fig.~\ref{fig:autonomous_rsrp_simulation} shows the simulated RSRP measurement for an autonomous UAV trajectory with respect to individual BSs. Strong RSRP levels are observed near LW1, LW2, and LW3 as the UAV trajectory passes close to these BSs. In contrast, lower RSRP values are observed for LW4, since the UAV remains farther away and is restricted from approaching LW4 due to geofencing constraints. Additional trajectory details for both fixed and autonomous cases, along with complementary results such as distance-versus-RSRP and throughput measurements, are presented in \cite{Hossen2025}.

\subsection{Possible Uses of Dataset}
This dataset can be used for the following research tasks:

\noindent\textbf{Simulation and Digital Twin Calibration:}
The dataset supports calibration, validation, and tuning of simulation and emulation environments by enabling direct comparison between emulated, simulated, and field-measured data. Such comparisons allow researchers to assess the fidelity of digital twin and network emulation platforms and improve their predictive accuracy for real-world UAV deployments.

\noindent\textbf{A2G Propagation and Coverage Analysis:}
The dataset enables empirical analysis of air-to-ground signal propagation, including the impact of altitude, trajectory, and environment on RSRP, SNR, and throughput. These measurements support the construction and calibration of A2G path loss models, spatial radio maps, and two-dimensional and three-dimensional coverage representations along complex UAV flight paths.

\noindent\textbf{Trajectory-Aware Algorithm Evaluation:}
The availability of emulated and simulated UAV trajectory data supports the development and benchmarking of trajectory-aware algorithms, including path planning, handover management, and resource allocation strategies. By coupling mobility information with measured and emulated signal metrics, researchers can evaluate algorithm performance under realistic aerial operating conditions.

\noindent\textbf{Machine Learning for Link Quality Prediction:}
The dataset supports data-driven modeling of link quality by enabling training and evaluation of machine learning models that predict RSRP or throughput using spatial, environmental, and mobility-related features. Such models can be used to complement analytical propagation approaches and inform adaptive UAV networking strategies.

\section{Ray Tracing Simulation and Measurement Comparison Dataset}\label{sec:ray_tracing}

In this section, we present a dataset that enables direct comparison between ray tracing (RT) simulation and real-world measurement of RSS measured at the AERPAW testbed. The dataset includes UAV trajectory and altitude information, and RSS data collected at each tower. To investigate realistic propagation characteristics in the RT simulation, we implement forest areas with a simple tree model and also incorporate geographical information, including buildings.

\subsection{Description of Hardware and Software}

\begin{figure}[t!]
    \centering
    \subfloat[Trajectory]{
        \includegraphics[trim={0.4cm 0.1cm 1.1cm 0.6cm}, clip, width=0.97\linewidth]{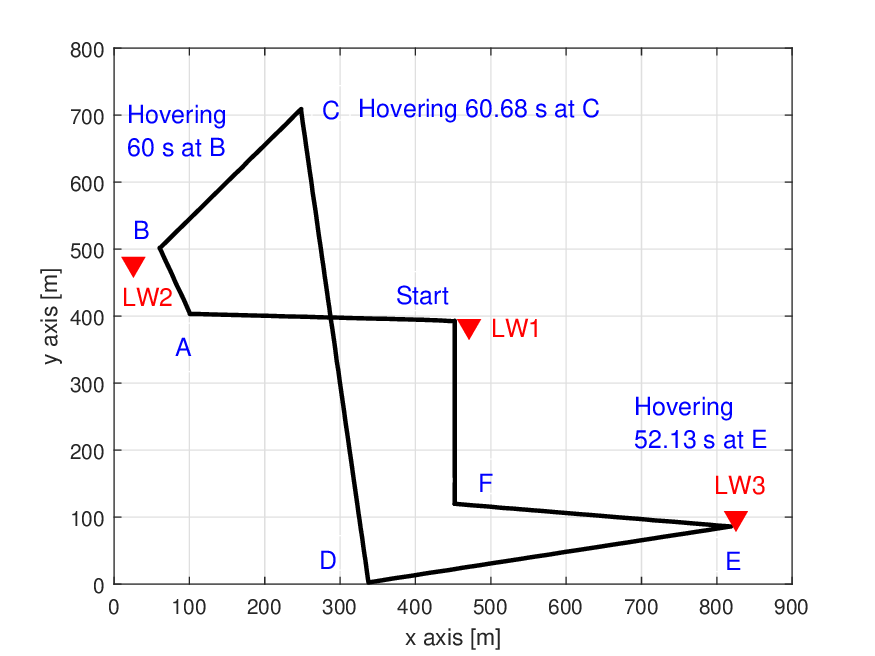}
        \label{fig:DLee_trajectory_measurement_RSS}
    }\\ 
    \subfloat[Altitude]{
        \includegraphics[trim={0.4cm 0.1cm 1.3cm 0.6cm}, clip, width=0.97\linewidth]{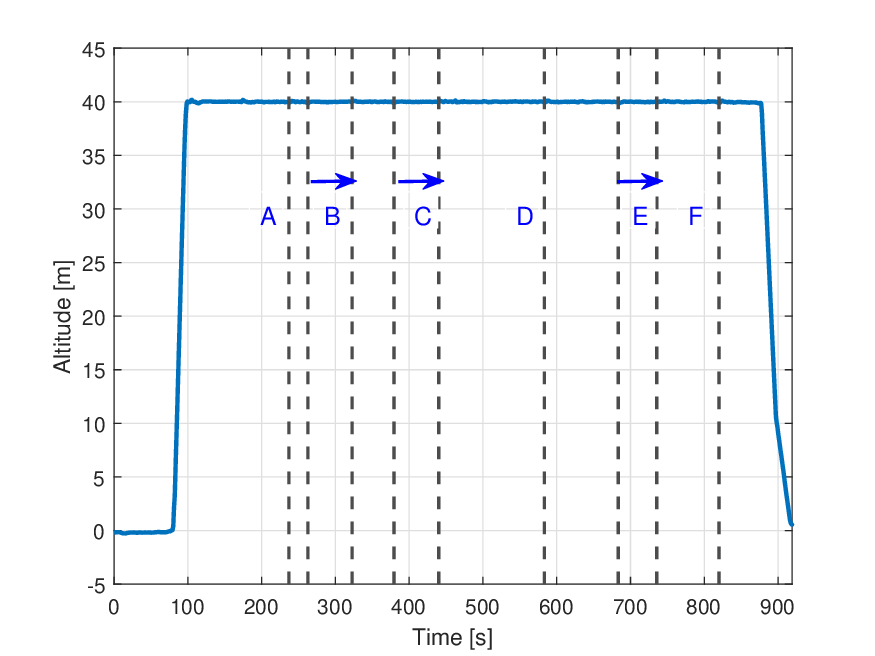}
        \label{fig:DLee_altitude_measurement_RSS}
    }
    \caption{\textbf{(a)} Trajectory and \textbf{(b)} altitude of the signal coverage measurements and RT simulation. The blue horizontal arrows between the vertical lines highlight the hovering duration of the UAV at each waypoint.}
    \label{fig:DLee_trajectory_altitude_measurement_RSS}
\end{figure}

The predefined trajectory and altitude over time of the UAV are demonstrated in Fig.~\ref{fig:DLee_trajectory_altitude_measurement_RSS}. The UAV takes off near the LW1 tower. After takeoff, the UAV sweeps LW2 and LW3 towers. The UAV returns to LW1 and lands on the ground. Each waypoint of the trajectory is highlighted with a letter, which corresponds to vertical lines in Fig.~\ref{fig:DLee_altitude_measurement_RSS}, and the blue horizontal arrows between the vertical lines highlight the hovering duration at each waypoint. The UAV and LW towers have SISO antenna setups with $3.3$~GHz carrier frequency. The RSS is recorded by a dual-channel USRP B210 and GNU Radio at each tower while the UAV transmits signals. The RSS is measured for $20$~ms for every $100$~ms intervals.

For the RT, NVIDIA Sionna~\cite{Donggu_nvidia_sionna} is used with support of the Open Street Map (OSM) database~\cite{Donggu_osm} for geographic and building information and Blender~\cite{Donggu_blender} for 3D modeling. To consider the realistic effects of the trees in the LWRFL areas, we implement a simple tree model and populate it in the Lake Wheeler area, as shown in Fig.~2 of~\cite{Donggu_OJVT}. We adopt predefined material settings for RT simulation from Sionna. Specifically, the surface materials of the buildings are set to concrete and medium dry ground, which are defined as ``itu\_concrete'' and ``itu\_medium\_dry\_ground'', respectively. On the other hand, a tree model consists of a wooden cylinder with ``itu\_wood'' and a cone on top of the cylinder with custom foliage material constants, which is calculated under~\cite{Donggu_itu_R_vegetation}.

The RT simulation is conducted at each GPS coordinate of the UAV along the predefined trajectory to allow for direct comparison with the measurement data. For calibration purposes, offsets that have minimum root mean squared error (RMSE) are searched within the range of [$-50$:$50$]~dB by the unit of $0.1$~dB. Moreover, the altitude over the predefined trajectory below $0.5$~m is rounded up to $0.5$~m for the RT simulation.

  \begin{figure*}[t!]
     \centering
     \subfloat[LW1]{\includegraphics[trim={0.3cm 0.1cm 1.35cm 0.6cm},clip,width=0.64\columnwidth]{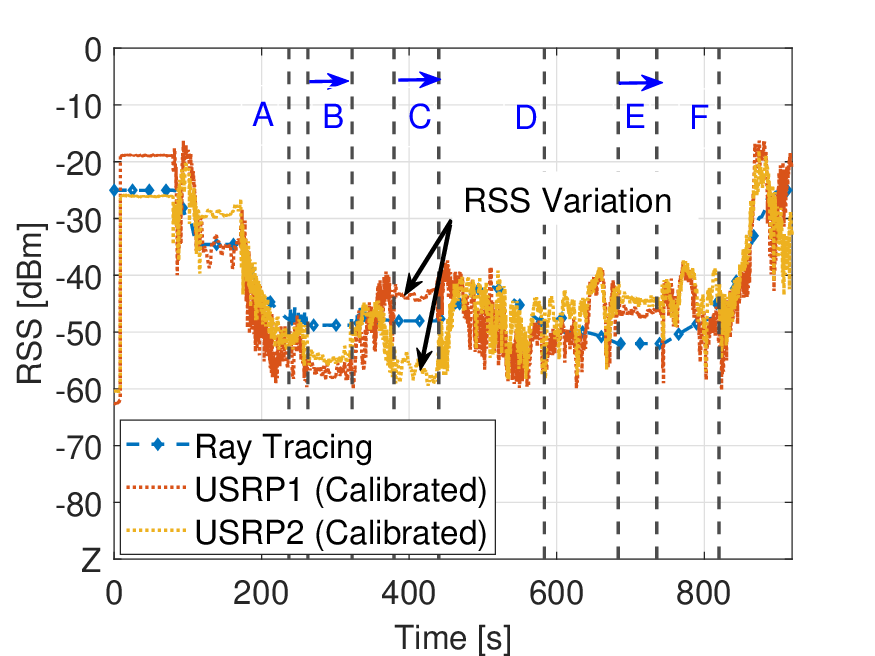}
     \label{fig:DLee_measurement_vs_sionna_LW1}
     }
     \subfloat[LW2]{\includegraphics[trim={0.3cm 0.1cm 1.35cm 0.6cm},clip,width=0.64\columnwidth]{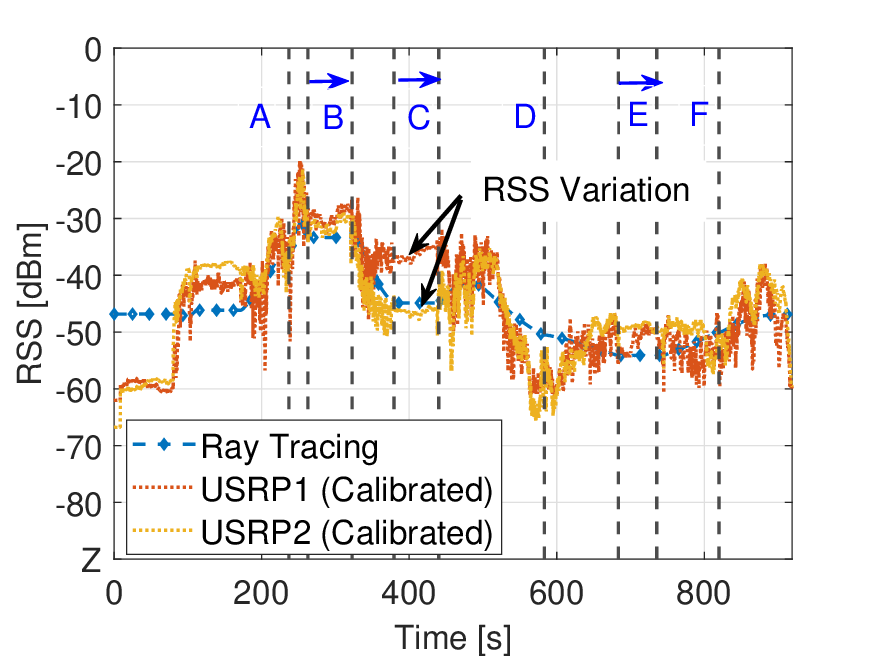}
     \label{fig:DLee_measurement_vs_sionna_LW2}
     }
     \subfloat[LW3]{\includegraphics[trim={0.3cm 0.1cm 1.35cm 0.6cm},clip,width=0.64\columnwidth]{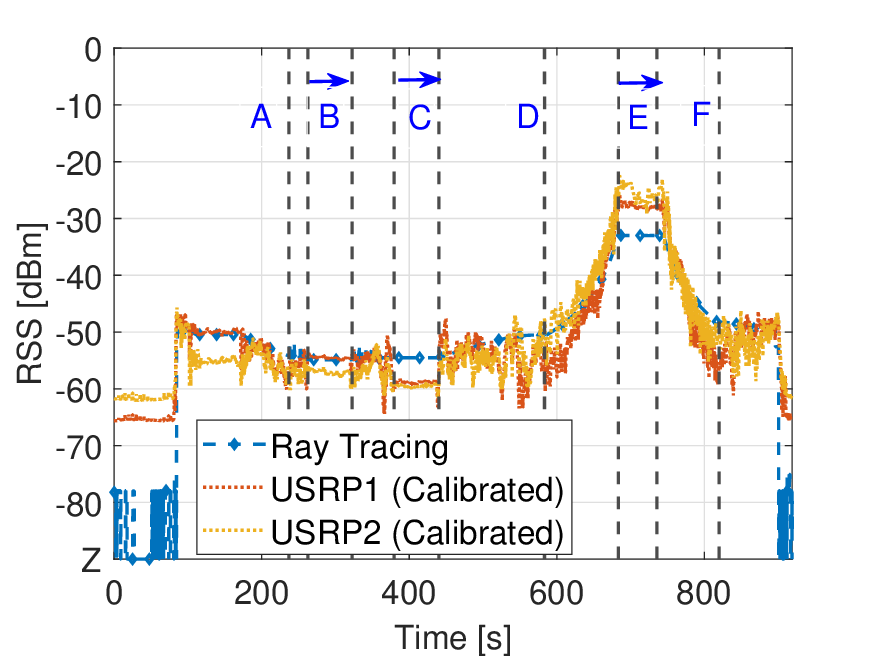}
     \label{fig:DLee_measurement_vs_sionna_LW3}
     } 
     \caption{RSS measurement and RT simulation results with the predefined trajectory for the specific UAV locations {A}-{F} in Fig.~\ref{fig:DLee_trajectory_altitude_measurement_RSS}. \textbf{(a)} LW1, \textbf{(b)} LW2, and \textbf{(c)} LW3. The blue horizontal arrows between the vertical lines highlight the hovering duration of the UAV at each waypoint. Agreement between measured and simulated RSS is location dependent, highlighting where propagation models may require calibration.}
     \label{fig:DLee_measurement_vs_sionna_LW1_LW5}
 \end{figure*}

\subsection{Dataset Format}

This dataset provides a direct comparison between measured RSS along UAV trajectories and corresponding RT simulation results generated using the Sionna RT framework. The dataset supports systematic evaluation of RT model fidelity by enabling point-by-point comparison of simulated and measured RSS values across multiple base station locations.

The released data include measured RSS collected along a predefined UAV trajectory, RT-simulated propagation parameters such as path coefficients and delays, and calibrated RSS results that align simulation outputs with real-world measurements. Together, these components enable validation of RT-based propagation modeling and facilitate reproducible benchmarking of simulation accuracy in aerial wireless environments.

The dataset is accompanied by post-processing utilities that support trajectory visualization, RSS calibration, and comparison between measured and simulated results. All data and scripts are publicly available through an open-access repository~\cite{Donggu_dryad_dataset}. Detailed file inventories, directory structure, and script-level descriptions used to reproduce the reported results are provided in Appendix~\ref{app:file_structure}.

\subsection{Representative Results}
The RT simulation results of RSS and measurement from each tower are shown in Fig.~\ref{fig:DLee_measurement_vs_sionna_LW1_LW5}, where $\mathrm{Z}$ indicates out-of-coverage area. Here, the measurements from different antennas of the dual-channel USRP at each tower are labeled as USRP1 and USRP2, respectively. The RT simulation results are consistent with the measurements at all tower cases. Fluctuations are observed in the measurement over the trajectory due to changes in the direction (roll/yaw/pitch) of the UAV at each waypoint, fading effects, and other factors. Notably, during the hovering periods highlighted by the blue arrows, the measurements tend to show a mismatch. During the hovering, small variations of orientation can cause noticeable RSS fluctuation, while the fixed orientation is assumed in the RT simulation. It is also worthwhile to note that a $10$~dB variation between USRP1 and USRP2 is observed in the LW1 and LW2 cases, which are highlighted in the figure with arrows. This variation can be attributed to LoS blockage and channel conditions due to the antenna orientations of USRP1 and USRP2, facing the UAV. Moreover, altitude-dependent out-of-coverage areas can be found in the RT simulation for LW3 at lower altitudes during takeoff and landing (time interval before $100$~s and after $850$~s). Overall, while the RT simulations capture the large-scale trend of RSS, the hovering intervals show that UAV orientation changes can cause RSS mismatch. This highlights the need for more realistic modeling of antenna radiation patterns and orientation to improve the accuracy of RT simulations.

\subsection{Possible Uses of Dataset}

Given the RT-based RSS results and dual-channel USRP measured dataset at the towers, this dataset can be used for the following purposes or analysis scenarios.

\noindent\textbf{Benchmarking RT Algorithms:} The dataset provides RSS measurements from the towers and simulated results from the NVIDIA Sionna RT. Thus, the dataset can be used for validation and benchmarking of different RT approaches by comparing the simulation results.

\noindent\textbf{RF Coverage Analysis and Trajectory Planning:} Based on a rural area with dense foliage, the dataset enables the prediction and design of reliable RF coverage by comparing with the desired performance requirements, such as required SNR levels, etc.

\noindent\textbf{UAV Communication Link Analysis:} The dataset captures fluctuation in RSS over the predefined trajectory, allowing for detailed analysis of link performance, e.g., throughput, under realistic propagation conditions.

\noindent\textbf{Statistical Propagation Channel Modeling:} Since the RT results include propagation delay and path coefficient information, the dataset can be used for propagation channel characterization by statistically analyzing this information.

\section{Synthesis, Impact, and Research Outlook}
\label{sec:synthesis_outlook}
This section synthesizes the broader insights, demonstrated impact, and forward-looking implications that emerge from the consolidated analysis of aerial wireless datasets presented in this work. Rather than introducing new datasets, the discussion focuses on what can be learned from jointly examining existing measurements, how these datasets have already informed state-of-the-art research, and how they can guide future experimental design and data-driven studies. Together, these perspectives position the curated datasets not only as archival resources, but as active enablers of reproducible and impactful aerial wireless research.

\subsection{State-of-the-Art Insights from Aerial Wireless Datasets}\label{subsec:SOTA_AERPAW}

By consolidating and jointly examining a broad collection of publicly available aerial wireless datasets, this work provides several state-of-the-art insights that are not apparent when datasets are considered in isolation.

First, the datasets collectively illustrate the increasing diversity of aerial wireless measurement modalities, spanning raw signal-level captures, power-level spectrum measurements, and higher-layer key performance indicators. This diversity reflects a maturation of the field from early propagation-centric studies toward cross-layer and system-level evaluation of aerial networks.

Second, the surveyed datasets highlight the expanding range of platforms and deployment scenarios used in state-of-the-art aerial wireless research, including untethered UAVs, tethered aerial platforms, ground-based sensor networks, and digital twin environments. This breadth underscores the importance of interoperable datasets that support comparative analysis across platforms and environments.

Third, the integration of datasets across multiple frequency bands, from sub-GHz LoRaWAN to mid-band 5G and cellular spectrum, reveals a growing emphasis on multi-band and heterogeneous network evaluation, which is essential for future 5G-Advanced and 6G aerial systems.

Finally, the presence of datasets explicitly designed for machine learning, localization, and ray-tracing validation reflects a state-of-the-art shift toward data-driven and hybrid modeling approaches, where empirical data, simulation, and learning-based methods are jointly employed.

\subsection{Research Enabled by AERPAW Datasets}\label{sec:AERPAW_enabled}
Beyond serving as archival resources, the datasets curated through AERPAW have already enabled a broad range of peer-reviewed research spanning wireless architectures, A2G propagation, localization and sensing, machine learning, security, and cyber-physical system orchestration~\cite{grote2023flypaw, morel2023flynet, kudyba2024uav}.

At the physical and protocol layers, multiple studies have leveraged AERPAW measurement datasets to characterize A2G cellular performance, including full-stack LTE and 5G-NSA experiments with UAV-mounted radios. These datasets have supported analyses of altitude-dependent RSRP, SINR, and throughput behavior, spectrum reuse opportunities, and the limitations of terrestrial cellular infrastructure when serving aerial users, providing empirical grounding for UAV-aware network design and standardization efforts \cite{Drago2023FullStack,asokan2024aerial}.

AERPAW datasets have also been instrumental in digital twin-driven experimentation, enabling controlled, repeatable studies that bridge simulation and field measurements. Several works have utilized AERPAW's digital twin to validate propagation models, emulate UAV-assisted base station deployments, and study discrepancies between emulated and real-world conditions, thereby establishing a workflow for dataset-informed experimentation prior to live flights \cite{Moore2024VTCDT,Gurses2023MILCOM, Javed2026AERPAW}.

In the area of localization, sensing, and data-driven optimization, AERPAW datasets have supported UAV-based RF source localization, trajectory-aware throughput optimization, and machine learning-assisted data collection. These datasets have been used both to benchmark algorithmic performance and to train learning-based models under realistic wireless conditions, including trajectory optimization and adaptive hovering strategies informed by measured channel quality \cite{Sadique2025WiNTECH,grote2023flypaw}.

A growing body of work further demonstrates the use of AERPAW datasets in machine learning and security-focused studies, including federated learning-enabled anomaly detection for drone swarms and secure drone video analytics. These studies rely on realistic, distributed measurement data to evaluate learning accuracy, communication overhead, and robustness against adversarial conditions, highlighting the importance of reproducible, system-level datasets for emerging AI-native aerial networks \cite{Kostage2025ICDCN,DroneVideoAERPAW}.

Finally, AERPAW datasets have enabled cross-domain applications, such as autonomous agricultural monitoring with RF energy-harvesting sensor tags and UAV-assisted IoT data collection. These studies illustrate how aerial wireless datasets extend beyond communications research to support cyber-physical systems, precision agriculture, and edge-enabled sensing workflows \cite{Kudyba2025AgriRF, morel2025task}.

Collectively, these works demonstrate that AERPAW datasets already function as benchmarks, training data, and empirical validation tools across multiple layers of the wireless stack and application domains. This existing body of research underscores the practical impact of the datasets consolidated in this paper and motivates the need for a unified, accessible dataset descriptor to further accelerate reproducible and data-driven aerial wireless research.

\subsection{Future Research Directions Enabled by Aerial Wireless Datasets}\label{subsec:future_AERPAW}

The integrated analysis of existing aerial wireless datasets highlights several directions for future research and dataset development. First, there is a clear need for cross-layer datasets that jointly capture raw signal measurements, intermediate physical-layer metrics, and higher-layer performance indicators within a single, coherent framework. Such datasets would enable holistic evaluation of aerial networks and facilitate cross-layer optimization.

Second, while current datasets span a wide range of frequency bands and platforms, future efforts would benefit from coordinated multi-band and multi-technology measurements, allowing systematic comparison across sub-GHz, mid-band, and emerging high-frequency aerial links. This is particularly relevant as aerial communications evolve toward heterogeneous 5G-Advanced and 6G architectures.

Third, most publicly available datasets are limited in temporal scope. Long-term and large-scale data collection, capturing seasonal, environmental, and mobility-induced variations, remains an open challenge and an important direction for improving model robustness and generalization.

Fourth, the growing adoption of data-driven and machine learning-based techniques motivates the development of ML-ready datasets, including standardized labeling, synchronized metadata, and well-documented training and evaluation splits. Such design considerations would significantly lower the barrier to reproducible and comparative learning-based research.

Finally, future datasets can further strengthen the connection between measurement and modeling by enabling tighter integration between empirical data, ray-tracing simulations, and digital twin environments, supporting hybrid validation workflows and more realistic system evaluation.

\section{Concluding Remarks}\label{sec:conclusion}
Aerial wireless connectivity is becoming an essential enabler of next-generation communication systems, including 5G-Advanced and 6G. This paper presented a diverse collection of open and well-documented datasets from the NSF AERPAW testbed, covering various radio technologies such as 5G, Wi-Fi, and LoRa, and captured using UAVs, helikites, programmable SDR nodes, and commercial UE.
We discussed the technical and regulatory challenges associated with developing programmable aerial wireless platforms to collect such datasets, including the integration of SDRs, real-time localization with centimeter-level precision, testbed-wide time synchronization, and compliance with FAA and FCC requirements. The resulting datasets offer high spatial and temporal resolution, supporting a wide range of research activities in wireless communications, signal processing, and machine learning.

By explicitly addressing the fragmentation, lack of standardization, and limited reproducible access that characterize many existing aerial wireless datasets, this dataset descriptor fills a critical gap between individual measurement campaigns and the broader research community's need for integrated, reusable data resources.
These datasets are curated according to FAIR principles and are intended to support the academic and industrial research community. They enable rigorous evaluation of propagation models, data-driven algorithm design, and performance benchmarking in altitude-varying environments. Beyond cataloging datasets, our work provides a state-of-the-art perspective on how aerial wireless research has evolved toward integrated, data-driven methodologies. The collective analysis of existing datasets reveals emerging trends in platform diversity, frequency utilization, and cross-layer measurement, offering insight into how future datasets and experiments can be designed to better support reproducible and comparative research. Future work will involve expanding the dataset scope to include cooperative and mobile scenarios, incorporating advanced networking features, and aligning with emerging standards to inform data-driven regulatory policy.

\appendices
\section{File structure}\label{app:file_structure}
This document provides detailed descriptions of dataset directory structures, file formats, and supporting conversion and post-processing utilities associated with the datasets presented in this paper. These implementation-level details are consolidated here to preserve full reproducibility while maintaining a clear separation from the main text, which emphasizes dataset scope, measurement content, and representative results. The information in this section enables users to efficiently navigate the released datasets, interpret stored variables, and reproduce the reported analyses using the provided scripts and tools.

\subsection{Wireless I/Q Dataset File Organization}
\label{app:iq_structure}

\noindent\textbf{Directory hierarchy:}
The dataset is organized by measurement altitude, with one top-level directory per UAV altitude. Each altitude-specific directory contains separate subdirectories for I/Q samples and GPS logs, along with optional post-processing outputs generated by the provided scripts.

\begin{verbatim}
<altitude_folder>/
  IQ_samples/
    *.sigmf-data
    *.sigmf-meta
    matfile/        (generated MATLAB files)
  GPS_logs/
    *.sigmf-data
    *.sigmf-meta
    csvfile/        (generated CSV files)
\end{verbatim}

\noindent\textbf{I/Q sample files:}
The \path{IQ_samples} directory contains SigMF-formatted raw I/Q recordings. Each file corresponds to a $20$~ms capture window and is named according to the Unix timestamp representing the measurement start time. Consecutive captures are separated by $100$~ms, resulting in approximately $40\,000$ I/Q samples per file at a $2$~MHz sampling rate. Measurement timestamps are also embedded in the SigMF metadata.

\noindent\textbf{GPS log files:}
The \path{GPS_logs} directory contains SigMF-formatted GPS measurements. When converted to CSV, each record includes latitude, longitude, altitude (in meters), and the corresponding Unix timestamp. GPS measurements are logged at a one-second interval.

\noindent\textbf{Conversion utilities:}
To facilitate post-processing, the repository provides Python scripts that convert SigMF files into commonly used analysis formats:
\begin{itemize}
    \item \path{sigMF2mat_IQ.py}: Converts SigMF I/Q recordings into MATLAB (\path{.mat}) format.
    \item \path{sigMF2csv_GPS.py}: Converts SigMF GPS logs into comma-separated values (\path{.csv}) format.
\end{itemize}

\begin{figure}[!t]
        \centering
        \subfloat[I/Q samples of $20$~milliseconds, converted to MATLAB file using provided script: \colorbox{gray!20}{\texttt{results\_2022\_04\_02\_12\_13\_18\_008.mat}}.]{\fbox{\includegraphics[width=0.4\textwidth]{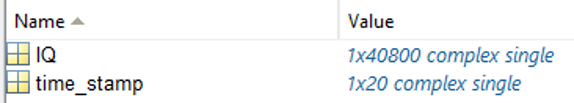}}}
        \vspace{-0.01in}
        \subfloat[GPS Trajectory Data for at \(90\)~m altitude, converted to CSV using provided script: \colorbox{gray!20}{\texttt{2022-04-02\_12\_57\_03\_vehicleOut.csv}}.]{ \fbox{\includegraphics[width=0.4\textwidth]{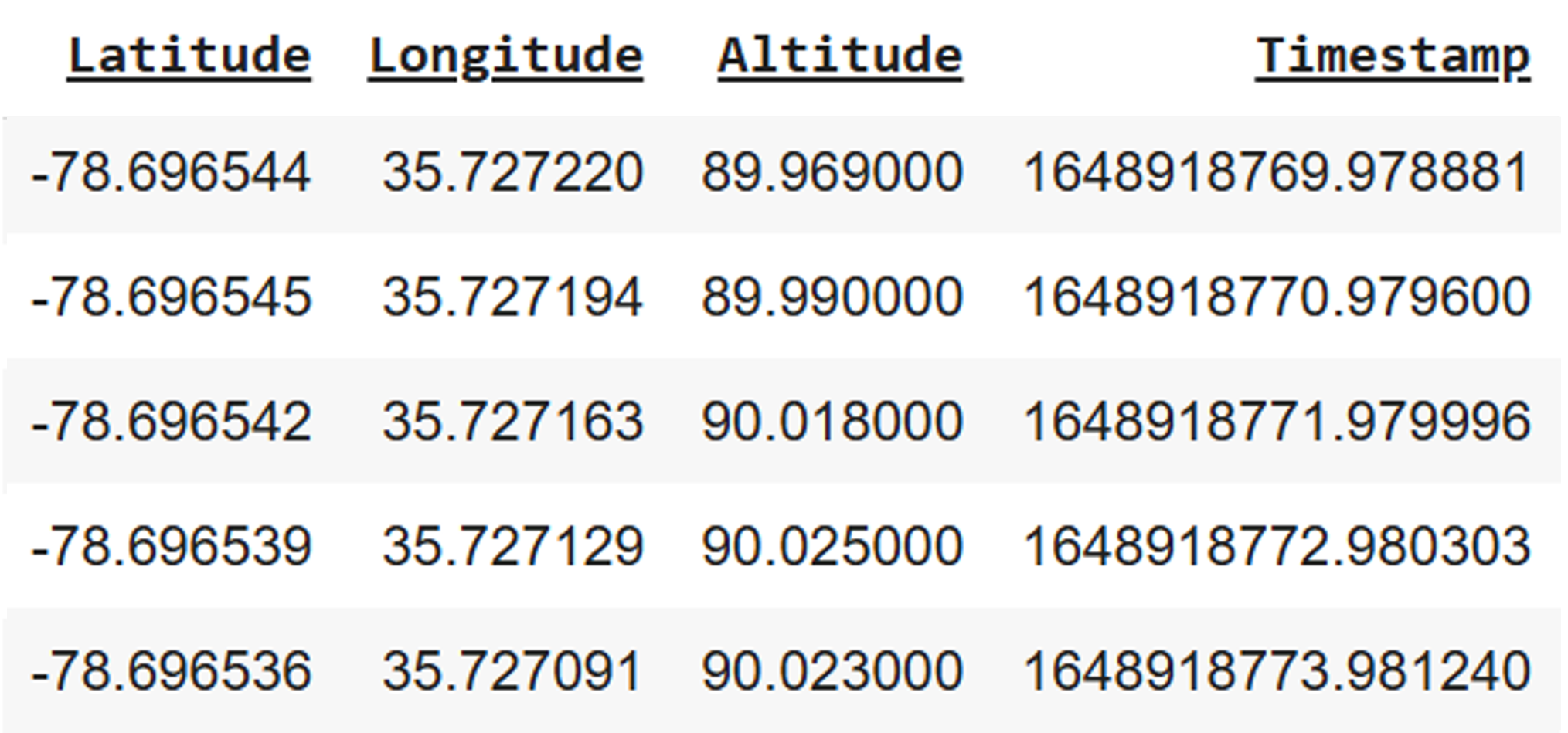}}}
        \vspace{-0.01in}
	\caption{Snapshot of I/Q samples and GPS logs for I/Q Measurement Dataset.}\label{fig:snapshot_lte_iq}
\end{figure}

Fig.~\ref{fig:snapshot_lte_iq} shows a snapshot of a MATLAB file after conversion representing a 20~milliseconds I/Q measurement, along with five rows of the generated CSV file depicting the UAV trajectory over a five-second period.

\noindent A \path{README} file is included in the dataset repository, summarizing dataset usage, measurement methodology, equipment configuration, environment description, and references to post-processing workflows for A2G propagation modeling~\cite{maeng2023lte,maeng2023lte_codeocean}.

\subsection{Wireless Spectrum Dataset File Organization}
\label{app:spectrum_structure}

\noindent\textbf{Directory hierarchy:}
The dataset repository is organized into dedicated directories for power spectrum measurements and GPS logs. Each directory contains SigMF-formatted files indexed by measurement timestamps, with optional subdirectories created when conversion utilities are executed.

\begin{verbatim}
pow_spec/
  *.sigmf-data
  *.sigmf-meta
  matfile/        (generated MATLAB files)
GPS_logs/
  *.sigmf-data
  *.sigmf-meta
  csvfile/        (generated CSV files)
\end{verbatim}

\noindent\textbf{Power spectrum files:}
The \path{pow_spec} directory contains SigMF-formatted power spectrum measurements. When converted to MATLAB format, each file includes frequency vectors (in MHz) and corresponding power values (in dBm). File names correspond to Unix timestamps that indicate the time of spectrum acquisition.

\noindent\textbf{GPS log files:}
The \path{GPS_logs} directory contains SigMF-formatted GPS measurements. After conversion to CSV format, each record includes longitude, latitude, altitude (in meters), and the associated Unix epoch timestamp, enabling time alignment with spectrum measurements.

\begin{figure}[!t]
        \centering
        \subfloat[GPS Trajectory Data for the Packapalooza 2024 Event: \texttt{2024-08-24\_11\_48\_21\_vehicleOut.csv}.]{ \fbox{\includegraphics[width=0.48\textwidth]{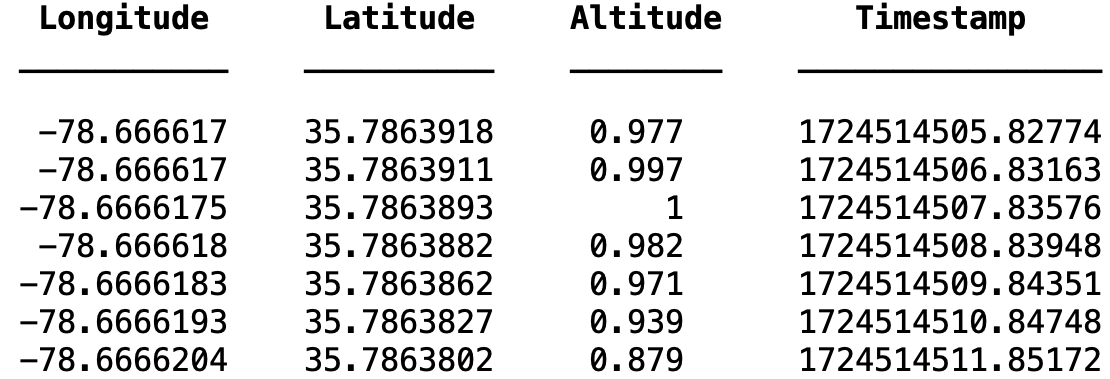}}\label{fig:GPS_S_Pack24}}
        \vspace{-0.01in}
        \subfloat[Frequency and Power Data for the Packapalooza 2024 Event: \texttt{spec\_results\_20240824\_140512.mat}.]{ \fbox{\includegraphics[width=0.48\textwidth]{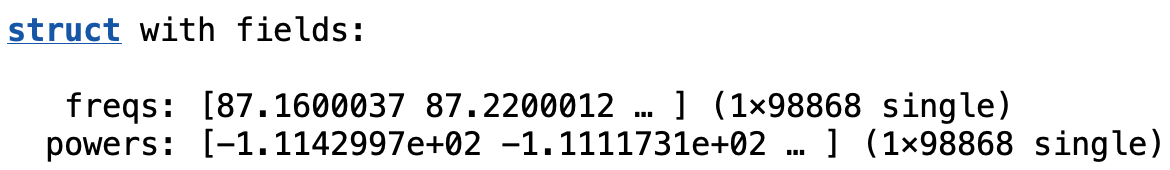}}\label{fig:Freq_pow_S_pack24}}
        \vspace{-0.01in}
	\caption{Snapshot of GPS Log, Frequency, and Power Data for the Packapalooza 2024 Dataset.}\label{fig:snapshot_pack24}
\end{figure}

\noindent\textbf{Conversion utilities:}
To support post-processing and data reuse, the repository provides Python scripts for format conversion:
\begin{itemize}
    \item \path{sigMF2mat_PW.py}: Converts SigMF-formatted power spectrum data into MATLAB (\path{.mat}) files.
    \item \path{sigMF2csv_GPS.py}: Converts SigMF-formatted GPS logs into comma-separated values (\path{.csv}) files.
\end{itemize}

\noindent Fig.~\ref{fig:snapshot_pack24} shows a snapshot of the GPS trajectory data from Packapalooza 2024, along with a MATLAB file generated after conversion that represents the frequency and corresponding measured power for a single sweep.

\noindent A \path{README} file accompanies the dataset and provides an overview of dataset scope, measurement methodology, equipment configuration, usage instructions, metadata description, and curator contact information. This documentation supports transparent reuse and reproducible spectrum analysis.

\subsection{5G Modem-Based SPN Dataset File Organization}
\label{app:5g_modem_structure}

The 5G modem-based dataset provides comprehensive RF/PHY-layer and throughput measurements collected from an Ericsson 5G-NSA network on the AERPAW platform during UAV flight missions conducted on September 15 and October 26, 2023. A portable node equipped with a Quectel 5G modem was mounted on a UAV for these experiments.

The UAV followed a predefined zigzag flight trajectory at a fixed altitude of 30~m, covering the north-west quadrant relative to the serving base station tower. Measurements were repeated for UAV yaw orientations of 315° and 45° to evaluate the impact of antenna orientation. Downlink traffic was generated using an \path{iperf3} application between the portable node and an AERPAW fixed node.

\noindent\textbf{Raw measurement logs:}
\begin{itemize}
    \item \path{Serving_cell_Params_ENDC.csv}: Serving cell configuration and RF measurements including Cell ID, RSRP, SINR, eARFCN, operating band, and subcarrier spacing for both LTE and NR carriers.
    \item \path{Basic_and_Other_Params.csv}: PHY-layer parameters including downlink MCS values and reported CSI metrics such as CQI, PMI, and RI.
    \item \path{<date>_vehicleOut.txt}: UAV telemetry log containing geolocation, orientation, speed, and timestamps.
    \item \path{<date>_iperfclient_log.txt}: Raw \path{iperf3} client log capturing downlink throughput statistics.
\end{itemize}

\noindent\textbf{Post-processed parameter files:}
\begin{itemize}
    \item \path{inputf1_cellid_with_header.csv}: LTE serving cell ID.
    \item \path{inputf2_cellid_with_header.csv}: NR serving cell ID.
    \item \path{inputf1_rsrp_with_header.csv}: LTE RSRP (dBm).
    \item \path{inputf2_rsrp_with_header.csv}: NR RSRP (dBm).
    \item \path{inputf1_sinr_with_header.csv}: LTE SINR (dB).
    \item \path{inputf2_sinr_with_header.csv}: NR SINR (dB).
    \item \path{inputf2_cqi_with_header.csv}: NR CQI values.
    \item \path{inputf2_mcs_with_header.csv}: NR MCS values.
    \item \path{inputf2_ri_with_header.csv}: NR rank indicator values.
    \item \path{input_throughput_with_header.csv}: Downlink throughput (Mbps).
\end{itemize}

All post-processed files include synchronized timestamps and UAV geolocation (longitude, latitude, altitude). The datasets are publicly available via the AERPAW datasets webpage and the Dryad digital repository.

\subsection{Nemo and PawPrints Dataset File Organization}
\label{app:nemo_pawprints_structure}

Depending on the experimental configuration, the Nemo and PawPrints datasets contain up to three logs per measurement campaign: LTE logs, NR logs, and throughput logs. Within each log type, radio or throughput KPIs are merged with UAV location information based on timestamps into a single CSV file.

\noindent\textbf{PawPrints datasets:}
The PawPrints datasets are obtained using Android APIs and include LTE and NR KPIs along with throughput measurements when available.
\begin{itemize}
    \item \path{pawprints_4G_LTE.csv}: LTE KPIs including RSRP, RSRQ, RSSI, Physical Cell Identity~(PCI), TAC, and cell ID.
    \item \path{pawprints_5G_NR.csv}: NR KPIs including synchronization signal RSRP, RSRQ, and RSSI.
    \item \path{pawprints_iperf_throughput.csv}: Downlink throughput measurements obtained using \path{iperf}.
\end{itemize}

\noindent\textbf{Nemo datasets:}
The Nemo datasets provide more comprehensive PHY-layer measurements and include:
\begin{itemize}
    \item \path{nemo_4G_LTE.csv}: LTE RF and cell-level KPIs.
    \item \path{nemo_5G_NR.csv}: NR KPIs including MCS, CQI, and channel rank.
\end{itemize}

All Nemo and PawPrints files include synchronized timestamps and UAV geolocation (latitude, longitude, altitude), enabling spatially resolved analysis of aerial LTE and NR performance.

\subsection{LoRa Propagation Dataset File Organization}
\label{app:lora_structure}

\noindent\textbf{Transmitter-side logs:}
The transmitter logs record packet-level metadata and vehicle state information associated with each LoRa transmission.

\begin{itemize}
    \item Packet identifiers and sequence numbers.
    \item LoRaWAN transmission parameters, including data rate, bandwidth, code rate, spreading factor, and carrier frequency.
    \item Transmission timestamps.
    \item Vehicle geographic location (latitude, longitude, altitude), orientation (yaw, pitch, roll), and velocity.
    \item GPS metadata, including the number of satellites in view.
\end{itemize}

\noindent\textbf{Gateway-side logs:}
Logs collected at the LoRaWAN gateways record reception metrics for successfully received packets.

\begin{itemize}
    \item RSSI.
    \item SNR.
    \item Packet reception timestamps.
    \item Reception frequency channel and RF chain identifier.
    \item Gateway identifier and geographic location.
\end{itemize}

\noindent\textbf{Derived and auxiliary files:}
In addition to raw transmitter and gateway logs, the dataset includes auxiliary CSV files generated during post-processing:

\begin{itemize}
    \item \path{failed_tx_packages.csv}: Records of failed or missing packet transmissions.
    \item \path{gateway-dataRate-Table.csv}: Summary of data rates grouped by receiving gateway.
\end{itemize}

\noindent\textbf{Signal-level data and utilities:}
Raw signal measurements, including RSSI and SNR, are provided in SigMF format. Example Python scripts are included in the repository to convert SigMF files into CSV format and to generate visualization outputs from the processed data.

\subsection{A2G Channel Sounding Dataset File Organization}
\label{app:multipath_structure}

\noindent\textbf{Raw signal files:}
Each channel sounding capture is stored using the SigMF format and consists of two files:

\begin{itemize}
    \item \path{.sigmf-data}: A binary file containing complex I/Q samples stored as 32-bit floating-point values in little-endian format.
    \item \path{.sigmf-meta}: A JSON-formatted metadata file compliant with the SigMF specification.
\end{itemize}

\noindent\textbf{Metadata contents:}
In addition to standard SigMF fields, the metadata files include custom entries describing:
\begin{itemize}
    \item UAV GPS coordinates and altitude at the time of capture.
    \item Zadoff--Chu waveform parameters, including sequence length and root index.
    \item Measurement frequency, timestamp alignment, and synchronization parameters.
    \item Flight configuration and experimental context.
\end{itemize}

\noindent\textbf{Post-processing utilities:}
The associated GitHub repository provides post-processing software for channel analysis. This includes Python scripts and a Jupyter notebook that operate either directly on SigMF-formatted files or on compressed intermediate data representations.

\begin{itemize}
    \item \path{PostProcess.ipynb}: Implements correlation-based channel impulse response extraction, path loss computation, and visualization routines used to generate the published results.
\end{itemize}

\noindent
These utilities enable users to reproduce the representative figures and analyses presented in the main text~\cite{gurses_sichitiu_a2g_uav}.

\subsection{TDOA-Based UAV Localization Dataset File Organization}
\label{app:tdoa_structure}

\noindent\textbf{Primary dataset archive:}
The datasets reported in~\cite{dickerson2025tdoa,dickerson2025aerpaw} are distributed as a compressed archive containing one CSV file per UAV flight. File names encode the flight altitude, signal bandwidth, and recording date.

\begin{itemize}
    \item Example filename: \path{40m_1.25MHz_7.15.24.csv}, corresponding to a flight at 40~m altitude with 1.25~MHz signal bandwidth recorded on July~15,~2024.
\end{itemize}

\noindent\textbf{CSV file contents:}
Each CSV file includes the following fields:
\begin{itemize}
    \item Center frequency of the transmitted signal.
    \item Estimated UAV latitude and longitude.
    \item Ground-truth (GT) latitude, longitude, and altitude.
    \item Localization performance metrics, including the degree of cross-correlation (RHO) and circular error probability (CEP).
    \item Timestamps corresponding to each localization estimate.
    \item Binary LoS indicators (\path{LOStoLW2-5}) describing LoS visibility to AERPAW towers LW2 through LW5, where a value of \path{1} denotes LoS and \path{0} denotes NLoS.
\end{itemize}

\noindent
If both latitude and longitude estimates are recorded as zero, the corresponding entry indicates that the localization algorithm failed to produce a valid position estimate for that measurement instance.

\noindent\textbf{Processing utilities:}
The dataset archive includes a MATLAB helper script:
\begin{itemize}
    \item \path{KeysightRTDOALocalizationforFlights.m}: Processes a selected CSV file to generate UAV trajectory visualizations and localization performance metrics.
\end{itemize}

\noindent\textbf{Related datasets:}
The datasets reported in~\cite{bhattacherjee2022experimental,uditalocalization} follow a similar format but differ in two aspects: (i) LoS indicator variables are not included, and (ii) measurement and ground-truth data are provided in separate CSV files (\path{Inspiron_backup.csv} and \path{GPS_Flight1_backup.csv}, respectively).

\subsection{AFAR Challenge Dataset File Organization}
\label{app:afar_structure}

The AFAR dataset is organized hierarchically to reflect team participation, experimental environment, and UGV transmitter placement. For each of the five finalist teams, data are provided for three UGV locations (Loc-1, Loc-2, Loc-3) and for both the DT simulation environment and the real-world AERPAW testbed.

\noindent\textbf{Directory hierarchy:}
\begin{itemize}
    \item Top-level directories correspond to team identifiers (e.g., \path{288}, \path{300}, \path{301}).
    \item Each team directory contains two subfolders:
    \begin{itemize}
        \item \path{development}: Digital twin simulation data.
        \item \path{testbed}: Real-world AERPAW testbed data.
    \end{itemize}
    \item Within each environment folder, location-specific subfolders are provided:
    \begin{itemize}
        \item \path{loc-1}, \path{loc-2}, and \path{loc-3}.
    \end{itemize}
\end{itemize}

\noindent\textbf{Core files in each testbed location folder:}
Each location-specific folder within the \path{testbed} directory contains the following files:

\begin{itemize}
    \item \path{power_log.txt}: Contains RSS measurements captured by the UAV receiver. Each row includes a timestamp (microsecond resolution), a sample index, and the measured RSS in dB. The sample index column is not used for signal analysis.
    \item \path{quality_log.txt}: Contains RSQ measurements with the same format as \path{power_log.txt}.
    \item \path{log.csv}: Time-synchronized UAV navigation data, including GPS coordinates (latitude, longitude, altitude), speed, heading, and satellite metadata.
    \item \path{angles.mat}: UAV orientation data, including roll, pitch, and yaw angles.
\end{itemize}

\noindent
Table~\ref{tab:power-log-format} provides an example snippet of the \path{power_log.txt} file format, and Table~\ref{tab:log-sample} illustrates representative entries from the \path{log.csv} file.

\begin{table}[t]
\caption{Sample Format of \texttt{power\_log.txt}}
\label{tab:power-log-format}
\centering
\begin{tabular}{@{}l c r@{}}
\toprule
\textbf{Timestamp} & \textbf{Index} & \textbf{RSS (dB)} \\
\midrule
2023-12-13 13:45:34.041027 & 0000000 & -34.9675 \\
2023-12-13 13:45:34.072347 & 0000004 & -40.7695 \\
2023-12-13 13:45:34.105068 & 0000010 & -48.0318 \\
\bottomrule
\end{tabular}
\end{table}


\begin{table*}[tbh!]
\centering
\caption{Sample entries from \texttt{log.csv} showing UAV navigation and GPS metadata.}
\label{tab:log-sample}
\setlength{\tabcolsep}{6pt}
\renewcommand{\arraystretch}{1.15}
\begin{tabular}{cccccccccccccc}
\hline
TimeUS & Status & GMS & GWk & NSats & HDop & Lat & Lng & Alt & Spd & GCRs & VZ & Yaw & U \\
\hline
2166080919 & 5 & 3.27E+08 & 2292 & 14 & 0.77 & 35.727371 & -78.6962127 & 112.65 & 0.03985 & 72.47443 & 0.007 & 0 & 1 \\
2166280903 & 5 & 3.27E+08 & 2292 & 14 & 0.77 & 35.727371 & -78.6962128 & 112.66 & 0.028284 & 261.8699 & -0.059 & 0 & 1 \\
2166460908 & 5 & 3.27E+08 & 2292 & 14 & 0.77 & 35.7273709 & -78.6962131 & 112.66 & 0.04639 & 172.5686 & 0.007 & 0 & 1 \\
2166660988 & 5 & 3.27E+08 & 2292 & 14 & 0.77 & 35.727371 & -78.6962132 & 112.65 & 0.061294 & 5.61758 & 0.21 & 0 & 1 \\
2166860968 & 5 & 3.27E+08 & 2292 & 14 & 0.77 & 35.727371 & -78.6962131 & 112.66 & 0.023087 & 252.3499 & -0.189 & 0 & 1 \\
2167060890 & 5 & 3.27E+08 & 2292 & 14 & 0.77 & 35.7273709 & -78.6962132 & 112.68 & 0.011664 & 329.0363 & -0.148 & 0 & 1 \\
2167280929 & 5 & 3.27E+08 & 2292 & 14 & 0.77 & 35.7273708 & -78.6962132 & 112.69 & 0.008485 & 315 & -0.13 & 0 & 1 \\
\hline
\end{tabular}
\end{table*}

\subsection{UAV Signal Classification Dataset Organization and Utilities}
\label{app:uas_classification_structure}

\noindent\textbf{Dataset contents:}
The UAV signal classification dataset consists of processed RF signal recordings and associated metadata stored in MATLAB \path{.mat} files. The dataset includes RF signals from 17 commercial UAV controllers representing eight manufacturers, as summarized in Table~\ref{tab:UAV_Catalogue}. Metadata for each RF sample, including controller identity and acquisition parameters, are summarized in Table~\ref{tab:UAV_Metadata}.

\begin{table}[t]
\centering
\caption{UAV platforms and remote controllers included in the drone signal classification dataset.}
\label{tab:UAV_Catalogue}
\begin{tabular}{|c|c|c|c|}
\hline
\textbf{Make} & \textbf{Model} & \textbf{Make} & \textbf{Model} \\
\hline
 & Inspire 1 Pro & & DX5e \\
 & Matrice 100 & & DX6e \\
DJI & Matrice 600\textsuperscript{\textbf{}} & Spektrum & DX6i \\
 & Phantom 4 Pro & & JR X9303 \\
 & Phantom 3 & & \\
\hline
Futaba & T8FG & Graupner & MC-32 \\
\hline
 HobbyKing & HK-T6A & FlySky & FS-T6 \\
\hline
Turnigy & 9X & Jeti Duplex & DC-16 \\
\hline
\end{tabular}
\end{table}

\begin{table}[t]
\centering
\caption{Metadata of UAS signal classification dataset.}
\label{tab:UAV_Metadata}
\begin{tabular}{|l|r|}
\hline
\textbf{Description} & \textbf{Value} \\
\hline
Number of drone controllers & 17 \\
\hline
Sampling frequency & 20 GSa/s \\
\hline
Center frequency & 2.4 GHz \\
\hline
Number of signals/drone RC & $\sim$1000 \\
\hline
Number of samples/signal & 5 million \\
\hline
Time duration/signal & 0.25 ms \\
\hline
Average data size/signal & 7 MB \\
\hline
Dataset size & 124 GB \\
\hline
Data format & .mat \\
\hline
\end{tabular}
\end{table}

\begin{figure}
\includegraphics[width=\linewidth]{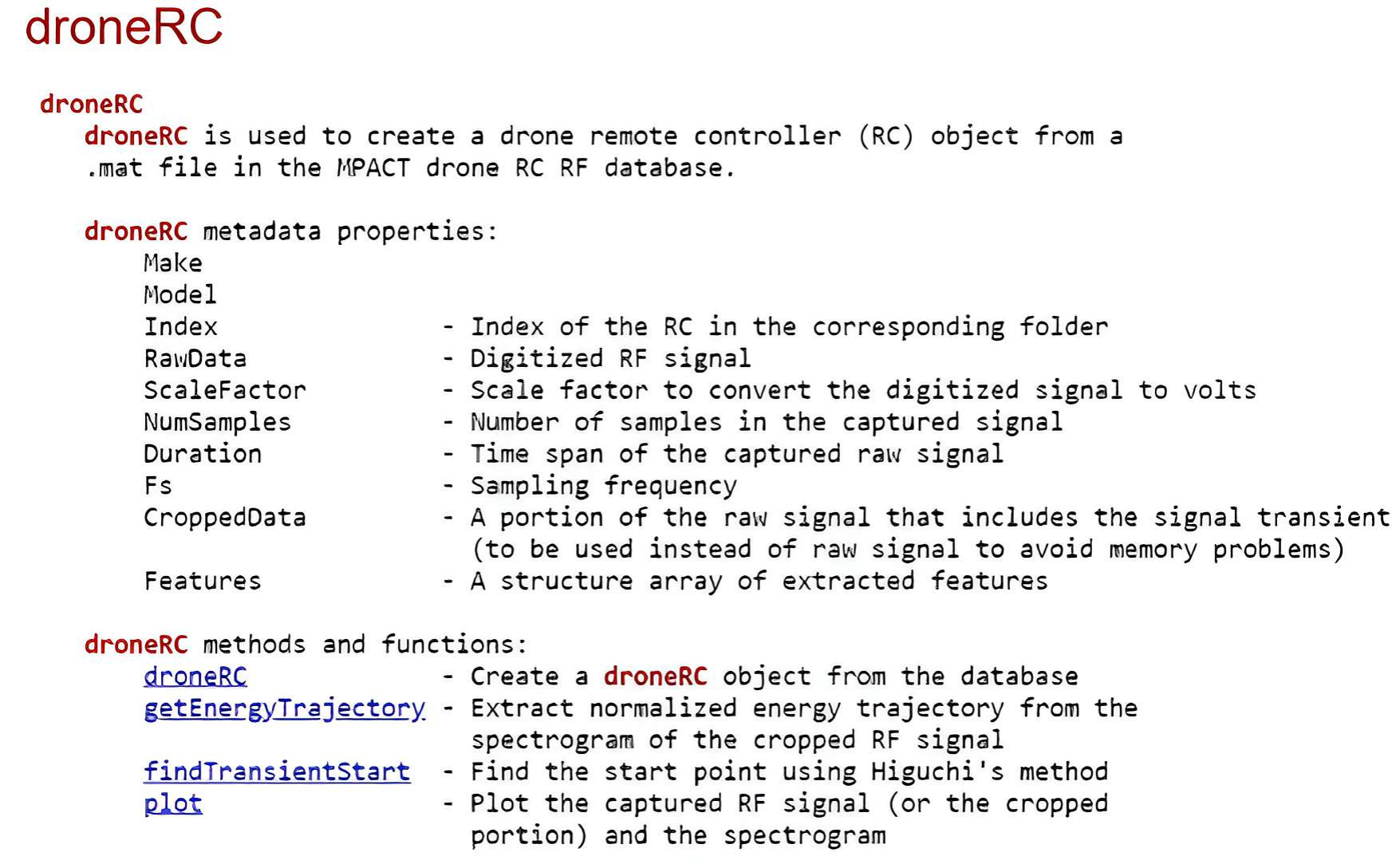}
	\centering
	\caption{The \textit{droneRC.m} class is used to extract both RF signal data and associated metadata from specific .mat files in the UAS signal classification dataset~\cite{UAV_RF_Dataset_IEEEPortal}.}
	\label{fig1:DroneRCm}
\vspace{-0.3cm}
\end{figure}

\noindent\textbf{Signal extraction and visualization:}
The dataset provides a MATLAB class definition, \path{droneRC.m}, which enables object-oriented access to individual UAV controller RF signals and their associated metadata. This class supports extraction, visualization, and preprocessing of RF samples stored in the dataset. Fig.~\ref{fig1:DroneRCm} illustrates the use of the \path{droneRC} class for accessing RF signals and metadata.

\begin{figure}
\includegraphics[width=\linewidth]{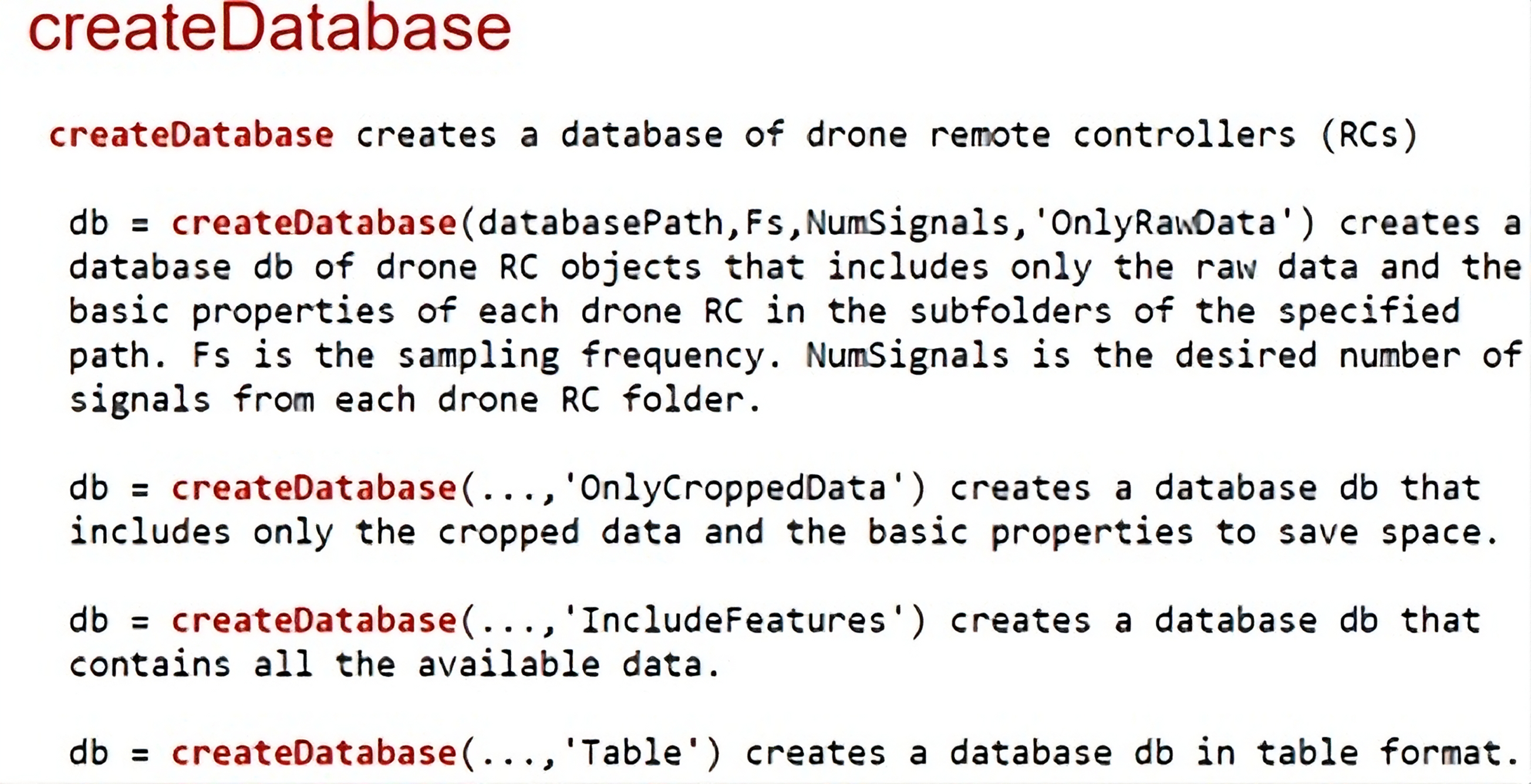}
	\centering
	\caption{The \textit{createDatabase.m} function is used to create a table or matrix of droneRC objects from the UAS signal classification dataset. The database generated can be used to benchmark different UAV RF signal detection and classification algorithms in the presence or absence of interference~\cite{UAV_RF_Dataset_IEEEPortal}.}
	\label{fig1:createDatabasem}
\end{figure}

\noindent\textbf{Dataset construction for experiments:}
To support flexible experimental design, the dataset includes a MATLAB utility function, \path{createDatabase.m}, which enables construction of custom datasets from selected UAV controller signals. The function can generate databases in matrix or table format and allows users to specify the number of samples, selected controllers, and feature representations. Fig.~\ref{fig1:createDatabasem} illustrates the database construction process.

\noindent
These utilities enable reproducible benchmarking of UAV RF signal detection and classification algorithms under controlled experimental conditions, including evaluation across SNR levels, controller diversity, and interference scenarios.

\subsection{Trajectory-Aware RSRP and Throughput Dataset File Organization}
\label{app:trajectory_structure}

\noindent\textbf{Directory organization:}
The dataset is organized into three primary directories corresponding to the measurement methodology and trajectory type:

\begin{itemize}
    \item \path{fixed_trajectory_rsrp_emulation}: Emulated RSRP measurements collected along predefined UAV trajectories. Each record includes timestamps, UAV position (latitude, longitude, altitude), orientation (pitch, roll, yaw), and RSRP values.
    \item \path{fixed_trajectory_rsrp_simulation}: Simulated RSRP measurements generated for the same fixed UAV trajectories. Each record includes timestamps, UAV position, and RSRP values.
    \item \path{autonomous_trajectory_rsrp_simulation}: Simulated RSRP measurements collected along autonomous UAV trajectories subject to geofencing constraints. Each record includes timestamps, UAV position, and RSRP values.
\end{itemize}

\noindent\textbf{CSV file structure:}
Each CSV file corresponds to a UAV flight and contains time-indexed measurements including UAV position (longitude, latitude, altitude), radio metrics such as RSRP and SNR, and application-layer throughput where applicable. Table~\ref{tab:uav_measurement_sample} illustrates a representative sample of the CSV structure.

\noindent\textbf{Post-processing utilities:}
The dataset repository includes MATLAB and Python scripts used to generate the figures and analyses presented in the paper. These scripts support visualization of RSRP as a function of UAV location, distance to base stations, and time, as well as throughput analysis for fixed and autonomous trajectories.

\noindent
These utilities enable reproducible preparation of all representative figures and statistical results reported in~\cite{Hossen2025}.

\setlength{\tabcolsep}{2pt} 

\begin{table}[t]
\caption{UAV measurement dataset sample with RSRP, SNR, and data rate.}
\label{tab:uav_measurement_sample}
\centering
\scalebox{0.92}{
\begin{tabular}{@{}l r r r r r r@{}}
\toprule
\makecell{\textbf{Time} \\ \textbf{(s)}} & 
\makecell{\textbf{Lon.} \\ \textbf{(deg)}} & 
\makecell{\textbf{Lat.} \\ \textbf{(deg)}} & 
\makecell{\textbf{Alt.} \\ \textbf{(m)}} & 
\makecell{\textbf{RSRP} \\ \textbf{(dBm)}} & 
\makecell{\textbf{SNR} \\ \textbf{(dB)}} & 
\makecell{\textbf{Rate} \\ \textbf{(Mbps)}} \\
\midrule
25-03-28 03:36:50 & -78.69627 & 35.72748 & 0.00  & -53.00 & 37.00 & 7.35 \\
25-03-28 03:36:51 & -78.69627 & 35.72748 & 0.47  & -53.00 & 37.00 & 7.35 \\
\multicolumn{7}{c}{\ldots} \\
25-03-28 03:37:05 & -78.69627 & 35.72748 & 30.0  & -54.00 & 36.00 & 7.29 \\
25-03-28 03:37:06 & -78.69622 & 35.72748 & 30.0  & -53.00 & 37.00 & 7.35 \\
25-03-28 03:37:07 & -78.69615 & 35.72745 & 30.0  & -52.00 & 38.00 & 7.40 \\
\bottomrule
\end{tabular}
}
\end{table}
    
\subsection{Ray-Tracing Simulation and Measurement Comparison Dataset File Organization}
\label{app:raytracing_structure}

The dataset is distributed as a collection of data folders and MATLAB scripts labeled using the prefixes ``D'' (data) and ``C'' (code). The organization supports reproducible comparison between measured RSS and ray-tracing simulation results.

\noindent\textbf{Data components:}
\begin{itemize}
    \item \path{D1_RSS_Measurements}: Contains subfolders corresponding to RSS measurements (in dB) collected at five base station locations (LW1--LW5). Each subfolder also includes the GPS information associated with the predefined UAV trajectory.
    \item \path{D2_RT_RSS_results_data.mat}: Stores RT simulation outputs for the predefined UAV trajectory, including path coefficients and propagation delays at each UAV position.
    \item \path{D3_RT_Measurement_RSS_postprocessed_data.mat}: Contains the time duration of the measurement campaign, RSS values derived from RT simulation, and calibrated RSS values obtained from measurements.
\end{itemize}

\noindent\textbf{Post-processing scripts:}
\begin{itemize}
    \item \path{C1_trajectory_altitude.m}: Plots the predefined UAV trajectory and altitude as a function of time.
    \item \path{C2_RSS_postprocessing.m}: Performs post-processing and calibration of RT and measured RSS data. This step is optional for reproducing the published results, as the calibrated outputs are included in \path{D3_RT_Measurement_RSS_postprocessed_data.mat}.
    \item \path{C3_RSS_comparison.m}: Generates comparative plots of measured and RT-simulated RSS values for each base station, corresponding to the results presented in the main text.
\end{itemize}

\noindent
These components enable full reproduction of the reported ray-tracing comparison results.

\bibliographystyle{IEEEtran}
\bibliography{references.bib, aerpaw.bib}

\end{document}